\documentclass[12pt,letterpaper]{article}
\usepackage{comment}
\usepackage[T1]{fontenc}
\input{glyphtounicode}
\usepackage[utf8]{inputenc}
\usepackage[defaultsups]{newtxtext}
\usepackage[margin=1in]{geometry}
\usepackage{amsmath,amsthm,mathtools}
\usepackage{newtxmath}
\usepackage{booktabs,array,tabularx,enumitem,microtype,graphicx}
\usepackage[round,authoryear]{natbib}
\usepackage[hidelinks,bookmarksnumbered,hypertexnames=false]{hyperref}
\usepackage{caption,float}
\usepackage{xcolor}
\usepackage{etoolbox,titlesec}
\usepackage{tikz}
\usetikzlibrary{arrows.meta}
\setlist[enumerate]{itemsep=0pt,parsep=0pt,topsep=2pt,partopsep=0pt,leftmargin=*}
\setlist[itemize]{itemsep=0pt,parsep=0pt,topsep=2pt,partopsep=0pt,leftmargin=*}
\titlespacing{\section}{0pt}{4pt plus 2pt minus 1pt}{4pt plus 1pt minus 1pt}
\titlespacing{\subsection}{0pt}{4pt plus 2pt minus 1pt}{4pt plus 1pt minus 1pt}
\titlespacing{\subsubsection}{0pt}{4pt plus 2pt minus 1pt}{4pt plus 1pt minus 1pt}
\makeatletter
\g@addto@macro\th@plain{\thm@preskip3pt\relax\thm@postskip3pt\relax}
\g@addto@macro\th@definition{\thm@preskip3pt\relax\thm@postskip3pt\relax}
\renewenvironment{proof}[1][\proofname]{%
  \par\pushQED{\qed}\normalfont\topsep3pt\relax\trivlist
  \item[\hskip\labelsep\itshape #1\@addpunct{.}]\ignorespaces
}{\popQED\endtrivlist\@endpefalse}
\makeatother
\appto\normalsize{%
  \abovedisplayskip=4pt plus 2pt minus 1pt
  \belowdisplayskip=4pt plus 2pt minus 1pt
  \abovedisplayshortskip=0pt plus 2pt
  \belowdisplayshortskip=2pt plus 1pt minus 1pt}
\AtBeginDocument{\normalsize\setlength{\jot}{2pt}}
\apptocmd{\titlepage}{\thispagestyle{plain}\setcounter{page}{1}\linespread{1}\selectfont}{}{\PackageError{preamble}{Could not adapt the title page}{Check the document class.}}
\newtheorem{theorem}{Theorem}
\newtheorem{proposition}{Proposition}
\newtheorem{appendixlemma}{Lemma}[section]
\newtheorem{lemma}{Lemma}
\newtheorem{corollary}{Corollary}
\newtheorem{claim}{Claim}
\theoremstyle{definition}
\newtheorem{definition}{Definition}
\newtheorem{example}{Example}
\newtheorem{assumption}{Assumption}
\newcommand{\OO}{\emptyset}
\newcommand{\MG}{\operatorname{MG}}
\newcommand{\OA}{\operatorname{OA}}
\hypersetup{pdftitle={Reserve Systems in Dynamic Settings: Account Design and Inherited Rights},pdfauthor={},pdfsubject={Reserve system design, version 181},pdfcreator={},pdfproducer={}}
\newtheorem{appendixdefinition}{Definition}[section]
\newcommand{\TopSet}{\operatorname{Top}}
\newcommand{\CC}{\mathrm{C}}
\makeatletter
\newcommand{\choiceheading}[2]{
  \par\addvspace{\medskipamount}
  \noindent\textbf{#1} #2\par\nobreak
  \@afterheading
}
\makeatother

\newcounter{choiceprocedure}
\newcounter{choicestep}[choiceprocedure]
\newenvironment{choicesteps}{%
  \par\stepcounter{choiceprocedure}\begin{list}{}{%
    \setlength{\leftmargin}{1.25em}\setlength{\rightmargin}{0pt}%
    \setlength{\itemindent}{-1.25em}\setlength{\listparindent}{0pt}%
    \setlength{\labelwidth}{0pt}\setlength{\labelsep}{0pt}%
    \setlength{\topsep}{2pt}\setlength{\partopsep}{0pt}%
    \setlength{\itemsep}{3pt}\setlength{\parsep}{2pt plus 1pt minus 1pt}}%
}{\end{list}}
\newcommand{\choicestep}{\item\refstepcounter{choicestep}\textbf{Step \thechoicestep:}\enspace}

\makeatletter
\newcommand{\medsetoperator}[2]{\vcenter{\hbox{\scalebox{1.2}{$\m@th#1#2$}}}}
\DeclareRobustCommand{\medcap}{\mathbin{\mathpalette\medsetoperator\cap}}
\DeclareRobustCommand{\medcup}{\mathbin{\mathpalette\medsetoperator\cup}}
\makeatother

\title{Reserve Systems in Dynamic Settings:\\[.10in]
Account Design and Inherited Rights\thanks{We thank Parag A. Pathak and Edwin Song for initial discussions.\\ 
\emph{Use of AI tools.} We began using AI tools for this paper in September 2026, when we already had a fairly mature first draft containing versions of all the main theorems and their proofs. Of the applications, only the Chicago Public Schools waiting lists and allocation of COVID-19 vaccines and monoclonal antibodies were worked out in detail. We also knew several affirmative action systems in Chile, India, and Brazil fit the framework. We then used OpenAI's ChatGPT Astra and Anthropic's Claude Fable to search for further direct applications and possible new results. To guide this search, we uploaded \citet{Sonmez2023MMD} and prompted the tools to follow the minimalist market design framework. 
The search identified dozens of additional direct applications, including many from non-English sources, and some new application-specific formal results. Many of these applications are in the main text. Most of the new results, and some further applications, are in the Supplemental Appendix. The formal results proposed by these tools needed substantial work. We checked and revised their statements and proofs, included missing arguments, and expanded explanations for an economics readership. We verified every formal result in the paper ourselves. For the applications, we checked the institutional claims against the cited primary sources.
We also used these tools to edit the text, to revise the exposition as the applications were added, and to locate and cross-check official sources.
Throughout, the tools worked under our direction, much as a capable research assistant would. We take full responsibility for the content of the paper.}}

\author{Tayfun S\"onmez\thanks{Department of Economics, Boston College, email:~\href{mailto:sonmezt@bc.edu}{\texttt{sonmezt@bc.edu}}.}\qquad M. Utku \"Unver\thanks{Department of Economics, Boston College, email:~\href{mailto:unver@bc.edu}{\texttt{unver@bc.edu}}.}}

\date{November 16, 2026}

\begin{document}
\hypersetup{pageanchor=false}
\begin{titlepage}
\makeatletter
\def\@maketitle{%
  \begingroup\centering\vspace*{.18in}
  {\LARGE\@title\par}\vspace{.24in}
  {\large\@author\par}\vspace{.22in}
  {\@date\par}\vspace{.18in}
  \endgroup
}
\makeatother
\maketitle

\begin{abstract}
Research on reserve systems mainly considers static settings, formulating and studying choice rules for a given allocation problem. Dynamic settings are different in that the problem is not given. Over time, more positions become available, and each time the institution must build the current problem from its past. We distinguish the ledger, which stores information about earlier allocations, the account, which transforms these data into the current problem, and the choice rule, which selects an outcome for that problem. The same choice rule can implement very different policies under different accounts. If each new batch is assigned while holding earlier recipients and their categories fixed, then as batches shrink, over-and-above, the most generous reserve policy we study, can select nearly the same recipients as minimum guarantee, the most conservative one. Cumulative implementation uses a different account, which reconsiders the categories of earlier recipients, and prevents this compression. We characterize it by resource-release invariance. With arrivals or changes in priorities, the ordinary choice rule might reject an earlier recipient whose appointment has to be honored. We introduce and characterize retain, reassign, and boost (RRB) choice rules, under which every incumbent retains her appointment and can have her category changed. Combining RRB with individual-proposing deferred acceptance gives a strategy-proof mechanism under which each incumbent keeps her appointment or receives one she prefers. Applications include waiting lists in school and university admissions, public recruitment, and court-ordered corrections.
\end{abstract}
\end{titlepage}
\hypersetup{pageanchor=true}\setcounter{page}{2}

\section{Introduction}\label{sec:introduction}

Institutions use reserve policies to balance priority claims with affirmative action for disadvantaged groups.
Universities reserve admission places, public employers reserve positions, and health programs reserve access to scarce medical resources and services.
Reserve systems typically divide positions into an open category, where anyone is eligible, and reserved categories, where only members of the corresponding protected group are eligible.
With one reserved category, the same division of positions can implement different levels of protection depending on how the categories are processed. Filling reserved positions first implements a minimum guarantee \citep{HYY2013, EcheniqueYenmez2015}. If the group's high-priority members would obtain at least the reserved number of positions under selection by priority alone, this policy does not increase the group's total allocation. Filling open positions first instead allows eligible applicants to receive those positions on priority without using the reserve, leaving reserved positions for additional members of the group \citep{DurEtAl2018}. The reserve thus provides protection over and above the group's selection through the open category. These are the least and most generous policies among the priority-respecting reserve rules we consider, with intermediate processing arrangements providing protection between them \citep{DurEtAl2018}.

While the literature on reserve systems largely focuses on static settings,
reserve systems are commonly used in dynamic settings, where positions become available and applicants arrive at different times, 
and earlier allocations affect later decisions or create rights that must be honored.
And yet, little is known about how these systems behave in such settings, or whether the range of policies they offer, from minimum guarantee to over-and-above, carries over.
Perhaps one reason is that, in many real-world applications, a dynamic problem is simply treated as a sequence of stand-alone problems, with the announced choice rule applied at each decision.
Our starting point is to ask how this common practice influences outcomes. 
The answer is that it can move the implemented policy far from the intended one. 
Even when the institution applies its chosen rule correctly at every step, the eventual outcome need not deliver the policy for which decision makers chose that rule.
We study how the treatment of previous allocations causes such failures, and how reserve systems can be designed to avoid them without violating commitments made to past recipients.

Take the allocation of COVID-19 vaccines early in the pandemic. The National Academies of Sciences, Engineering, and Medicine (NASEM) recommended reserving a fraction of the federal vaccine supply for areas with high social vulnerability, in addition to the shares those areas would receive based on population \citep{NASEM2020}. The aim was to protect vulnerable areas by giving them a share over-and-above their proportional allotment.
Supply arrived over a period of months, yet the recommendation did not address how this policy should be implemented when a jurisdiction receives its supply over a period rather than all at once. In practice,
several jurisdictions applied the over-and-above rule to each shipment as it arrived, among the applicants not yet served. Can this practice preserve the intended boost?

The short answer is no. Repeating the choice rule in this way removes earlier recipients from consideration and fixes the categories through which they were selected. An applicant selected through a reserved category continues to be counted against her group's reserve even if her priority would entitle her to selection in the open category once everything released so far is taken into account. This prevents the reserve policy from providing an additional benefit to another member of the group. As a result, this treatment of earlier allocations largely eliminates the difference between the over-and-above and the minimum guarantee choice rules.
In Theorem~\ref{thm:compression}, we show that this anomaly is not peculiar to the over-and-above rule. With one reserved category, we bound the 
difference between any viable category-locked batch procedure and repeated minimum guarantee. As batches become smaller, the scope for differences between these policies narrows.

Our remedy is to apply the choice rule, at each decision, to the initial population and all capacity released so far. We call this \emph{cumulative implementation}.
As a key feature of this design, earlier category assignments are not locked. For example, an earlier recipient whose priority now suffices for the open category can be reassigned from the reserved category to the open category,
leaving a reserved place for another eligible applicant.
When the choice rule satisfies capacity monotonicity, this flexibility keeps every earlier recipient while maintaining the policy at the cumulative capacities.
The institution can allocate resources as supply becomes available without knowing when further releases will come.

The difference between repeated batch implementation and cumulative implementation is the input to the choice rule, that is, the problem it is applied to.
To make the construction of that problem an explicit part of the design, we introduce a \emph{ledger--account--choice} framework in which the ledger maintains a record of allocations and other events,
the account forms the problem for the choice rule from these events,
and the choice rule selects an allocation for that problem.
The implicit account used in repeated batch implementation excludes earlier recipients and their positions from current choice and holds their category assignments fixed.
In contrast, cumulative implementation uses the history to reconsider category assignments of earlier recipients.
This separation also allows us to study dynamic implementation through a sequence of static choice problems linked by institutional history.

We say that a process satisfies \emph{resource-release invariance} if the timing and division of resource releases do not affect the outcome. When it fails, an administrator can interfere with outcomes by altering the timing and division of releases without changing the stated policy. Theorem~\ref{thm:reattribution} shows that a procedure satisfies resource-release invariance if and only if it is the cumulative implementation of its one-shot choice rule, with capacity monotonicity ensuring that earlier recipients remain selected. 

France's Parcoursup admissions system, which we examine in Section~\ref{app:parcoursup}, shows cumulative implementation in practice. Each program turns its ranking of applicants into a calling order, the order in which offers are made. The calling order is built so that the minimum share for grant recipients is satisfied at every point. As a result, whatever number of offers the program ends up making, the applicants who receive them are exactly those the minimum guarantee choice rule would select for that number of places.
The calling order is computed once and is not revised when applicants refuse offers. Importantly, a grant recipient who refuses still counts toward the minimum. That is, the guarantee is about offers made, not enrollment. Whether the policy counts offers made or offers held is an account decision, separate from the choice rule.

In richer settings, earlier allocations can affect eligibility, available resources, and rights even when the current decision is considered on its own. Account design raises the following questions:
\begin{enumerate}[label=(\roman*)]
\item Which individuals remain relevant, and which are considered for a current allocation?
\item Which resources or allocation opportunities remain relevant, and which are available for use?
\item What does the policy count, over what assessment period, and can several assessments apply to the same allocation?
\item Is a category attached to an identified resource, an allocation opportunity, or an individual’s assignment?
\item Which rights from earlier allocations must still be honored?
\end{enumerate}

Even when two institutions use the same choice rule, different answers to these questions can reflect different policy objectives, as when an institution counts admission opportunities rather than offers currently held. An alternative would be to build the history and its consequences into a richer choice rule for each institution. We do not take this route. Such a rule would combine institution-specific account decisions with the choice rule. Specifying the account separately allows us to compare how institutions treat earlier allocations and to study constraints on the account and choice rule together.

The ledger also has a specific informational role.
For example, consider California's Assisted Living Waiver, a Medicaid program that limits the number of distinct participants it serves in a year. Suppose a participant leaves the program during the year. She is then no longer considered for service.
But she still counts toward the yearly limit. If she returns later in the year, the program has to know that she was already served.
So the ledger has to keep her identity. 
In contrast, the current selection does not need this information. It needs only the number of distinct participants served so far.

Of the questions above, the last, which rights from earlier allocations must still be honored, 
requires special care, because it may have to be addressed jointly in the account and the choice rule.
With population, eligibility, and priority fixed, cumulative implementation of a capacity-monotone choice rule retains earlier recipients as capacity expands. 
New arrivals or changes in priorities, by contrast, can cause the same rule to reject an earlier recipient who had been appointed in accordance with the institution's desired policy. 
The same can happen to appointments made under a previous policy or through a mistaken implementation that must nonetheless be respected. 
Suppose the institution must honor such an appointment. 
Then the account reports its holder as an incumbent in the current problem, and the choice rule has to retain her. 
In this way, something the institution did earlier constrains what it can choose now. In general, what has to be honored is the appointment, not its category. The incumbent keeps her position, and the institution may count it in a different category. An institution could fix the category as well, and some do. But as Theorem~\ref{thm:compression} shows, locking categories can move the implemented policy away from the one its choice rule was adopted to deliver.
The question is how to honor these appointments while remaining as faithful as possible to the policy that the institution's choice rule implements.

We address this question for reserve policies broadly and explore it in depth for the over-and-above policy, one of the most prominent worldwide. Its boost is the policy goal disrupted by category locking in Theorem~\ref{thm:compression}. Among other places, it is mandated by law, court ruling, or regulation in Brazil, India, and Turkey, and it is used in school admissions in Chicago and New South Wales and in university admissions in Chile. 
We introduce \emph{retain, reassign, and boost} (RRB) choice rules that preserve over-and-above protection subject to honoring inherited appointments.
Nonoverlapping reserves give rise to a single RRB rule that maintains the open-first design of the over-and-above choice rule but deviates from priority in selecting recipients of open positions
only when deviation is needed to retain an incumbent.
An incumbent who can receive an open position without preventing retention need not continue to occupy a position reserved for her protected group, leaving that position available to another eligible individual.
Category flexibility also permits the reverse reassignment, when accommodating an incumbent in a reserved category allows a higher-priority individual to take an open position.

Overlapping reserves lead to multiple variants of the RRB rule, thereby introducing a further policy lever.
An institution may choose to consider reserved categories jointly by priority or specify how they should be ordered.
We characterize both variants through priority and protection axioms subject to incumbent retention (Theorems~\ref{thm:flexcharacterization} and~\ref{thm:precedencecharacterization}).
Each RRB variant reproduces its ordinary benchmark whenever that benchmark already selects every incumbent, and otherwise departs from it only as far as retention requires.
The construction underlying precedence RRB also extends to other category orders. 
RRB preserves the additional benefit intended by over-and-above protection, whereas the guaranteed-enrollment approach of \citet{DIY2025} is better suited to minimum representation. 
For other reserve policies, the appropriate retaining extension depends on the institution’s protection objective (Section~\ref{subsec:otherprecedences}).

\begin{table}[t]
\centering\linespread{1.0}\fontsize{11}{13}\selectfont
\caption{Policy proposals and parallel applications.}\label{tab:applicationcontributions}
\renewcommand{\arraystretch}{1.10}\setlength{\tabcolsep}{4pt}
\begin{tabularx}{\textwidth}{@{}>{\raggedright\arraybackslash}p{.205\textwidth}>{\raggedright\arraybackslash}p{.255\textwidth}>{\raggedright\arraybackslash}X>{\raggedright\arraybackslash}p{.075\textwidth}@{}}
\toprule
\textbf{Main application} & \textbf{Parallel settings} & \textbf{Principal design contribution} & \textbf{Section}\\\midrule
\multicolumn{4}{@{}l}{\textbf{Single institution: over-and-above protection}}\\\addlinespace[4pt]
Chicago schools (US) & Dallas magnet schools (US) & Basic RRB retains admissions under nonoverlapping reserves (Corollary~\ref{cor:basicrrbcharacterization}) & \ref{app:basicrrb}\\\midrule[.25pt]
AMAZUL recruitment (Brazil) & NSW selective schools (Australia) & Joint RRB compares claims across overlapping reserves (Theorem~\ref{thm:flexcharacterization}) & \ref{app:jointrecruitment}\\\midrule[.25pt]
UFC admissions (Brazil) & & Precedence RRB retains appointments under prescribed open-first orders (Theorem~\ref{thm:precedencecharacterization}) & \ref{app:ceara}\\\midrule[.25pt]
Adolfo Ib\'a\~nez admissions (Chile) & CSAB/JoSAA reassignment (India); IFCE admissions (Brazil) & Category reassignment and continuation after withdrawals or capacity growth (Lemma~\ref{lem:rrbwithdrawals}; Theorem~\ref{thm:flexrelease}) & \ref{app:chilecontinuation}\\\midrule
\multicolumn{4}{@{}l}{\textbf{Multiple institutions: over-and-above protection}}\\\addlinespace[4pt]
Chicago schools (US) & Dallas schools (US); NSW selective schools (Australia); ENARE residency (Brazil) & RRB--DA coordinates protected participation and gives individual-optimal outcomes (Theorem~\ref{thm:rrbdacharacterization}) & \ref{app:chicagocoordination}\\\midrule[.25pt]
CPNU recruitment (Brazil) & & Retention without priority bonuses; benchmark recovery and welfare comparison (Proposition~\ref{prop:dabenchmark}; Corollary~\ref{cor:daharm}) & \ref{app:cpnu}\\\midrule[.25pt]
University admissions (Chile) & USP/FUVEST (Brazil) & Protected upgrades; release invariance as category capacities grow (Theorem~\ref{thm:dacontinuation}) & \ref{app:chilecoordination}\\\midrule
\multicolumn{4}{@{}l}{\textbf{Multiple institutions: different protection policies}}\\\addlinespace[4pt]
My School DC (US) & U. of Otago (New Zealand) & DA coordinates retaining choices under different policies and category orders (Lemma~\ref{lem:sequentialmenus}) & \ref{app:mixedpolicies}\\
\bottomrule\end{tabularx}
\par\smallskip
\end{table}

With the other inputs fixed, protecting the new recipients selected by RRB at later decisions adds nothing to the consequences of the original commitments: as capacity grows, the outcome is the same as under RRB with only the original incumbents protected (Theorem~\ref{thm:flexrelease}).

In a more general setting with multiple institutions, an incumbent may also want to move to another institution without giving up her existing appointment.
Integrating RRB with the celebrated individual-proposing deferred acceptance algorithm of \citet{GaleShapley1962} gives RRB--DA, which we characterize in Theorem~\ref{thm:rrbdacharacterization}. 
Together with the single-institution RRB choice rules, it is the basis of the reform proposals we present in Sections~\ref{sec:casestudies} and~\ref{sec:multiapplications} for waiting lists and other continuing allocations in school admissions, medical residency, public-service recruitment, and university admissions.
RRB--DA inherits the properties of RRB: it reproduces ordinary deferred acceptance whenever no adjustment is needed to honor inherited appointments, and, with the other inputs fixed, protecting later assignments does not change the outcome as capacities expand. The same coordination extends to the other retaining rules we study, so institutions with different reserve policies can take part in one mechanism.

Table~\ref{tab:applicationcontributions} groups the developed policy proposals with parallel applications in several countries and distinguishes over-and-above protection from the extension to other policies. 

\subsection{Related Literature and Organization of the Rest of the Paper}\label{subsec:introliterature}

The reserve system design literature studies choice rules that implement an institution's policy objectives. Assuming nonoverlapping reserves, \citet{HYY2013} introduce minority reserves as minimum guarantees, \citet{EcheniqueYenmez2015} characterize them, and \citet{DurEtAl2018} introduce the over-and-above policy. \citet{KominersSonmez2016} study precedence across positions in a general framework of slot-specific priorities.\footnote{A category with several positions can be represented by several slots sharing the same priority and eligibility restrictions.}
\citet{SY2022} characterize the meritorious horizontal choice rule under India's vertical and horizontal reservations. Their overlapping horizontal traits concern additional protection within categories; here the reserved categories themselves can overlap.\footnote{\citet{DoganErdil2025} study admissions with flexible access and intake targets. \citet{Celebi2023} studies the rationalizability of choice by diversity preferences and the role of intersectional identities in satisfying substitutes.}

In the spirit of our release comparison, \citet{AbizadaBo2021} study minimum guarantee choice rules in sequential hiring. They show that applying the guarantee round by round preserves minimum protection but can change who is hired and favor lower-priority minority workers even when the cumulative minimum has already been met. They characterize a rule that adjusts each round's reserve by the minority hires already made. This rule yields the same cumulative recipient sets as the Parcoursup calling order discussed in Section~\ref{app:parcoursup}, which coincide with those of cumulative implementation of the minimum guarantee (Corollary~\ref{cor:calling}).
Our compression result instead asks how much of the difference in protection between reserve policies survives repeated batch implementation.
Theorem~\ref{thm:compression} shows that small batches leave little scope for such differences under any viable category-locked batch procedure.\footnote{Sequential implementation can also change the properties of a static allocation rule in the settings studied by \citet{DurKesten2019}, \citet{KennesMonteTumennasan2014,KennesMonteTumennasan2019}, and, for the block-by-block processing of Chinese college admissions, \citet{ChenKesten2017}.}

Resource-release invariance is related to composition requirements in rationing, under which allocation in installments should agree with allocation of the total resource when remaining claims are adjusted appropriately \citep{Young1988,Moulin2000,Thomson2019}. Our comparison distinguishes preserving recipients from preserving their category assignments.

\citet{DIY2025} consider school choice with enrollment guarantees and overlapping reserves. Their guaranteed-enrollment choice rule protects returning students while implementing minimum-guarantee reservations; without returning students, it reduces to the meritorious horizontal choice rule of \citet{SY2022}. Our starting point is instead arbitrary feasible inherited appointments under an over-and-above policy, and we ask how to retain them while maintaining priority and protection as far as those commitments permit. The difference matters for how incumbents count: they can already fulfill a minimum guarantee, whereas over-and-above protection seeks to leave reserved positions for additional beneficiaries when incumbents can receive open positions. We compare the two retaining extensions in Section~\ref{subsec:otherprecedences}.
We use the characterization method of \citet{DIY2025} to extend our single-institution characterizations of RRB choice rules to characterizations of the corresponding individual-proposing deferred-acceptance mechanisms.\footnote{Their method generalizes the choice-to-mechanism argument used by \citet{SY2024} for deferred acceptance based on the two-step meritorious horizontal choice rule of \citet{SY2022}.}

\citet{PSUY2024} consider medical reserve allocation with category-specific priorities. In their Appendix~A.2, they discuss changes of category during treatment and grandfathering in ventilator allocation when withdrawal is not allowed. We study retention-adjusted priority and protection principles under a common priority.

In housing and teacher assignment, existing recipients can keep their assignment or receive one they prefer \citep{AbdulkadirogluSonmez1999,Kurino2014,Pereyra2013,CombeTercieuxTerrier2022}. Our single-institution category reassignment preserves the appointment while changing how it counts and the opportunities available to others.\footnote{Earlier assignments and continuing claims also have distinct roles in landing-slot allocation \citep{SchummerVohra2013,SchummerAbizada2017}.}
With multiple institutions, an incumbent can also obtain a preferred appointment; each institution uses the retaining choice rule appropriate to its policy.

The literature on dynamic matching markets examines how waiting, entry, and future opportunities shape allocations \citep[see][for a survey]{Doval2025}. 
Waiting changes exchange opportunities in kidney exchange \citep{Unver2010}, enters the definition of dynamic stability \citep{Doval2022}, and trades off against access to preferred objects on waiting lists \citep{Leshno2022}; \citet{AnderssonEhlersMartinello2018} study the dynamic assignment of asylum seekers to localities under quotas and category restrictions. 
Although dynamic reserve systems are one of our main subjects, our focus within this context is different. We study how repeated implementation of a static choice rule may change the policy that is implemented, how an appropriate specification of the problem at each date may prevent this, and how inherited appointments can be honored with minimal deviation from the primary policy objective.

The rest of the paper is organized as follows. Section~\ref{sec:diagnosis} introduces the ledger--account--choice framework and studies implementation across resource releases. Section~\ref{sec:state} describes how the account forms the current problem and which records later decisions require. Section~\ref{sec:rules} develops RRB and its extension to other category orders. Section~\ref{sec:multiinstitution} studies RRB--DA and coordination across institutions with different reserve policies. Sections~\ref{sec:casestudies} and~\ref{sec:multiapplications} present single-institution and multi-institution applications, respectively. We collect proofs of the single-institution and multi-institution results in Appendices~\ref{app:principalproofs} and~\ref{app:multiinstitutionproofs}, respectively. The Supplemental Appendix provides further results and institutional details.

\section{Ledgers, Accounts, and Dynamic Implementation}\label{sec:diagnosis}

Starting with \citet{HYY2013}, the literature on reserve systems has largely focused on the design and analysis of choice rules that map a given problem to an outcome; in its simplest form, a problem consists of a set of applicants and category capacities, and an outcome specifies which applicants receive positions and through which categories.

In many applications, however, the problem is not given but formed. 
A major case is when the environment is dynamic. Positions become available over time, applicants arrive and leave, and prior allocations affect future decisions. The institution needs to make decisions repeatedly, and each decision is made with the earlier ones in place. Every decision entails three questions. First, what does the institution know about its past? Second, what does it take its past to imply about the current decision? Third, who should receive positions based on the resulting problem? Research on reserve systems focuses on the third question and takes the problem as given.

We separate these questions for two reasons. A failure to implement the intended policy can then be traced back to the record, the assessment of prior allocations, or the choice on the resulting problem. Separation also helps with design, even though the three elements rely on each other. The record must preserve what is needed for later decisions.

The \emph{ledger} is the institution's memory of its past, the record of releases, category assignments, and subsequent events. The \emph{account} determines how these records affect the current problem: which recipients and positions to revisit, which category assignments to keep fixed, and which rights to honor. The \emph{choice rule} then selects recipients for the problem the account presents, as in the static literature.

California's Assisted Living Waiver, a Medicaid program that provides care in homes and communities, helps illustrate this separation. 
The program sets limits on how many people can receive service at any given date and how many different people can receive service over the year.\footnote{\label{fn:medicaid}California's \href{https://www.dhcs.ca.gov/file/application-for-1915c-hcbs-waiver-ca-0431-r04-01-mar-01-2025-pdf/}{2025 Assisted Living Waiver application} specifies the concurrent and annual participant limits (Appendix~B-3, pp.~27--28). The Centers for Medicare \& Medicaid Services' \href{https://wms-mmdl.cms.gov/WMS/help/35/Instructions_TechnicalGuide_V3.6.pdf}{technical guide} explains how distinct participants are counted over the year (2019, Appendix~B-3, pp.~82--84, and the glossary entry ``Unduplicated Participant'').} When a participant leaves, there is space for someone new, but the departed person is still counted as someone served that year. The ledger keeps track of when she was served and when she left. 
Through the account, the institution determines how much room remains under each limit and which applicants have already been served this year, since a returning participant takes a current place without adding to the annual count. The choice rule then selects participants subject to these two constraints.

We now formalize the distinction between the ledger, the account, and the choice rule.
Consider decision dates $t=0,1,\ldots,T$. Let $\Lambda_t$ be the ledger record, $\sigma_t$ the current problem, and $\mathrm G_t$ the outcome at date $t$. Suppose $\Lambda_0$ is an initial ledger record and $\mathrm G_0$ an initial outcome. Let $e_t$ be the set of events occurring after the allocation at date $t-1$ and before the allocation at date $t$. For $t=1,\ldots,T$, a \emph{ledger update rule} $U_t$ computes the new ledger record:
\[
\Lambda_t=U_t(\Lambda_{t-1},\mathrm G_{t-1},e_t).
\]
Next, an \emph{account rule} $A_t$ determines the current problem:
\[
\sigma_t=A_t(\Lambda_t).
\]
Finally, a \emph{choice rule} $\CC_t$ selects a feasible outcome for that problem:
\[
\mathrm G_t=\CC_t(\sigma_t).
\]
The ledger update and the account link the static choice problems faced at successive dates.
Central for our purposes, a policy can change through the account rule alone: a previous allocation is then counted differently, without changing the fact that it was made, and this in turn can affect how the institution operates in the future.

Different accounts can express different policy objectives, rather than correct and incorrect treatments of the same policy. By making the account explicit, we can study requirements on its treatment of earlier records jointly with requirements on choice.

\subsection{Choice Rules and Reserve Systems in Static Settings}\label{para:allocationmodel}
We begin by introducing choice rules and reserve systems in static settings where outcomes of different problems do not interact with each other. A set of identical indivisible objects is to be allocated to unit-demand individuals. In much of our analysis, we refer to these objects as positions, although, depending on the application, we also refer to them as particular goods or opportunities being allocated. A baseline priority ranking determines individuals' claims in the absence of protection policies. Our focus is on affirmative action policies implemented through reserve systems that channel their protections through restricted eligibility for reserved categories.

Let $\mathcal I$ be a finite universe of potential individuals and $\mathcal V=\{o\}\medcup\mathcal R$ be a finite set of position categories. Category $o$ is the \emph{open category}, and each category $c\in\mathcal R$ is a \emph{reserved category}. Positions in these categories are called open and reserved positions, respectively. For each $v\in\mathcal V$, let $\mathcal I^v\subseteq\mathcal I$ be the set of individuals eligible for category $v$. All individuals are eligible for the open category; thus $\mathcal I^o=\mathcal I$. For any set of individuals $I\subseteq\mathcal I$, let $I^v=I\medcap\mathcal I^v$ denote its members eligible for category $v\in\mathcal V$ and $I^\varnothing=I\setminus\bigcup_{c\in\mathcal R}I^c$ denote those who are ineligible for any reserved category.

In our general model, an individual can qualify for multiple reserved categories. Thus $\mathcal I^c\medcap\mathcal I^{c'}$ may be nonempty for distinct reserved categories $c,c'\in\mathcal R$. In this case, we say that reserved categories, or protections, are \emph{overlapping}. For some of our concepts and results, we assume that an individual can qualify for at most one reserved category.
That is, $\mathcal I^c\medcap\mathcal I^{c'}=\emptyset$ for all distinct $c,c'\in\mathcal R$. We refer to this case as \emph{nonoverlapping} reserved categories or protections.

Let $\pi$ be a strict baseline priority order on $\mathcal I$.\footnote{We take baseline priority as given. \citet{CelebiFlynn2022} study the coarsening of scores into priority classes, and \citet{CelebiFlynn2025} develop adaptive priority mechanisms for match quality and diversity under uncertainty about the applicant pool.} For $i,j\in\mathcal I$, let $i\mathrel{\pi}j$ mean that individual $i$ has higher priority than individual $j$. We suppress priority and eligibility in the notation for a problem when they are fixed. Displayed priority lists run from highest to lowest. For example, $\pi:i_1\ i_2\ i_3$ indicates $i_1\mathrel{\pi}i_2\mathrel{\pi}i_3$.

Let $\mathbb Z_+$ denote the set of nonnegative integers. With eligibility and priority fixed in our basic setting, the current problem $\sigma$ is specified by a population $I\subseteq\mathcal I$ and a list of category capacities $q=(q^v)_{v\in\mathcal V}\in\mathbb Z_+^{\mathcal V}$. We therefore denote it by $(I,q)$.

A \emph{category-assignment profile}, or simply an \emph{outcome}, $\mathrm G=(\mathrm G^v)_{v\in\mathcal V}$ specifies the recipients in each category. Here, we have $\mathrm G^v\subseteq\mathcal I$ for each $v\in\mathcal V$, and $\mathrm G^v\medcap\mathrm G^{v'}=\emptyset$ for distinct $v,v'\in\mathcal V$. Let $\widehat{\mathrm G}=\bigcup_{v\in\mathcal V}\mathrm G^v$ denote its aggregate recipient set.

\begin{definition}
Given a problem $(I,q)$, an outcome $\mathrm G$ is \emph{feasible} if, for each category $v\in\mathcal V$,
\[
\mathrm G^v\subseteq I^v\quad\mbox{and}\quad |\mathrm G^v|\le q^v.
\]
\end{definition}
Throughout the paper, outcomes are understood to be feasible.
An eligibility restriction therefore defines a \emph{hard reserve}: an individual outside $I^c$ cannot receive a category-$c$ position.\footnote{Some systems use \emph{soft reserves}, where applicants outside the protected group may fill unused reserved positions. Others transfer unfilled positions to another category. Our characterizations use hard reserves at the capacities reported by the account. Appendix~\ref{app:softsequential} gives an extension of the retention construction to soft reserves.}

\begin{definition}\label{def:staticchoice}
A \emph{choice rule} $\CC$ selects an outcome $\CC(I,q)=\bigl(\CC^v(I,q)\bigr)_{v\in\mathcal V}$ for each problem $(I,q)$ in its domain.
\end{definition}
In a static setting, a reserve system is represented by a choice rule. Let $\widehat\CC(I,q)=\bigcup_{v\in\mathcal V}\CC^v(I,q)$ denote its aggregate recipient set.

\begin{definition}\label{def:nonwastefulness}
Given a problem $(I,q)$, an outcome $\mathrm G$ satisfies \emph{non-wastefulness} if, for each category $v\in\mathcal V$,
\[
|\mathrm G^v|<q^v\quad\Longrightarrow\quad
I^v\setminus\widehat{\mathrm G}=\emptyset.
\]
\end{definition}
That is, a position can remain unfilled only if no unselected individual is eligible for it.

\begin{definition}\label{def:staticNJE}
Given a problem $(I,q)$, an outcome $\mathrm G$ satisfies \emph{no justified envy} if, for each category $v\in\mathcal V$ and each pair of individuals $i,j\in I$,
\[
i\in\mathrm G^v\quad\mbox{and}\quad j\in I^v\setminus\widehat{\mathrm G}
\quad\Longrightarrow\quad i\mathrel{\pi}j.
\]
\end{definition}
Each recipient must have higher priority than every unselected individual eligible for her category.

Using the terminology of  \citet{Thomson2023},
non-wastefulness and no justified envy are \emph{punctual axioms}, defined over outcomes for a fixed problem.
A choice rule satisfies a punctual axiom if its outcome satisfies that axiom at every problem in its domain. A \emph{relational axiom}, in contrast, imposes requirements on outcomes across problems. Our next axiom is relational.

\begin{definition}\label{def:resourcemonotonicity}
A choice rule $\CC$ satisfies \emph{capacity monotonicity} if, for any set of individuals $I\subseteq\mathcal I$ and pair of capacity vectors $q,q'$,
\[
q\le q'\quad\Longrightarrow\quad
\widehat\CC(I,q)\subseteq\widehat\CC(I,q').
\]
\end{definition}
Holding the population, eligibility, and priority fixed, capacity monotonicity requires that weakly increasing capacities for each category preserve the recipients chosen at the smaller capacities, although not necessarily their category assignments.\footnote{\citet{DoganDoganYildiz2021} study capacity changes in lexicographic choice rules.}

For each $J\subseteq\mathcal I$ and $\ell\in\mathbb Z_+$, define
\[
\TopSet_\ell^\pi(J)=\bigl\{i\in J:\bigl|\{j\in J:j\mathrel{\pi}i\}\bigr|<\ell\bigr\}.
\]
Thus $\TopSet_\ell^\pi(J)$ consists of the $\min\{\ell,|J|\}$ highest-priority individuals in $J$, the recipients of $\ell$ positions under selection by priority alone.
We suppress the superscript $\pi$ when the priority order is fixed.

Even with the same category capacities, eligibility sets, and baseline priority, two reserve systems can provide very different levels of protection \citep{DurEtAl2018}.
A central distinction is to what extent, if any, positions received through baseline priority count toward a group's protection. The minimum guarantee policy counts these positions toward the protection, whereas the over-and-above policy provides reserved positions in addition to those that could be received without the protection. The difference is implemented by processing reserved positions before open positions under minimum guarantee, and open positions before reserved positions under over-and-above.

The following two choice rules are defined for nonoverlapping reserved categories.\medskip

\choiceheading{The minimum guarantee choice rule $\CC_{\mathrm{MG}}$}{\citep{HYY2013}.}
\begin{choicesteps}
\choicestep For each $c\in\mathcal R$, assign the reserved positions of category $c$ to the highest-priority eligible individuals, up to its capacity:
\[
\CC_{\mathrm{MG}}^c(I,q)=\TopSet_{q^c}(I^c).
\]
\choicestep Assign the open positions to the highest-priority remaining individuals, up to capacity:
\[
\CC_{\mathrm{MG}}^o(I,q)=\TopSet_{q^o}\left(I\setminus\bigcup_{c\in\mathcal R}\CC_{\mathrm{MG}}^c(I,q)\right).
\]
\end{choicesteps} \smallskip

\choiceheading{The over-and-above (O\&A) choice rule $\CC_{\mathrm{OA}}$}{\citep{DurEtAl2018}.}

\begin{choicesteps}
\choicestep Assign the open positions to the highest-priority individuals, up to capacity:
\[
\CC_{\mathrm{OA}}^o(I,q)=\TopSet_{q^o}(I).
\]
\choicestep For each $c\in\mathcal R$, assign the reserved positions of category $c$ to the highest-priority remaining eligible individuals, up to its capacity:
\[
\CC_{\mathrm{OA}}^c(I,q)=\TopSet_{q^c}\bigl(I^c\setminus \CC_{\mathrm{OA}}^o(I,q)\bigr).
\]
\end{choicesteps} \smallskip

For both choice rules, the reserved categories can be processed in any precedence order or simultaneously because nonoverlapping eligibility makes their selections independent of one another. Both choice rules satisfy non-wastefulness, no justified envy, and capacity monotonicity.

Our next axiom requires that reserved positions add as many beneficiaries as possible beyond those selected on priority alone.
It is satisfied by the O\&A choice rule but not by the minimum guarantee choice rule.

\begin{definition}\label{def:baseVR}
An outcome $\mathrm G$ for a problem $(I,q)$ satisfies \emph{maximal reserve protection} if, for each $c\in\mathcal R$ and $i\in\mathrm G^c$,
\[
|\mathrm G^o|=q^o\quad\mbox{and}\quad j\mathrel{\pi}i\ \text{for each }j\in\mathrm G^o.
\]
\end{definition}
An individual who can obtain an open position based on priority alone should not occupy a position reserved for her group. This principle is explicit in India's \emph{Indra Sawhney} judgment; in Sri Lanka's university admissions, where district-based places are allocated in addition to the places district residents obtain through national merit; and in Brazil's federal employment law, which excludes appointments within the open allocation from the reserved-vacancy count.\footnote{\emph{Indra Sawhney v.\ Union of India} (Supreme Court of India, November~16, 1992, paragraphs~811--812) distinguishes India's two kinds of reservations by how positions obtained without affirmative action are counted. Under horizontal reservations they count toward the protection, so the reservation is satisfied once enough group members are selected on their own. Under vertical reservations a group member selected in open competition is not counted against the group's reserved positions. Over-and-above protection also arises in H-1B immigration lotteries studied by \citet{PathakReesJonesSonmez2025}. For Sri Lanka, see University Grants Commission, \href{https://www.ugc.ac.lk/downloads/admissions/Handbook_2024_25/student_handbook_english.pdf}{\emph{University Admissions Handbook: Academic Year 2024/2025}}, Section~1.1, pp.~8--9. For Brazil, see \href{https://www.planalto.gov.br/ccivil_03/_ato2023-2026/2025/lei/l15142.htm}{Lei~15.142, June~3, 2025}, Article~7, paragraph~2.} Assuming reserved categories are nonoverlapping, non-wastefulness, no justified envy, and maximal reserve protection characterize the O\&A choice rule \citep{SY2022}.

So far, we have considered static implementation. We next study implementation in dynamic settings.

\subsection{Repeated Batch Implementation and Policy Compression}\label{subsec:compression}

Returning to the vaccine allocation question in the Introduction, we study successive releases of supply with fixed priority and eligibility.
We also hold the population fixed, so that we can examine whether repetition of a given static choice rule maintains the desired reserve policy.

More generally, an institution can adopt a choice rule and category capacities for the protection they would provide if all positions were allocated together, but may have to fill positions as they become available. We call the positions released together at a decision date a \emph{batch}.
A common approach in practice is to apply the chosen rule to each batch and the applicants not yet assigned a position.
An applicant who receives a position plays no role in further allocations. Her category remains unchanged.

Although this practice is natural, it comes with an implicit account. When a choice rule is repeated in this way, the earlier recipients drop out of the problem and their categories are locked.
Decision makers might not notice this account, which is a consequence of repeating the rule.
Our next example illustrates how this practice can lead to a policy that is very different from the one the underlying static rule was adopted to deliver.

\begin{example}\label{ex:200}
Let $\mathcal V=\{o,c\}$. There are 100 positions, of which 80 are open and 20 are reserved for category $c$. The population $I$ consists of 500 applicants, with $I^c=\{r_1,\ldots,r_{100}\}$ and $I^\varnothing=\{u_1,\ldots,u_{400}\}$. A member of $I^c$ occupies every fifth priority rank, beginning with the first:
\[
\pi:\quad r_1\quad u_1\quad u_2\quad u_3\quad u_4\quad r_2\quad\cdots\quad r_{100}\quad u_{397}\quad u_{398}\quad u_{399}\quad u_{400}.
\]
If all 100 positions are allocated together in one batch, the minimum guarantee choice rule selects the 20 highest-priority members of $I^c$ and the 80 highest-priority members of $I^\varnothing$. The O\&A choice rule selects 16 members of $I^c$ among the 80 recipients of open positions and another 20 through category $c$, for 36 in total.

Now divide the positions into two batches, each with 40 open and 10 reserved positions. We compare a procedure that applies the minimum guarantee choice rule in each batch with one that applies the O\&A choice rule in each batch. Each procedure first applies its choice rule to all 500 applicants and the first 50 positions. It then removes its 50 recipients from consideration and applies the same choice rule to the remaining 450 applicants and the second 50 positions. Earlier appointments and their categories remain fixed.

Under minimum guarantee, the first application selects $r_1,\ldots,r_{10}$ for reserved positions and $u_1,\ldots,u_{40}$ for open positions. The second selects $r_{11},\ldots,r_{20}$ for reserved positions and $u_{41},\ldots,u_{80}$ for open positions. The two applications select 20 reserve-eligible applicants in total.

In the first application of the O\&A choice rule, $r_1,\ldots,r_8$ and $u_1,\ldots,u_{32}$ are selected for open positions, followed by $r_9,\ldots,r_{18}$ for reserved positions. The highest-priority reserve-eligible applicant still available for the second application is $r_{19}$. All 40 applicants $u_{33},\ldots,u_{72}$ rank above her, so they take the open positions of the second batch, and $r_{19},\ldots,r_{28}$ take its reserved positions. The first application selects 18 reserve-eligible applicants and the second adds ten, for a total of 28.
The applicants $r_9,\ldots,r_{16}$  could have received open positions when the second batch arrived if their category assignments had been reconsidered. Instead, their categories remain fixed, so their reserved positions cannot benefit other eligible applicants.

Next, split the positions into five batches of 16 open and four reserved positions, and apply the same choice rule to each batch and its remaining applicants. With minimum guarantee, four reserve-eligible applicants are selected in each batch, 20 in total. With the O\&A choice rule, four are selected for open positions and another four for reserved positions in the first batch. In each of the four later batches, the 16 highest-priority remaining applicants are reserve-ineligible, so only the four reserved positions go to eligible applicants. After the successive batches there are 8, 12, 16, 20, and 24 reserve-eligible recipients.

Finally, divide the positions into twenty batches, each with four open and one reserved position, and apply the rule in each batch individually. Minimum guarantee selects one reserve-eligible applicant per batch, 20 in total. The O\&A choice rule assigns the first four open positions to $r_1,u_1,u_2,u_3$ and the first reserved position to $r_2$.
In every later batch, the four open positions go to reserve-ineligible applicants, and the reserved position goes to a reserve-eligible applicant. 
So the first batch has two reserve-eligible recipients and each of the other nineteen batches has one, for a total of 21.
Figure~\ref{fig:compression} shows the number of reserve-eligible recipients after each batch under the four schedules.

\begin{figure}[!tp]
\centering
\begin{tikzpicture}[x=8.61pt,y=21pt,font=\small]
\draw[line width=1.2pt] (0,7.25)--(3.5,7.25);
\node[anchor=west] at (4.3,7.25) {Over-and-above};
\draw[line width=0.45pt] (23,7.25)--(26.5,7.25);
\node[anchor=west] at (27.3,7.25) {Minimum guarantee};
\node[anchor=east] at (-1,6.4) {One batch of 100};
\draw[line width=1.2pt] (0,6.66)--(36,6.66);
\draw[fill=black,draw=black] (36,6.66) circle[radius=2.25pt];
\node[anchor=west,inner sep=0pt,xshift=4pt,font=\footnotesize] at (36,6.66) {36};
\draw[line width=0.45pt] (0,6.14)--(20,6.14);
\draw[fill=white,draw=black,line width=0.55pt] (20,6.14) circle[radius=2.25pt];
\node[anchor=west,inner sep=0pt,xshift=4pt,font=\footnotesize] at (20,6.14) {20};
\node[anchor=east] at (-1,4.65) {Two batches of 50};
\draw[line width=1.2pt] (0,4.91)--(28,4.91);
\draw[fill=black,draw=black] (18,4.91) circle[radius=1.4375pt];
\node[anchor=south,inner sep=0pt,yshift=2.2pt,font=\fontsize{7}{8}\selectfont] at (18,4.91) {18};
\draw[fill=black,draw=black] (28,4.91) circle[radius=2.25pt];
\node[anchor=west,inner sep=0pt,xshift=4pt,font=\footnotesize] at (28,4.91) {28};
\draw[line width=0.45pt] (0,4.39)--(20,4.39);
\draw[fill=white,draw=black,line width=0.55pt] (10,4.39) circle[radius=1.4375pt];
\node[anchor=north,inner sep=0pt,yshift=-2.2pt,font=\fontsize{7}{8}\selectfont] at (10,4.39) {10};
\draw[fill=white,draw=black,line width=0.55pt] (20,4.39) circle[radius=2.25pt];
\node[anchor=west,inner sep=0pt,xshift=4pt,font=\footnotesize] at (20,4.39) {20};
\node[anchor=east] at (-1,2.9) {Five batches of 20};
\draw[line width=1.2pt] (0,3.16)--(24,3.16);
\draw[fill=black,draw=black] (8,3.16) circle[radius=1.4375pt];
\node[anchor=south,inner sep=0pt,yshift=2.2pt,font=\fontsize{7}{8}\selectfont] at (8,3.16) {8};
\draw[fill=black,draw=black] (12,3.16) circle[radius=1.4375pt];
\node[anchor=south,inner sep=0pt,yshift=2.2pt,font=\fontsize{7}{8}\selectfont] at (12,3.16) {12};
\draw[fill=black,draw=black] (16,3.16) circle[radius=1.4375pt];
\node[anchor=south,inner sep=0pt,yshift=2.2pt,font=\fontsize{7}{8}\selectfont] at (16,3.16) {16};
\draw[fill=black,draw=black] (20,3.16) circle[radius=1.4375pt];
\node[anchor=south,inner sep=0pt,yshift=2.2pt,font=\fontsize{7}{8}\selectfont] at (20,3.16) {20};
\draw[fill=black,draw=black] (24,3.16) circle[radius=2.25pt];
\node[anchor=west,inner sep=0pt,xshift=4pt,font=\footnotesize] at (24,3.16) {24};
\draw[line width=0.45pt] (0,2.64)--(20,2.64);
\draw[fill=white,draw=black,line width=0.55pt] (4,2.64) circle[radius=1.4375pt];
\node[anchor=north,inner sep=0pt,yshift=-2.2pt,font=\fontsize{7}{8}\selectfont] at (4,2.64) {4};
\draw[fill=white,draw=black,line width=0.55pt] (8,2.64) circle[radius=1.4375pt];
\node[anchor=north,inner sep=0pt,yshift=-2.2pt,font=\fontsize{7}{8}\selectfont] at (8,2.64) {8};
\draw[fill=white,draw=black,line width=0.55pt] (12,2.64) circle[radius=1.4375pt];
\node[anchor=north,inner sep=0pt,yshift=-2.2pt,font=\fontsize{7}{8}\selectfont] at (12,2.64) {12};
\draw[fill=white,draw=black,line width=0.55pt] (16,2.64) circle[radius=1.4375pt];
\node[anchor=north,inner sep=0pt,yshift=-2.2pt,font=\fontsize{7}{8}\selectfont] at (16,2.64) {16};
\draw[fill=white,draw=black,line width=0.55pt] (20,2.64) circle[radius=2.25pt];
\node[anchor=west,inner sep=0pt,xshift=4pt,font=\footnotesize] at (20,2.64) {20};
\node[anchor=east] at (-1,1.15) {Twenty batches of 5};
\draw[line width=1.2pt] (0,1.41)--(21,1.41);
\draw[fill=black,draw=black] (2,1.41) circle[radius=1.4375pt];
\node[anchor=south,inner sep=0pt,yshift=2.2pt,font=\fontsize{7}{8}\selectfont] at (2,1.41) {2};
\draw[fill=black,draw=black] (3,1.41) circle[radius=1.4375pt];
\node[anchor=south,inner sep=0pt,yshift=2.2pt,font=\fontsize{7}{8}\selectfont] at (3,1.41) {3};
\draw[fill=black,draw=black] (4,1.41) circle[radius=1.4375pt];
\node[anchor=south,inner sep=0pt,yshift=2.2pt,font=\fontsize{7}{8}\selectfont] at (4,1.41) {4};
\draw[fill=black,draw=black] (5,1.41) circle[radius=1.4375pt];
\node[anchor=south,inner sep=0pt,yshift=2.2pt,font=\fontsize{7}{8}\selectfont] at (5,1.41) {5};
\draw[fill=black,draw=black] (6,1.41) circle[radius=1.4375pt];
\node[anchor=south,inner sep=0pt,yshift=2.2pt,font=\fontsize{7}{8}\selectfont] at (6,1.41) {6};
\draw[fill=black,draw=black] (7,1.41) circle[radius=1.4375pt];
\node[anchor=south,inner sep=0pt,yshift=2.2pt,font=\fontsize{7}{8}\selectfont] at (7,1.41) {7};
\draw[fill=black,draw=black] (8,1.41) circle[radius=1.4375pt];
\node[anchor=south,inner sep=0pt,yshift=2.2pt,font=\fontsize{7}{8}\selectfont] at (8,1.41) {8};
\draw[fill=black,draw=black] (9,1.41) circle[radius=1.4375pt];
\node[anchor=south,inner sep=0pt,yshift=2.2pt,font=\fontsize{7}{8}\selectfont] at (9,1.41) {9};
\draw[fill=black,draw=black] (10,1.41) circle[radius=1.4375pt];
\node[anchor=south,inner sep=0pt,yshift=2.2pt,font=\fontsize{7}{8}\selectfont] at (10,1.41) {10};
\draw[fill=black,draw=black] (11,1.41) circle[radius=1.4375pt];
\node[anchor=south,inner sep=0pt,yshift=2.2pt,font=\fontsize{7}{8}\selectfont] at (11,1.41) {11};
\draw[fill=black,draw=black] (12,1.41) circle[radius=1.4375pt];
\node[anchor=south,inner sep=0pt,yshift=2.2pt,font=\fontsize{7}{8}\selectfont] at (12,1.41) {12};
\draw[fill=black,draw=black] (13,1.41) circle[radius=1.4375pt];
\node[anchor=south,inner sep=0pt,yshift=2.2pt,font=\fontsize{7}{8}\selectfont] at (13,1.41) {13};
\draw[fill=black,draw=black] (14,1.41) circle[radius=1.4375pt];
\node[anchor=south,inner sep=0pt,yshift=2.2pt,font=\fontsize{7}{8}\selectfont] at (14,1.41) {14};
\draw[fill=black,draw=black] (15,1.41) circle[radius=1.4375pt];
\node[anchor=south,inner sep=0pt,yshift=2.2pt,font=\fontsize{7}{8}\selectfont] at (15,1.41) {15};
\draw[fill=black,draw=black] (16,1.41) circle[radius=1.4375pt];
\node[anchor=south,inner sep=0pt,yshift=2.2pt,font=\fontsize{7}{8}\selectfont] at (16,1.41) {16};
\draw[fill=black,draw=black] (17,1.41) circle[radius=1.4375pt];
\node[anchor=south,inner sep=0pt,yshift=2.2pt,font=\fontsize{7}{8}\selectfont] at (17,1.41) {17};
\draw[fill=black,draw=black] (18,1.41) circle[radius=1.4375pt];
\node[anchor=south,inner sep=0pt,yshift=2.2pt,font=\fontsize{7}{8}\selectfont] at (18,1.41) {18};
\draw[fill=black,draw=black] (19,1.41) circle[radius=1.4375pt];
\node[anchor=south,inner sep=0pt,yshift=2.2pt,font=\fontsize{7}{8}\selectfont] at (19,1.41) {19};
\draw[fill=black,draw=black] (20,1.41) circle[radius=1.4375pt];
\node[anchor=south,inner sep=0pt,yshift=2.2pt,font=\fontsize{7}{8}\selectfont] at (20,1.41) {20};
\draw[fill=black,draw=black] (21,1.41) circle[radius=2.25pt];
\node[anchor=west,inner sep=0pt,xshift=4pt,font=\footnotesize] at (21,1.41) {21};
\draw[line width=0.45pt] (0,0.89)--(20,0.89);
\draw[fill=white,draw=black,line width=0.55pt] (1,0.89) circle[radius=1.4375pt];
\node[anchor=north,inner sep=0pt,yshift=-2.2pt,font=\fontsize{7}{8}\selectfont] at (1,0.89) {1};
\draw[fill=white,draw=black,line width=0.55pt] (2,0.89) circle[radius=1.4375pt];
\node[anchor=north,inner sep=0pt,yshift=-2.2pt,font=\fontsize{7}{8}\selectfont] at (2,0.89) {2};
\draw[fill=white,draw=black,line width=0.55pt] (3,0.89) circle[radius=1.4375pt];
\node[anchor=north,inner sep=0pt,yshift=-2.2pt,font=\fontsize{7}{8}\selectfont] at (3,0.89) {3};
\draw[fill=white,draw=black,line width=0.55pt] (4,0.89) circle[radius=1.4375pt];
\node[anchor=north,inner sep=0pt,yshift=-2.2pt,font=\fontsize{7}{8}\selectfont] at (4,0.89) {4};
\draw[fill=white,draw=black,line width=0.55pt] (5,0.89) circle[radius=1.4375pt];
\node[anchor=north,inner sep=0pt,yshift=-2.2pt,font=\fontsize{7}{8}\selectfont] at (5,0.89) {5};
\draw[fill=white,draw=black,line width=0.55pt] (6,0.89) circle[radius=1.4375pt];
\node[anchor=north,inner sep=0pt,yshift=-2.2pt,font=\fontsize{7}{8}\selectfont] at (6,0.89) {6};
\draw[fill=white,draw=black,line width=0.55pt] (7,0.89) circle[radius=1.4375pt];
\node[anchor=north,inner sep=0pt,yshift=-2.2pt,font=\fontsize{7}{8}\selectfont] at (7,0.89) {7};
\draw[fill=white,draw=black,line width=0.55pt] (8,0.89) circle[radius=1.4375pt];
\node[anchor=north,inner sep=0pt,yshift=-2.2pt,font=\fontsize{7}{8}\selectfont] at (8,0.89) {8};
\draw[fill=white,draw=black,line width=0.55pt] (9,0.89) circle[radius=1.4375pt];
\node[anchor=north,inner sep=0pt,yshift=-2.2pt,font=\fontsize{7}{8}\selectfont] at (9,0.89) {9};
\draw[fill=white,draw=black,line width=0.55pt] (10,0.89) circle[radius=1.4375pt];
\node[anchor=north,inner sep=0pt,yshift=-2.2pt,font=\fontsize{7}{8}\selectfont] at (10,0.89) {10};
\draw[fill=white,draw=black,line width=0.55pt] (11,0.89) circle[radius=1.4375pt];
\node[anchor=north,inner sep=0pt,yshift=-2.2pt,font=\fontsize{7}{8}\selectfont] at (11,0.89) {11};
\draw[fill=white,draw=black,line width=0.55pt] (12,0.89) circle[radius=1.4375pt];
\node[anchor=north,inner sep=0pt,yshift=-2.2pt,font=\fontsize{7}{8}\selectfont] at (12,0.89) {12};
\draw[fill=white,draw=black,line width=0.55pt] (13,0.89) circle[radius=1.4375pt];
\node[anchor=north,inner sep=0pt,yshift=-2.2pt,font=\fontsize{7}{8}\selectfont] at (13,0.89) {13};
\draw[fill=white,draw=black,line width=0.55pt] (14,0.89) circle[radius=1.4375pt];
\node[anchor=north,inner sep=0pt,yshift=-2.2pt,font=\fontsize{7}{8}\selectfont] at (14,0.89) {14};
\draw[fill=white,draw=black,line width=0.55pt] (15,0.89) circle[radius=1.4375pt];
\node[anchor=north,inner sep=0pt,yshift=-2.2pt,font=\fontsize{7}{8}\selectfont] at (15,0.89) {15};
\draw[fill=white,draw=black,line width=0.55pt] (16,0.89) circle[radius=1.4375pt];
\node[anchor=north,inner sep=0pt,yshift=-2.2pt,font=\fontsize{7}{8}\selectfont] at (16,0.89) {16};
\draw[fill=white,draw=black,line width=0.55pt] (17,0.89) circle[radius=1.4375pt];
\node[anchor=north,inner sep=0pt,yshift=-2.2pt,font=\fontsize{7}{8}\selectfont] at (17,0.89) {17};
\draw[fill=white,draw=black,line width=0.55pt] (18,0.89) circle[radius=1.4375pt];
\node[anchor=north,inner sep=0pt,yshift=-2.2pt,font=\fontsize{7}{8}\selectfont] at (18,0.89) {18};
\draw[fill=white,draw=black,line width=0.55pt] (19,0.89) circle[radius=1.4375pt];
\node[anchor=north,inner sep=0pt,yshift=-2.2pt,font=\fontsize{7}{8}\selectfont] at (19,0.89) {19};
\draw[fill=white,draw=black,line width=0.55pt] (20,0.89) circle[radius=2.25pt];
\node[anchor=west,inner sep=0pt,xshift=4pt,font=\footnotesize] at (20,0.89) {20};
\draw[line width=0.4pt] (0,0)--(40,0);
\foreach \x in {0,10,20,30,40}{
\draw (\x,0.08)--(\x,-0.08);
\node[anchor=north,inner sep=0pt,yshift=-4pt,font=\footnotesize] at (\x,0) {\x};
}
\node[anchor=north,font=\footnotesize] at (20,-0.65) {Reserve-eligible recipients};
\end{tikzpicture}
\caption{Policy compression under category-locked batch implementation. For each release schedule, the two lines trace the cumulative number of reserve-eligible recipients in Example~\ref{ex:200} after each batch, including recipients of open positions. For the single-batch schedule, each rule has only a final total. Small markers show totals after intermediate batches; large markers show final totals. Filled markers denote over-and-above and hollow markers denote minimum guarantee. Each schedule allocates 80 open and 20 reserved positions, with four open positions per reserved position in every batch. In each batch, the procedure fixes earlier assignments and their categories and applies the rule to the newly released positions and the applicants not yet assigned a position.}

\label{fig:compression}
\end{figure}
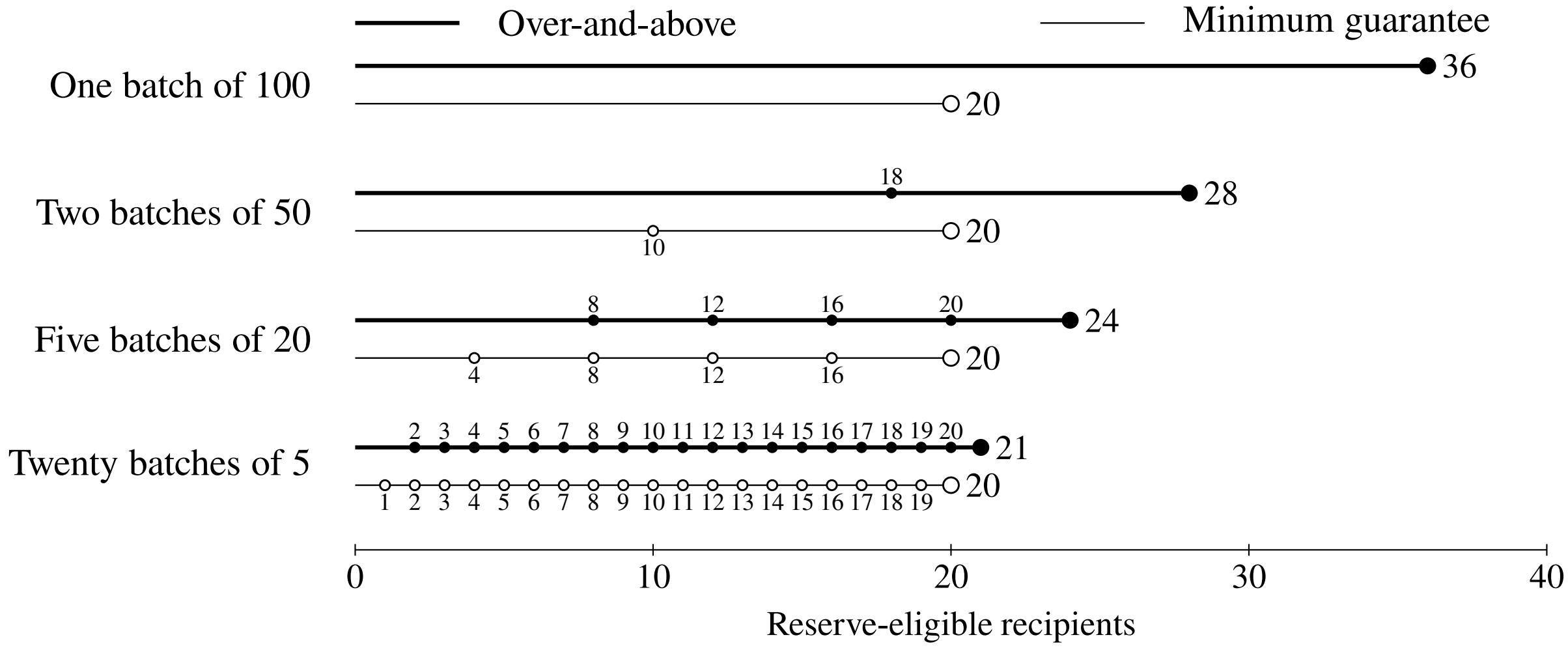
\end{example}

The minimum guarantee choice rule selects the same recipients under all four schedules in this example. These are also exactly the 100 highest-priority applicants, $r_1,\ldots,r_{20}$ and $u_1,\ldots,u_{80}$. 
So in this example, repeated minimum guarantee selects the same people as priority alone, with no reserve at all. In contrast, the extra protection of the over-and-above policy shrinks as the batches become smaller. 
With one batch of 100 positions, it selects 16 more reserve-eligible applicants than minimum guarantee. With twenty batches of 5 positions, it selects only one more. In that case, it differs from selection by priority alone in a single recipient. Theorem~\ref{thm:compression} below generalizes this compression.

Repeated application of the minimum guarantee choice rule can also change who is selected, but it preserves the promised minimum protection.\footnote{See Proposition~\ref{prop:mgasymmetry} in the Supplemental Appendix.}
Under over-and-above, repetition can largely eliminate the extra boost that distinguishes the policy from minimum guarantee.

To formalize the observation in Example~\ref{ex:200}, we next introduce some dynamic features, solution concepts, and axioms into the model. Let $t=1,2,\ldots$ index decision dates, with an empty allocation before date 1. To compare release histories in this section, fix a population $I\subseteq\mathcal I$, its eligibility sets $(I^v)_{v\in\mathcal V}$, and priority. Every individual in $I$ is available at date 1, no new individual arrives, and every unselected individual remains available at subsequent dates.

A \emph{release history} is a finite sequence $h=(\Delta q_1,\ldots,\Delta q_T)$ of any length $T\ge1$, where $\Delta q_t=(\Delta q_t^v)_{v\in\mathcal V}\in\mathbb Z_+^{\mathcal V}$ gives the positions released at date $t$. For each $t=1,\ldots,T$, the history through date $t$ is $h_t=(\Delta q_1,\ldots,\Delta q_t)$, where $h_0=\emptyset$ and $q(\emptyset)=0$. For any such history $h$, let
\[
q(h)=\sum_{t'=1}^T\Delta q_{t'}
\]
denote aggregate capacities. In particular, $q(h_t)=\sum_{t'=1}^t\Delta q_{t'}$.
For finite release histories $h=(\Delta q_1,\ldots,\Delta q_T)$ and $h'=(\Delta q'_1,\ldots,\Delta q'_{T'})$, let $h\preceq h'$ mean that $T\le T'$ and $\Delta q_{t'}=\Delta q'_{t'}$ for each $t'\le T$. Thus $h'$ extends $h$ by adding releases after date $T$.

\begin{definition}\label{def:continuation}
A \emph{dynamic reserve procedure} $\Phi$ assigns an outcome $\Phi(h)$ for the problem $\bigl(I,q(h)\bigr)$ to each finite release history $h$, including the empty history.
\end{definition}
Let $\widehat\Phi(h)=\bigcup_{v\in\mathcal V}\Phi^v(h)$ denote the aggregate recipient set. The outcome $\Phi(h_t)$ records the aggregate category assignments through date $t$.

A dynamic reserve procedure satisfies a punctual axiom if $\Phi(h)$ satisfies it for the problem $\bigl(I,q(h)\bigr)$ at every history $h$. The next axiom protects the appointments made at earlier dates.

\begin{definition}\label{def:recipientretention}
A dynamic reserve procedure $\Phi$ satisfies \emph{recipient retention} if, for each pair of finite release histories $h,h'$,
\[
h\preceq h'\quad\Longrightarrow\quad\widehat\Phi(h)\subseteq\widehat\Phi(h').
\]
It satisfies \emph{retention with category locking} if, for each pair of finite release histories $h,h'$ and each category $v\in\mathcal V$,
\[
h\preceq h'\quad\Longrightarrow\quad\Phi^v(h)\subseteq\Phi^v(h').
\]
\end{definition}
Recipient retention protects an appointment. Category locking additionally fixes its category.

For a dynamic reserve procedure $\Psi$ that retains earlier recipients in their categories, the \emph{residual problem} at date $t$
\[
\bigl(I\setminus\widehat\Psi(h_{t-1}),\Delta q_t\bigr)
\]
consists of the unselected individuals and the newly released positions, with their eligibility and priority unchanged.

\begin{definition}\label{def:residualcontinuation}
A dynamic reserve procedure $\Psi$ is a \emph{viable category-locked batch procedure} if it satisfies the following conditions:
\begin{enumerate}[label=(\roman*)]
\item It satisfies retention with category locking.
\item For each history $h=(\Delta q_1,\ldots,\Delta q_T)$ and date $t=1,\ldots,T$, the new assignments
\[
\bigl(\Psi^v(h_t)\setminus\Psi^v(h_{t-1})\bigr)_{v\in\mathcal V}
\]
form a feasible outcome satisfying non-wastefulness and no justified envy for the residual problem
\[
\bigl(I\setminus\widehat\Psi(h_{t-1}),\Delta q_t\bigr).
\]
\end{enumerate}
\end{definition}

Given a choice rule $\CC$, define its category-locked batch implementation by $\Psi^{\CC}(\emptyset)=(\emptyset)_{v\in\mathcal V}$ and
\[
\Psi^{\CC,v}(h_t)=\Psi^{\CC,v}(h_{t-1})\medcup \CC^v\bigl(I\setminus\widehat\Psi^{\CC}(h_{t-1}),\Delta q_t\bigr),
\qquad\text{for each }v\in\mathcal V.
\]
The account $A_t$ presents the residual problem $\bigl(I\setminus\widehat\Psi^{\CC}(h_{t-1}),\Delta q_t\bigr)$ together with the earlier profile, whose assignments are fixed. The rule $\CC$ determines the new assignments; adding them to the fixed profile gives the aggregate outcome. If $\CC$ satisfies non-wastefulness and no justified envy, $\Psi^{\CC}$ is a viable category-locked batch procedure. In particular, let $\Psi_{\mathrm{MG}}=\Psi^{\CC_{\mathrm{MG}}}$ and $\Psi_{\mathrm{OA}}=\Psi^{\CC_{\mathrm{OA}}}$ denote these repeated batch implementations of the minimum guarantee and O\&A choice rules. More generally, a viable category-locked batch procedure can use different choice rules across batches, whether fixed in advance or chosen in response to earlier history, provided each batch satisfies the stated conditions.

This form of repeated batch implementation formalizes the practice in Example~\ref{ex:200} and the implementation question raised by vaccine supplies arriving over time. We next ask how much extra protection it can preserve, allowing more generally any viable category-locked batch procedure.

For Theorem~\ref{thm:compression}, let $\mathcal R=\{c\}$. Define
\[
q^{\Sigma}(h)=q^o(h)+q^c(h)\quad\mbox{and}\quad
b(h)=\max_{1\le t'\le T}\min\{\Delta q_{t'}^o,\Delta q_{t'}^c\},
\]
with $q^{\Sigma}(\emptyset)=b(\emptyset)=0$. Thus $q^{\Sigma}(h)$ is total capacity and $b(h)$ is the largest, across batches, of the smaller of the open and reserved capacities in a batch.

\begin{assumption}\label{ass:abundantdemand}
The history $h$ satisfies $|I^c|\ge q^{\Sigma}(h)$.
\end{assumption}
Under Assumption~\ref{ass:abundantdemand}, after each date $t$ the remaining reserve-eligible applicants can fill all positions released at dates $t+1,\ldots,T$. Every viable category-locked batch procedure therefore fills every batch.

\begin{theorem}\label{thm:compression}
Let $\mathcal V=\{o,c\}$. Fix a finite population $I\subseteq\mathcal I$ with fixed eligibility and priority, available at date 1 with no arrivals or departures of unselected individuals through date $T$, and a release history $h=(\Delta q_1,\ldots,\Delta q_T)$ satisfying Assumption~\ref{ass:abundantdemand}. For any viable category-locked batch procedure $\Psi$ and each $t=1,\ldots,T$,
\[
\bigl|\widehat\Psi(h_t)\setminus\widehat\Psi_{\mathrm{MG}}(h_t)\bigr|
\le
\bigl|\widehat\Psi_{\mathrm{OA}}(h_t)\setminus\widehat\Psi_{\mathrm{MG}}(h_t)\bigr|
\le b(h_t).
\]
\end{theorem}

Among viable category-locked batch procedures, repeated application of the O\&A choice rule selects the highest number of reserve-eligible applicants. Yet its total boost over repeated minimum guarantee, counting all recipients selected through date $t$, is at most $b(h_t)$. Under the twenty-batch schedule in Example~\ref{ex:200}, the bound permits at most one additional reserve-eligible recipient across all twenty batches. Every such procedure therefore selects either 20 or 21 reserve-eligible applicants, although the O\&A choice rule selects 36 when all positions are allocated together.
Policies intended to provide different levels of protection can thus select nearly the same recipients as the repeated minimum guarantee policy.\footnote{The extreme compression in Theorem~\ref{thm:compression} depends on there being a single reserved category, and the exact bound is not our main point. The point is rather that the range of protection a reserve system can provide can be sharply narrowed by its treatment of earlier allocations. Appendix~\ref{app:multigroup} gives a bound for the general case of multiple nonoverlapping reserved categories. Example~\ref{ex:entry} shows why the fixed-population assumption matters.}

\subsection{Cumulative Implementation and Category Reassignment}\label{subsec:cumulative}

Because Theorem~\ref{thm:compression} applies to every viable category-locked batch procedure, choosing another rule for the residual problems cannot escape this policy compression. Instead, the institution can keep its static choice rule and change the problem it faces, applying the rule to the original population and to all positions released so far. We refer to this as \emph{cumulative implementation}. Earlier recipients are considered again along with those still waiting, and their category assignments may change. In the ledger--account--choice framework, this changes the account's handling of earlier allocations but leaves the choice rule in place.

\begin{definition}\label{def:cumulativeimplementation}
The \emph{cumulative implementation} of a choice rule $\CC$ is the dynamic reserve procedure $\Phi^{\CC}$ that applies $\CC$ to the fixed population $I$ and all positions released so far. For each release history $h$,
\[
\Phi^{\CC}(h)=\CC\bigl(I,q(h)\bigr).
\]
\end{definition}
Here the account presents $A_t(\Lambda_t)=(I,q(h_t))$, and the choice rule determines the allocation of all positions released so far. Cumulative implementation alone does not assure that earlier recipients remain selected. Capacity monotonicity of the choice rule provides this assurance. The cumulative implementations of the O\&A and minimum guarantee choice rules therefore retain earlier recipients.

Continuing Example~\ref{ex:200}, we next show how category reassignments under cumulative implementation preserve the over-and-above boost.

\begin{figure}[!tp]
\centering
\begin{tikzpicture}[x=8.61pt,y=21pt,font=\small]
\draw[line width=1.2pt] (0,7.25)--(3.5,7.25);
\node[anchor=west] at (4.3,7.25) {Over-and-above};
\draw[line width=0.45pt] (23,7.25)--(26.5,7.25);
\node[anchor=west] at (27.3,7.25) {Minimum guarantee};
\node[anchor=east] at (-1,6.4) {One batch of 100};
\draw[line width=1.2pt] (0,6.66)--(36,6.66);
\draw[fill=black,draw=black] (36,6.66) circle[radius=2.25pt];
\node[anchor=west,inner sep=0pt,xshift=4pt,font=\footnotesize] at (36,6.66) {36};
\draw[line width=0.45pt] (0,6.14)--(20,6.14);
\draw[fill=white,draw=black,line width=0.55pt] (20,6.14) circle[radius=2.25pt];
\node[anchor=west,inner sep=0pt,xshift=4pt,font=\footnotesize] at (20,6.14) {20};
\node[anchor=east] at (-1,4.65) {Two batches of 50};
\draw[line width=1.2pt] (0,4.91)--(36,4.91);
\draw[fill=black,draw=black] (18,4.91) circle[radius=1.4375pt];
\node[anchor=south,inner sep=0pt,yshift=2.2pt,font=\fontsize{7}{8}\selectfont] at (18,4.91) {18};
\draw[fill=black,draw=black] (36,4.91) circle[radius=2.25pt];
\node[anchor=west,inner sep=0pt,xshift=4pt,font=\footnotesize] at (36,4.91) {36};
\draw[line width=0.45pt] (0,4.39)--(20,4.39);
\draw[fill=white,draw=black,line width=0.55pt] (10,4.39) circle[radius=1.4375pt];
\node[anchor=north,inner sep=0pt,yshift=-2.2pt,font=\fontsize{7}{8}\selectfont] at (10,4.39) {10};
\draw[fill=white,draw=black,line width=0.55pt] (20,4.39) circle[radius=2.25pt];
\node[anchor=west,inner sep=0pt,xshift=4pt,font=\footnotesize] at (20,4.39) {20};
\node[anchor=east] at (-1,2.9) {Five batches of 20};
\draw[line width=1.2pt] (0,3.16)--(36,3.16);
\draw[fill=black,draw=black] (8,3.16) circle[radius=1.4375pt];
\node[anchor=south,inner sep=0pt,yshift=2.2pt,font=\fontsize{7}{8}\selectfont] at (8,3.16) {8};
\draw[fill=black,draw=black] (15,3.16) circle[radius=1.4375pt];
\node[anchor=south,inner sep=0pt,yshift=2.2pt,font=\fontsize{7}{8}\selectfont] at (15,3.16) {15};
\draw[fill=black,draw=black] (22,3.16) circle[radius=1.4375pt];
\node[anchor=south,inner sep=0pt,yshift=2.2pt,font=\fontsize{7}{8}\selectfont] at (22,3.16) {22};
\draw[fill=black,draw=black] (29,3.16) circle[radius=1.4375pt];
\node[anchor=south,inner sep=0pt,yshift=2.2pt,font=\fontsize{7}{8}\selectfont] at (29,3.16) {29};
\draw[fill=black,draw=black] (36,3.16) circle[radius=2.25pt];
\node[anchor=west,inner sep=0pt,xshift=4pt,font=\footnotesize] at (36,3.16) {36};
\draw[line width=0.45pt] (0,2.64)--(20,2.64);
\draw[fill=white,draw=black,line width=0.55pt] (4,2.64) circle[radius=1.4375pt];
\node[anchor=north,inner sep=0pt,yshift=-2.2pt,font=\fontsize{7}{8}\selectfont] at (4,2.64) {4};
\draw[fill=white,draw=black,line width=0.55pt] (8,2.64) circle[radius=1.4375pt];
\node[anchor=north,inner sep=0pt,yshift=-2.2pt,font=\fontsize{7}{8}\selectfont] at (8,2.64) {8};
\draw[fill=white,draw=black,line width=0.55pt] (12,2.64) circle[radius=1.4375pt];
\node[anchor=north,inner sep=0pt,yshift=-2.2pt,font=\fontsize{7}{8}\selectfont] at (12,2.64) {12};
\draw[fill=white,draw=black,line width=0.55pt] (16,2.64) circle[radius=1.4375pt];
\node[anchor=north,inner sep=0pt,yshift=-2.2pt,font=\fontsize{7}{8}\selectfont] at (16,2.64) {16};
\draw[fill=white,draw=black,line width=0.55pt] (20,2.64) circle[radius=2.25pt];
\node[anchor=west,inner sep=0pt,xshift=4pt,font=\footnotesize] at (20,2.64) {20};
\node[anchor=east] at (-1,1.15) {Twenty batches of 5};
\draw[line width=1.2pt] (0,1.41)--(36,1.41);
\draw[fill=black,draw=black] (2,1.41) circle[radius=1.4375pt];
\node[anchor=south,inner sep=0pt,yshift=2.2pt,font=\fontsize{7}{8}\selectfont] at (2,1.41) {2};
\draw[fill=black,draw=black] (4,1.41) circle[radius=1.4375pt];
\node[anchor=south,inner sep=0pt,yshift=2.2pt,font=\fontsize{7}{8}\selectfont] at (4,1.41) {4};
\draw[fill=black,draw=black] (6,1.41) circle[radius=1.4375pt];
\node[anchor=south,inner sep=0pt,yshift=2.2pt,font=\fontsize{7}{8}\selectfont] at (6,1.41) {6};
\draw[fill=black,draw=black] (8,1.41) circle[radius=1.4375pt];
\node[anchor=south,inner sep=0pt,yshift=2.2pt,font=\fontsize{7}{8}\selectfont] at (8,1.41) {8};
\draw[fill=black,draw=black] (9,1.41) circle[radius=1.4375pt];
\node[anchor=south,inner sep=0pt,yshift=2.2pt,font=\fontsize{7}{8}\selectfont] at (9,1.41) {9};
\draw[fill=black,draw=black] (11,1.41) circle[radius=1.4375pt];
\node[anchor=south,inner sep=0pt,yshift=2.2pt,font=\fontsize{7}{8}\selectfont] at (11,1.41) {11};
\draw[fill=black,draw=black] (13,1.41) circle[radius=1.4375pt];
\node[anchor=south,inner sep=0pt,yshift=2.2pt,font=\fontsize{7}{8}\selectfont] at (13,1.41) {13};
\draw[fill=black,draw=black] (15,1.41) circle[radius=1.4375pt];
\node[anchor=south,inner sep=0pt,yshift=2.2pt,font=\fontsize{7}{8}\selectfont] at (15,1.41) {15};
\draw[fill=black,draw=black] (17,1.41) circle[radius=1.4375pt];
\node[anchor=south,inner sep=0pt,yshift=2.2pt,font=\fontsize{7}{8}\selectfont] at (17,1.41) {17};
\draw[fill=black,draw=black] (18,1.41) circle[radius=1.4375pt];
\node[anchor=south,inner sep=0pt,yshift=2.2pt,font=\fontsize{7}{8}\selectfont] at (18,1.41) {18};
\draw[fill=black,draw=black] (20,1.41) circle[radius=1.4375pt];
\node[anchor=south,inner sep=0pt,yshift=2.2pt,font=\fontsize{7}{8}\selectfont] at (20,1.41) {20};
\draw[fill=black,draw=black] (22,1.41) circle[radius=1.4375pt];
\node[anchor=south,inner sep=0pt,yshift=2.2pt,font=\fontsize{7}{8}\selectfont] at (22,1.41) {22};
\draw[fill=black,draw=black] (24,1.41) circle[radius=1.4375pt];
\node[anchor=south,inner sep=0pt,yshift=2.2pt,font=\fontsize{7}{8}\selectfont] at (24,1.41) {24};
\draw[fill=black,draw=black] (26,1.41) circle[radius=1.4375pt];
\node[anchor=south,inner sep=0pt,yshift=2.2pt,font=\fontsize{7}{8}\selectfont] at (26,1.41) {26};
\draw[fill=black,draw=black] (27,1.41) circle[radius=1.4375pt];
\node[anchor=south,inner sep=0pt,yshift=2.2pt,font=\fontsize{7}{8}\selectfont] at (27,1.41) {27};
\draw[fill=black,draw=black] (29,1.41) circle[radius=1.4375pt];
\node[anchor=south,inner sep=0pt,yshift=2.2pt,font=\fontsize{7}{8}\selectfont] at (29,1.41) {29};
\draw[fill=black,draw=black] (31,1.41) circle[radius=1.4375pt];
\node[anchor=south,inner sep=0pt,yshift=2.2pt,font=\fontsize{7}{8}\selectfont] at (31,1.41) {31};
\draw[fill=black,draw=black] (33,1.41) circle[radius=1.4375pt];
\node[anchor=south,inner sep=0pt,yshift=2.2pt,font=\fontsize{7}{8}\selectfont] at (33,1.41) {33};
\draw[fill=black,draw=black] (35,1.41) circle[radius=1.4375pt];
\node[anchor=south,inner sep=0pt,yshift=2.2pt,font=\fontsize{7}{8}\selectfont] at (35,1.41) {35};
\draw[fill=black,draw=black] (36,1.41) circle[radius=2.25pt];
\node[anchor=west,inner sep=0pt,xshift=4pt,font=\footnotesize] at (36,1.41) {36};
\draw[line width=0.45pt] (0,0.89)--(20,0.89);
\draw[fill=white,draw=black,line width=0.55pt] (1,0.89) circle[radius=1.4375pt];
\node[anchor=north,inner sep=0pt,yshift=-2.2pt,font=\fontsize{7}{8}\selectfont] at (1,0.89) {1};
\draw[fill=white,draw=black,line width=0.55pt] (2,0.89) circle[radius=1.4375pt];
\node[anchor=north,inner sep=0pt,yshift=-2.2pt,font=\fontsize{7}{8}\selectfont] at (2,0.89) {2};
\draw[fill=white,draw=black,line width=0.55pt] (3,0.89) circle[radius=1.4375pt];
\node[anchor=north,inner sep=0pt,yshift=-2.2pt,font=\fontsize{7}{8}\selectfont] at (3,0.89) {3};
\draw[fill=white,draw=black,line width=0.55pt] (4,0.89) circle[radius=1.4375pt];
\node[anchor=north,inner sep=0pt,yshift=-2.2pt,font=\fontsize{7}{8}\selectfont] at (4,0.89) {4};
\draw[fill=white,draw=black,line width=0.55pt] (5,0.89) circle[radius=1.4375pt];
\node[anchor=north,inner sep=0pt,yshift=-2.2pt,font=\fontsize{7}{8}\selectfont] at (5,0.89) {5};
\draw[fill=white,draw=black,line width=0.55pt] (6,0.89) circle[radius=1.4375pt];
\node[anchor=north,inner sep=0pt,yshift=-2.2pt,font=\fontsize{7}{8}\selectfont] at (6,0.89) {6};
\draw[fill=white,draw=black,line width=0.55pt] (7,0.89) circle[radius=1.4375pt];
\node[anchor=north,inner sep=0pt,yshift=-2.2pt,font=\fontsize{7}{8}\selectfont] at (7,0.89) {7};
\draw[fill=white,draw=black,line width=0.55pt] (8,0.89) circle[radius=1.4375pt];
\node[anchor=north,inner sep=0pt,yshift=-2.2pt,font=\fontsize{7}{8}\selectfont] at (8,0.89) {8};
\draw[fill=white,draw=black,line width=0.55pt] (9,0.89) circle[radius=1.4375pt];
\node[anchor=north,inner sep=0pt,yshift=-2.2pt,font=\fontsize{7}{8}\selectfont] at (9,0.89) {9};
\draw[fill=white,draw=black,line width=0.55pt] (10,0.89) circle[radius=1.4375pt];
\node[anchor=north,inner sep=0pt,yshift=-2.2pt,font=\fontsize{7}{8}\selectfont] at (10,0.89) {10};
\draw[fill=white,draw=black,line width=0.55pt] (11,0.89) circle[radius=1.4375pt];
\node[anchor=north,inner sep=0pt,yshift=-2.2pt,font=\fontsize{7}{8}\selectfont] at (11,0.89) {11};
\draw[fill=white,draw=black,line width=0.55pt] (12,0.89) circle[radius=1.4375pt];
\node[anchor=north,inner sep=0pt,yshift=-2.2pt,font=\fontsize{7}{8}\selectfont] at (12,0.89) {12};
\draw[fill=white,draw=black,line width=0.55pt] (13,0.89) circle[radius=1.4375pt];
\node[anchor=north,inner sep=0pt,yshift=-2.2pt,font=\fontsize{7}{8}\selectfont] at (13,0.89) {13};
\draw[fill=white,draw=black,line width=0.55pt] (14,0.89) circle[radius=1.4375pt];
\node[anchor=north,inner sep=0pt,yshift=-2.2pt,font=\fontsize{7}{8}\selectfont] at (14,0.89) {14};
\draw[fill=white,draw=black,line width=0.55pt] (15,0.89) circle[radius=1.4375pt];
\node[anchor=north,inner sep=0pt,yshift=-2.2pt,font=\fontsize{7}{8}\selectfont] at (15,0.89) {15};
\draw[fill=white,draw=black,line width=0.55pt] (16,0.89) circle[radius=1.4375pt];
\node[anchor=north,inner sep=0pt,yshift=-2.2pt,font=\fontsize{7}{8}\selectfont] at (16,0.89) {16};
\draw[fill=white,draw=black,line width=0.55pt] (17,0.89) circle[radius=1.4375pt];
\node[anchor=north,inner sep=0pt,yshift=-2.2pt,font=\fontsize{7}{8}\selectfont] at (17,0.89) {17};
\draw[fill=white,draw=black,line width=0.55pt] (18,0.89) circle[radius=1.4375pt];
\node[anchor=north,inner sep=0pt,yshift=-2.2pt,font=\fontsize{7}{8}\selectfont] at (18,0.89) {18};
\draw[fill=white,draw=black,line width=0.55pt] (19,0.89) circle[radius=1.4375pt];
\node[anchor=north,inner sep=0pt,yshift=-2.2pt,font=\fontsize{7}{8}\selectfont] at (19,0.89) {19};
\draw[fill=white,draw=black,line width=0.55pt] (20,0.89) circle[radius=2.25pt];
\node[anchor=west,inner sep=0pt,xshift=4pt,font=\footnotesize] at (20,0.89) {20};
\draw[line width=0.4pt] (0,0)--(40,0);
\foreach \x in {0,10,20,30,40}{
\draw (\x,0.08)--(\x,-0.08);
\node[anchor=north,inner sep=0pt,yshift=-4pt,font=\footnotesize] at (\x,0) {\x};
}
\node[anchor=north,font=\footnotesize] at (20,-0.65) {Reserve-eligible recipients};
\end{tikzpicture}
\caption{Preserving the reserve policy through cumulative implementation. For each release schedule, the two lines trace the cumulative number of reserve-eligible recipients in Example~\ref{ex:200} after each batch, as in Figure~\ref{fig:compression}, and the markers are read in the same way. 
Here the rule is applied again after each batch, to all the positions released so far and all the original applicants. Earlier recipients keep their positions, while their categories can change.}
\label{fig:cumulative}
\end{figure}
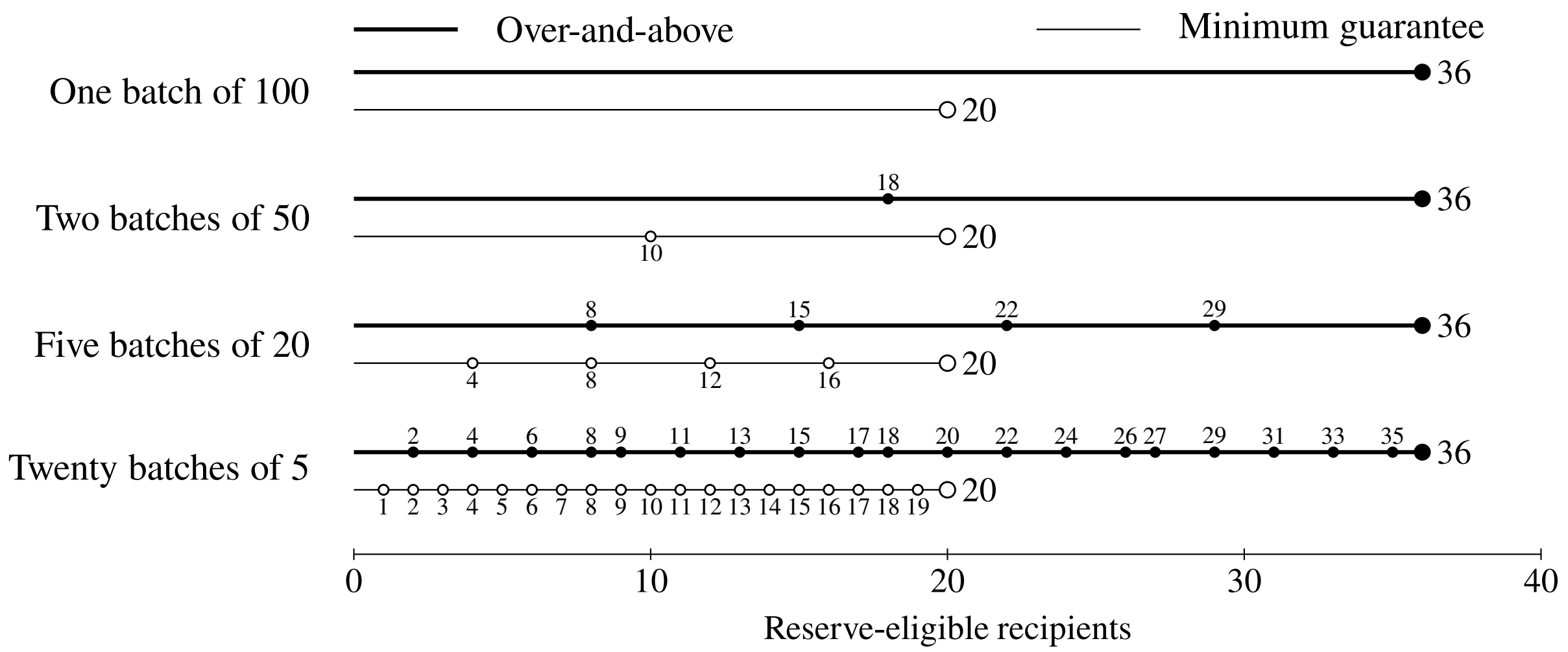

\medskip\noindent\textbf{Example~\ref{ex:200} (continued).}
Consider the 100 positions, 500 applicants, and priority order in Example~\ref{ex:200}. Suppose the positions are released in two batches, each containing 40 open and 10 reserved positions. At the first release, category-locked batch implementation and cumulative implementation agree, since neither has earlier assignments to carry forward. Under the O\&A choice rule, open positions go to $r_1,\ldots,r_8$ and $u_1,\ldots,u_{32}$, and reserved positions go to $r_9,\ldots,r_{18}$.

At the second release, the choice rule considers all 80 open and 20 reserved positions. The open positions go to $r_1,\ldots,r_{16}$ and $u_1,\ldots,u_{64}$, and the reserved positions go to $r_{17},\ldots,r_{36}$. Thus $r_9,\ldots,r_{16}$ move from reserved to open positions. The ten new reserved positions and the eight reserved positions freed by these reassignments admit $r_{19},\ldots,r_{36}$. Every first-round recipient remains appointed. The final recipient set includes 36 reserve-eligible applicants, compared with 28 under the category-locked batch procedure.\footnote{Such reassignment occurs in practice, sometimes on a large scale. In \emph{Pramod Kumar Singh v.\ State of Uttar Pradesh} (2021), 1,252 earlier appointees kept their jobs while being counted in the open category, freeing reserved positions for other eligible applicants. See Appendix~\ref{subsec:otherretention} and the applications in Section~\ref{sec:casestudies}.}

Now compare the final outcomes of the two implementations. Recall that with the category-locked batch procedure, the open positions in the second batch go to $u_{33},\ldots,u_{72}$. With cumulative implementation, $r_{29},\ldots,r_{36}$ are chosen instead of $u_{65},\ldots,u_{72}$. The latter eight applicants receive positions under the category-locked batch procedure but not under cumulative implementation. Since none of them was selected at the first release, the difference between the two outcomes involves no withdrawal of a first-release appointment.

Cumulative implementation of the minimum guarantee choice rule also preserves the first-round recipients. After the first release, $r_1,\ldots,r_{10}$ hold reserved positions and $u_1,\ldots,u_{40}$ hold open positions. At the final capacities, these sets expand to $r_1,\ldots,r_{20}$ and $u_1,\ldots,u_{80}$. Figure~\ref{fig:cumulative} shows the selections under cumulative implementation for all four release schedules.

\subsection{Resource-Release Invariance}\label{subsec:batching}

Under cumulative implementation, the outcome depends only on how many positions have been released in each category, not on when or in what batches they were released. We next explore whether this independence from scheduling requires cumulative implementation, and whether it can be achieved by a procedure that locks the categories of earlier assignments.

Consider the release histories from Section~\ref{subsec:compression}, holding the population, eligibility, and priority fixed. Since all that changes is the capacity of each category, we let $\CC(q)$ denote $\CC(I,q)$.

\begin{definition}\label{def:releaseinvariance}
A dynamic reserve procedure $\Phi$ satisfies \emph{resource-release invariance} if, for each pair of finite release histories $h,h'$,
\[
q(h)=q(h')\quad\Longrightarrow\quad \Phi(h)=\Phi(h').
\]
\end{definition}
The condition requires that, for fixed cumulative capacity in each category, splitting, consolidating, or reordering releases not affect the outcome. This axiom also prohibits an administrator from manipulating the allocation through the release schedule without changing the announced reserve policy. Our next result concerns whether this independence from release timing can be achieved under category locking.

\begin{proposition}\label{prop:lockingimpossibility}
There exist populations and priority orders for which no dynamic reserve procedure jointly satisfies retention with category locking, non-wastefulness, no justified envy, and resource-release invariance, even when total capacity is scarce and reserve-eligible demand satisfies Assumption~\ref{ass:abundantdemand}.
\end{proposition}

Protecting earlier appointments does not need to result in this failure if category assignments can be changed.
Next, we characterize the procedures that satisfy resource-release invariance.

Given a dynamic reserve procedure $\Phi$, we define its \emph{one-shot choice rule} by $\CC_\Phi(q)=\Phi((q))$, where $(q)$ denotes the single-batch release history in which all positions in $q$ are released together.

\begin{theorem}\label{thm:reattribution}
Fix a population $I\subseteq\mathcal I$, its eligibility and priority, and a dynamic reserve procedure $\Phi$ defined on all finite release histories.
\begin{enumerate}[label=(\roman*)]
\item The procedure $\Phi$ satisfies resource-release invariance if and only if it is the cumulative implementation of its one-shot choice rule:
\[
\Phi(h)=\CC_\Phi(q(h))\qquad\text{for each finite release history }h.
\]
\item A release-invariant procedure $\Phi$ additionally satisfies non-wastefulness, no justified envy, and recipient retention if and only if $\CC_\Phi$ satisfies non-wastefulness, no justified envy, and capacity monotonicity.
\end{enumerate}
\end{theorem}

Theorem~\ref{thm:reattribution} shows that cumulative implementation is necessary and sufficient for resource-release invariance. 
It also carries the underlying static rule's allocation properties over to each date. If that rule satisfies capacity monotonicity, earlier recipients remain selected. 
We next present a real-world example of cumulative implementation.
\subsection{Case Study: Cumulative Implementation in Parcoursup Admissions}\label{app:parcoursup}

Parcoursup is the French centralized system for applications to programs in higher education. Applicants who have received a national secondary-school grant based on family income, called \emph{boursiers}, receive positive discrimination in admission offers. Programs rank their applicants by their own admission criteria. The platform uses a minimum guarantee for grant recipients to decide who receives offers.\footnote{\label{fn:parcoursupalgorithms}French Ministry of Higher Education and Research, \href{https://services.dgesip.fr/fichiers/presentation_algorithmes_parcoursup_2025_finale.pdf}{\emph{Document de pr\'esentation des algorithmes de Parcoursup}}, May~2025, Sections~2.2--2.4.}

To implement the minimum guarantee, the platform computes a \emph{calling order} from the program ranking before making any offers. It places the highest-ranked remaining applicant next if doing so does not violate the minimum share of grant recipients; otherwise it places the highest-ranked remaining grant recipient next, if there is one. We focus on programs for which the minimum for grant recipients is the only additional restriction.\footnote{The official description cited in footnote~\ref{fn:parcoursupalgorithms} gives two calling-order rules. Section~4.1 covers programs that are subject only to the minimum rate for grant recipients. Section~4.2 covers programs that are also subject to a maximum rate of non-resident applicants. Our equivalence result is for the first rule.} For these programs, the calling order yields the same recipient sets as cumulative implementation of the minimum guarantee choice rule.

Fix the ranked list of applicants $I$, the set of grant recipients $I^c$, and the required share $\lambda\in[0,1]$. For $k=1,\ldots,|I|$ offers, the minimum requires
\[
m_k=\min\{\lceil\lambda k\rceil,|I^c|\}
\]
grant recipients, where $\lceil\lambda k\rceil$ is the smallest integer no less than $\lambda k$. The first $k$ applicants in the calling order are exactly the aggregate selection of $\CC_{\mathrm{MG}}$ at capacities $(k-m_k,m_k)$. 

The order thus defines in advance who would be chosen by cumulative implementation of the minimum guarantee choice rule for any possible number of offers. Both capacities are nondecreasing in $k$: every earlier recipient is still chosen as more offers become possible. The equivalence refers to the calling order constructed from a fixed list prior to the start of the live offer process.\footnote{See Appendix~\ref{app:callingresult} for details and Corollary~\ref{cor:calling} for the equivalence result.} It does not refer to who holds offers on a given date. 

The calling order concerns offers, not enrollment. When an applicant is offered a place, she can accept it, hold it while she waits for other offers, or refuse it. So a program typically makes offers to more applicants than it enrolls.
If an applicant refuses an offer, the calling order is not reconstructed. The reasoning behind this is explained by the Parcoursup Ethics and Scientific Committee. It points out that the designers read the law as setting the minimum for grant recipients on the initial calling list, at the start of the procedure: ``the initial calling list... is then no longer modified, regardless of withdrawals.''\footnote{Comit\'e \'ethique et scientifique de Parcoursup, \href{https://services.dgesip.fr/fichiers/Rapport_du_CESP_2019__janvier_2020_.pdf}{\emph{Rapport au Parlement}}, January~2020, Section~1.2, p.~8. We translate ``n'est ensuite plus modifi\'ee quels que soient les d\'esistements.''} The committee contrasts this interpretation with one where, if a grant recipient refuses her place, another grant recipient moves up the list. It does not recommend replacing the existing procedure, but asks that the interpretation and its consequences be made clear.\footnote{\citet{Schummer2021} examines the related question of whether an applicant's choice to reject an offer should maintain her relative standing on a waiting list. Here the question is whether a past offer still counts toward a group's protection.} The two interpretations are two different accounts. One counts the offers made and the other the offers still held; they need not give the same protection or the same next recipient.\footnote{\citet{Sokolov2026} develops an alternative offer procedure that retains offers still held, recalculates the remaining targets after refusals, and uses a choice rule for overlapping targets to fill the remaining places. \citet{CombeSokolov2026} discuss the proposal and its policy implications.} 

A similar contrast, between being offered a therapy and actually receiving it, is raised by \citet{RubinEtAl2021} with respect to Mass General Brigham's distribution of COVID-19 monoclonal antibodies.

They find that disproportionate refusals by socially vulnerable patients partly offset the intended increase in treatment, even though the reserve policy increased both offers and treatment compared to a lottery without the policy. 

Both applications illustrate how refusals can lower a group's share among those who enroll or receive treatment without changing the fact that its members were given these opportunities. 
An account evaluating treatment received would not treat a refused offer as treatment. An account evaluating opportunities could still count it.
In Parcoursup, the refusing applicant who no longer wants to enroll remains relevant because she received an offer before. Her refusal affects whether she is considered for a new offer, but need not take her out of the policy assessment.
The distinction between who is relevant to the policy and who is considered for current selection is one of several that the account needs to settle.  
In the next section, we specify the current problem in general terms and the records the ledger must keep to form it.

\section{Account Rules beyond Dynamic Resource Releases}\label{sec:state}

In Section~\ref{sec:diagnosis}, the account specified how earlier allocations enter the current problem when positions are released over time.
In general, its role is broader. In making their choices, institutions rely on existing records to decide who is eligible, what resources are available, and which claims must be honored.
These questions arise even in a single allocation decision, such as determining promotion eligibility from employment records or recognizing existing qualifications when implementing a reform.
In this section we describe what else the account may have to settle in forming the current problem, identify the records required for subsequent choices, and explain how policy requirements can restrict the account and choice rule jointly.

\subsection{The Current Allocation Problem}\label{subsec:publicstate}

Let $\mathcal I$ denote the universe of past, present, and potential future participants.
People can join, leave, and come back. The current population, eligibility, and priority may evolve over time. We use the date index $t$ for these inputs when necessary.

Fix a decision date and the assessment period, if any, specified by the policy. The \emph{set of relevant individuals} $\mathcal I_t^{\mathrm{rel}}\subseteq\mathcal I$ consists of those people whose applications, previous allocations, or ongoing claims enter the account. The \emph{current population} $I_t\subseteq\mathcal I_t^{\mathrm{rel}}$ consists of those being considered for current selection.
For instance, in Parcoursup, where protection is assessed by offers made (Section~\ref{app:parcoursup}), an applicant who rejected an offer remains relevant because she received an admission opportunity, even if she no longer seeks a new offer. Depending on the application, the current population can also include earlier recipients, as under cumulative implementation.

The \emph{set of relevant resources and opportunities} $\mathcal U_t^{\mathrm{rel}}$ contains the objects or allocation opportunities whose status enters the account.
Those available for current use form the subset $\mathcal U_t^{\mathrm{av}}\subseteq\mathcal U_t^{\mathrm{rel}}$.
What is allocated, how earlier allocations are counted, and what obligations there are determine whether a resource or opportunity is included in either set.

Hospital beds are durable and reusable. A bed is relevant while occupied and available afterwards. Vaccine doses are consumable and can be perishable: an injection uses one up, or a dose could go bad before being used.
Obligations can keep a consumed resource relevant. For a two-dose vaccination course, the record of the first dose matters because a second dose is still owed.
A resource is also different from the opportunities to use it.
When the holder of a position leaves, there is a new opportunity to fill it, even though no new position is created.
An admission offer is an opportunity to enroll but not enrollment itself. These features have implications for what the account counts and which resources are available for the current allocation.

Reconsidering an earlier allocation in the account does not mean that it will be allocated again. An account that counts earlier recipients under cumulative implementation includes them and the positions they hold, but a recipient whose category changes retains her appointment. Likewise, continuing an appointment does not allocate a second position. A cumulative account of treatments already given includes the previous recipients and the resources consumed in treating them. By contrast, completing a vaccination course requires an additional dose, so the account subtracts the doses owed to earlier recipients before the choice rule selects new recipients of identical courses. In every case, the objects or opportunities being allocated are interchangeable within an institution, so once the account determines what is available, their numbers can be represented by capacities. Category assignments do not refer to specific physical objects. If categories are attached to positions, a reassignment may mean moving to a different but interchangeable position,
subject to any binding constraints on the holder.\footnote{We develop completion obligations and post-based and vacancy-based recruitment in Appendices~\ref{app:completiondetails} and~\ref{app:postdetails}.}

\begingroup\clubpenalty=300\widowpenalty=300\relax
The specification of the current problem $\sigma_t$ consists of some or all of the following, depending on the application.
\begin{enumerate}[label=(\roman*),beginpenalty=0]
\item the sets of relevant individuals and resources (or opportunities), $\mathcal I_t^{\mathrm{rel}}$ and $\mathcal U_t^{\mathrm{rel}}$,
and the set of available resources (or opportunities) $\mathcal U_t^{\mathrm{av}} \subseteq \mathcal U_t^{\mathrm{rel}}$;
\item the current population $I_t$, assignment categories, recognized eligibility, and priority $\pi_t$;
\item the resource constraints, including category capacities;\footnote{\citet{KumanoKurino} treat school quotas as design variables within a set of implementable distributions. \citet{KhannaEvren2025} and \citet{EvrenKhanna2024} study how reserved positions are apportioned among categories over time in Indian rosters. \citet{DelacretazKominersTeytelboym2023} study refugee resettlement with multidimensional resource constraints reflecting families' needs and communities' capacities. Here capacities are among the inputs formed by the account from records and policy requirements, and are taken as given by the choice rule.}
\item the policy requirements, their assessment periods and counting conventions, and the earlier allocations that count toward them; and
\item the inherited rights that current assignments must preserve.
\end{enumerate}
\endgroup
These components determine the feasible current outcomes, along with the priority and protection requirements by which those outcomes are evaluated.

The choice rule $\CC_t$ is applied to this problem. We suppress components of $\sigma_t$ that are fixed or do not affect the analysis. An institutionally prescribed category precedence is a policy requirement in component~(iv), which we may hold fixed as a parameter of the choice rule. We formalize the inherited rights in component~(v) in Section~\ref{sec:rules}.

Policy requirements can jointly restrict the account and the choice rule. For example, if the account denies a qualified employee reserve eligibility for promotion because her earlier appointment was open,
the O\&A choice rule can promote a lower-priority member of her group while satisfying its allocation axioms at the reported inputs.
Requiring that an earlier open appointment not penalize her in promotion also restricts how the account recognizes eligibility.\footnote{This natural principle is also the judgment of the Punjab and Haryana High Court in \emph{Jugraj Singh v.\ State of Punjab}, which ruled that an earlier open appointment does not remove reserve eligibility for promotion. Appendix~\ref{app:promotionproof} presents the case and the joint characterization with its account restrictions.}

\subsection{Information Needed for Later Decisions}\label{subsubsec:ledgerinformation}

The inputs to the current choice rule may not include all the information needed for future problems. What the ledger should maintain therefore depends on the institution and the choices it will make. The ledger may record, for example, individual identities and qualifications, identities of objects or opportunities, past assignments and their dates and categories, applications, departures, rejections, refusals of offers, releases of resources, consumption, expirations, and outstanding obligations.

Let us go back to California's Assisted Living Waiver, which we described at the beginning of Section~\ref{sec:diagnosis}. Recall that the program limits both the number of people served at any one time and the number of distinct people served over the year, and that a participant who leaves still counts toward the yearly limit. The following scenario illustrates why the ledger needs to keep track of the identities of former participants, even when no former participant is asking for service.

Suppose the program can serve at most two people at a time and at most two distinct individuals during the year. Consider two histories. In the first, the program has served individuals $i$ and $j$ during the year, and $j$ has already left. In the second, it has served individuals $i$ and $k$ during the year, and $k$ has already left. In both histories, $i$ remains enrolled, one place is vacant, and no one is currently waiting. The number of distinct individuals served during the year is also the same. The histories differ only in the identity of the individual who has left.

Now suppose individual $j$ applies for admission in both histories. In the first history, $j$ is a returning participant and, hence, her admission does not increase the number of distinct individuals served this year. In the second history, $j$ has not previously participated during the year; therefore, her admission would exceed the annual limit of two distinct individuals.
Holding the other admission requirements fixed, the annual limit permits her admission in the first history but not in the second.
Her earlier participation affects this count without necessarily giving her any more priority.
If the system retained only the current population, available capacity, and current annual count, it would not have the information it needs to assess future returns. Appendix~\ref{app:medicaidchoice} gives the corresponding current problem and choice rule.

Records of resources can also play a similar role. The US EB-5 immigrant investor program allows unused reserved visa numbers to stay in their category for the following fiscal year. If they are still unused at the end of that year, they are made available to all EB-5 applicants.\footnote{The carryover rule is in the Immigration and Nationality Act, \href{https://www.govinfo.gov/content/pkg/USCODE-2024-title8/html/USCODE-2024-title8-chap12-subchapII-partI-sec1153.htm}{8 U.S.C.\ \S1153(b)(5)(B)}, as amended by the EB-5 Reform and Integrity Act of 2022. The State Department \href{https://travel.state.gov/content/travel/en/News/visas-news/annual-EB-5-unreserved-category-limit-reached-2025.html}{reported} in September~2025 that unused reserved numbers from fiscal year 2023 had been added to that year's unreserved supply.}
Two different records could therefore present the same capacity for reserved positions to the current choice rule, yet contain visa numbers from different years.
If those numbers stay unused, their categories need not change in the same way at the next fiscal year. The ledger contains the year in which the resources entered,
while the account uses those records to determine capacities for categories.

These examples illustrate why information sufficient for the current choice rule might not be sufficient to construct future problems.
The account can describe resources as capacities, while the ledger preserves identity and date information for evaluating a future return or a change in the category of a resource.

\section{Reserve System Design with Inherited Appointments}\label{sec:rules}

In Section~\ref{sec:state}, we discussed how earlier allocations shape the current problem through several components of the account. In the rest of the paper, we focus on one of them: the inherited appointments that current choice must respect. With population, eligibility, and priority fixed, cumulative implementation of a capacity-monotone choice rule preserves earlier appointments as capacities expand. With other changes, however, such as new applicants or changing priorities, the same choice rule, applied to the updated problem, may exclude a recipient whose appointment must be honored. In this case, the account reports the protected appointment, and the choice rule must be extended to accommodate it.
 We next formalize inherited rights and identify when all incumbents can be retained, and then introduce and study choice rules that maintain over-and-above protection subject to these commitments.

\subsection{Inherited Appointments and Retention Feasibility}\label{subsec:flexrule}

Consider one open position and one reserved position at category $c$. There are three individuals $i$, $j$, and $k$, ranked by priority in that order, where $j$ and $k$ are eligible for category $c$, whereas $i$ is not. Initially, only $k$ is available and the open position is released. The O\&A choice rule selects her. Individuals $i$ and $j$ then arrive when the reserved position becomes available. Individual $k$ is now an incumbent who holds the open position, and the reserved position is vacant. With no inherited rights, the O\&A choice rule assigns the open position to $i$ and the reserved position to $j$, rejecting $k$. 
Now suppose that incumbent $k$'s appointment is protected. If her category is also locked, non-wastefulness requires assigning the reserved position to $j$, and thus the highest-priority individual $i$ is rejected. On the other hand, if her category is not locked, incumbent $k$ can instead be appointed in category $c$, and the open position is then available for individual $i$. Every non-wasteful outcome that appoints $k$ in category $c$ assigns the open position to $i$ or to $j$, but only the former respects their priority order. Locking $k$'s category thus rejects $i$ in favor of $j$, who has lower priority, even though $k$'s appointment can be honored in category $c$.

The three cases have the same population, eligibility, priority, and capacities at the subsequent decision. The only difference between them is how the account treats the incumbent's prior appointment: it is either ignored, retained with its category locked, or retained with its category left flexible. The three outcomes exclude different people. Only the third both honors the appointment and respects priority among the others.

A commitment to honor an earlier appointment can emerge in various ways. Sometimes it is built into routine functioning. A hospital cannot take back a vaccine given to a lower-priority patient when a higher-priority patient comes along, and in continuing care it can opt to protect a patient's use of a ventilator.\footnote{In their Appendix~A.2, \citet{PSUY2024} discuss withdrawal and the protection of earlier patients in ventilator allocation. The analysis of this section applies to such cases whenever the resulting choice problem has a common priority and reserve eligibility of the form assumed here.} Other commitments arise because of appointments made under an earlier policy (or through a mistaken implementation) that nevertheless must be honored.  Which commitments have to be honored today? The account answers this, using the history and the policy in force. It enters the protected appointments into the current problem through the inherited profile. The reason for a protection is not part of the current problem, and the choice rule does not depend on it. Nor does protecting an appointment settle which category it counts in. That is a separate matter.

We now formalize these inherited rights. At each decision, the account provides the population $I$, which includes incumbents as well as applicants without protected appointments, the categories $\mathcal V$ with eligibility sets $(I^v)_{v\in\mathcal V}$, the capacities $q$, which count occupied positions as well as vacancies, the priority $\pi$, and the inherited rights. For each category $v\in\mathcal V$, let $\mathrm H^v\subseteq I$ be the set of incumbents whose appointments are currently attributed to category $v$. 
Let $\mathrm H=(\mathrm H^v)_{v\in\mathcal V}$ denote the \emph{inherited category-assignment profile} and $\widehat{\mathrm H}=\bigcup_{v\in\mathcal V}\mathrm H^v$ denote the \emph{set of incumbents}. This is a category-assignment profile as defined in Section~\ref{para:allocationmodel}, although one generated by the account rule as part of the input for the choice rule rather than an outcome. 
In the absence of additional structure, the current problem---the input to the current choice rule---is
\[
\sigma=\bigl(I,\mathcal V,(I^v)_{v\in\mathcal V},q,\pi,\mathrm H\bigr),
\]
which we write as $\sigma=(I,q,\mathrm H)$ when the other inputs are fixed.

\begin{definition}\label{def:currentretention}
A choice rule $\CC$ satisfies \emph{incumbent retention} if, at each problem $\sigma$,
\[
\widehat{\mathrm H}\subseteq\widehat\CC(\sigma).
\]
It satisfies \emph{incumbent category locking} if, at each problem $\sigma$ and each $v\in\mathcal V$,
\[
\mathrm H^v\subseteq\CC^v(\sigma).
\]
\end{definition}

The account lets us analyze each decision in a dynamic implementation as a static choice problem.
Under cumulative implementation in Section~\ref{sec:diagnosis}, it presents the fixed population at the cumulative capacities.
Here it also reports the appointments to be honored through the inherited profile.
Recipient retention (Definition~\ref{def:recipientretention}) is a relational axiom comparing outcomes across release histories. Its punctual counterpart at a given problem is incumbent retention.
Suppose the account reports all earlier recipients as incumbents. Then incumbent retention at each current problem yields recipient retention across dates.
A requirement linking successive outcomes thus becomes a requirement on current choice. We can impose such a requirement at the problems the account generates, and investigate the choice rule one problem at a time.

Retention constrains the current outcome through the incumbent set $\widehat{\mathrm H}$ alone; locking also makes the inherited category-assignment profile $\mathrm H$ relevant, since each incumbent's earlier category assignment must be preserved as well.

\begin{definition}\label{def:retention-feasibility}
A problem is \emph{retention-feasible} when there exists a feasible outcome that appoints every incumbent. 
\end{definition}

For each $J\subseteq I$, let $\rho_q(J)$ be the largest number of its members who can simultaneously be assigned to eligible reserved positions.\footnote{\label{fn:hall}Let individuals and positions form the two sides of a bipartite graph, 
with an edge for each eligible assignment.
By Hall's theorem \citep{Hall1935}, a set of individuals can be covered by a graph matching if and only if every subset of it is eligible, collectively, for at least as many positions as it has members.
Adding the open positions, for which every incumbent is eligible, turns this condition into the inequality in Lemma~\ref{lem:retentionfeasibility}.} 
Retaining all incumbents therefore requires at least $|\widehat{\mathrm H}|-\rho_q(\widehat{\mathrm H})$ open positions. Conversely, if this many open positions are available, we can assign a largest possible set of incumbents to eligible reserved positions and the remaining incumbents to open positions. We state this observation as a lemma.\footnote{A correction may require more than retaining earlier appointments. In \emph{BSNL v. Sandeep Choudhary}, the court protected existing appointments while requiring category reattribution and a new appointment. Retention was feasible at the advertised capacities, but the additional directions required more positions. We distinguish these resource demands and the adjustment to ordinary open selection in Appendix~\ref{app:bsnldetails}.}

\begin{lemma}\label{lem:retentionfeasibility}
A problem is retention-feasible if and only if
\begin{equation}\label{eq:flexfeasibility}
|\widehat{\mathrm H}|-\rho_q(\widehat{\mathrm H})\le q^o.
\end{equation}
\end{lemma}

\subsection{Over-and-Above Protection with Inherited Appointments}\label{subsec:retainedoa}

Under an over-and-above policy, beneficiaries whose priority suffices for an open position receive one without using the positions reserved for their groups. Leaving those reserved positions for additional beneficiaries provides the boost. 
In Section~\ref{sec:diagnosis}, we showed how category locking can largely eliminate the boost during repeated allocation.

Inherited appointments pose a different challenge in implementing this policy, and one that arises in the applications of RRB in Sections~\ref{sec:casestudies} and~\ref{sec:multiapplications}. 
We refer to the static choice rule the institution would use in the absence of inherited rights as its \emph{ordinary} choice rule; 
with nonoverlapping reserves under an over-and-above policy, this is the O\&A choice rule of Section~\ref{para:allocationmodel}. 
Applied to a problem with incumbents, the ordinary rule violates incumbent retention: it may reject an incumbent whose appointment must be honored.
We seek to retain such incumbents while maintaining non-wastefulness and honoring priority and protection claims as far as their appointments permit. Following minimalist market design \citep{Sonmez2023MMD}, we ask how an institution can meet these commitments with the smallest departure from its ordinary choice rule. The retention construction below also extends to other category precedences, but the principles to preserve depend on the policy; we return to this distinction in Section~\ref{subsec:otherprecedences}.

To this end, we introduce the \emph{retain, reassign, and boost (RRB)} choice rule. \emph{Retain} captures the protection of every incumbent's appointment. \emph{Reassign} captures the flexibility of an incumbent's category, so that honoring her appointment does not fix the category through which she was selected. \emph{Boost} captures consideration for open positions before the reserved ones, subject to retaining incumbents.

Suppose an incumbent was selected through a reserved category. RRB starts by evaluating her for open positions according to priority, subject to retaining all incumbents. If she receives one, she no longer counts toward the reserved capacity for her group, and another eligible person can benefit from the reserved position. Reassignments can also be made in the other direction. Moving an incumbent from the open category to an eligible reserved category could allow a higher-priority person to receive an open position.
Both types of reassignment keep the appointment, but their purposes differ. 
One restores an opportunity for members of the protected group, in the spirit of over-and-above protection.
The other prevents an unnecessary departure from priority.

\subsubsection{RRB Choice Rule with Nonoverlapping Reserves}\label{subsec:basicrrb}

We first introduce RRB for nonoverlapping reserved categories, the domain of the O\&A choice rule in Section~\ref{para:allocationmodel}. Let $\CC_{\mathrm{RRB}}$ denote the RRB choice rule under nonoverlapping reserves. On this domain, the largest number of members of $J\subseteq I$ who can simultaneously hold eligible reserved positions is
\[
\rho_q(J)=\sum_{c\in\mathcal R}\min\{|J\cap I^c|,q^c\},
\]
since each category can accommodate up to its capacity from the eligible members of $J$ and nonoverlapping eligibility makes these counts additive.
Consider a retention-feasible problem $\sigma$.

\choiceheading
{The RRB choice rule, $\CC_{\mathrm{RRB}}$.}{}
\begin{choicesteps}

\choicestep \emph{Open positions.}\label{step:basicopen} Start with $X=\emptyset$ and consider individuals in priority order. Select each individual if her selection leaves enough positions to retain all remaining incumbents, allowing their categories to change.

Specifically, while $|X|<q^o$, consider individual $i$ and let $K=\widehat{\mathrm H}\setminus(X\cup\{i\})$ be the incumbents who would remain outside the open category after selecting her. Add $i$ to $X$ if
\begin{equation}\label{eq:opentest}
|K|-\rho_q(K)\le q^o-|X|-1.
\end{equation}
The left side is the open capacity that the remaining incumbents need, by Inequality~\eqref{eq:flexfeasibility} applied to them; the right side is the open capacity that would remain. Continue until open positions are filled or every individual has been considered. Assign $X$ to open positions, so that $\CC_{\mathrm{RRB}}^o(\sigma)=X$.
\choicestep \emph{Reserved positions.} For each reserved category $c$, first assign positions to its eligible incumbents who were not selected for the open category. Fill the remaining category-$c$ positions with the highest-priority eligible individuals who remain unselected. Letting $K^c=(\widehat{\mathrm H}\medcap I^c)\setminus X$, the category-$c$ selection is
\[
\CC_{\mathrm{RRB}}^c(\sigma)=K^c\cup\TopSet_{q^c-|K^c|}\bigl(I^c\setminus(X\cup K^c)\bigr).
\]
\end{choicesteps} \smallskip

Without incumbents, Step~1 assigns the open positions in priority order, so the two steps give the O\&A choice rule. When Step~\ref{step:basicopen} ends, every incumbent still unassigned can be accommodated in her eligible reserved category in Step~2.

In the example of Section~\ref{subsec:flexrule}, $\CC_{\mathrm{RRB}}$ selects $i$ for the open category and retains $k$ in category $c$. Retaining $k$ requires departing from ordinary O\&A, which would select $i$ and $j$. It does not require excluding the highest-priority individual $i$, as locking $k$ in the open category would do.

\subsubsection{Variants of the RRB Choice Rule with Overlapping Reserves}\label{subsec:RRBvariants}

With overlapping eligibility for reserved categories, the same minimalist objective requires an additional decision about how reserved positions are allocated. \emph{Joint RRB} considers these positions jointly by priority. \emph{Precedence RRB} follows a prescribed precedence order. Both use the same open-selection step as the RRB choice rule, although the calculation of $\rho_q$ differs with overlapping eligibility, and both reduce to $\CC_{\mathrm{RRB}}$ when reserves do not overlap.

\choiceheading{Joint RRB choice rule, $\CC_{\parallel}$.}{}
\begin{choicesteps}
\choicestep \emph{Open positions.}\label{step:parallelopen} Apply Step~\ref{step:basicopen} of the RRB choice rule, 
where $\rho_q(K)$ is now the size of a maximum matching of $K$ into eligible reserved positions, as in footnote~\ref{fn:hall}, rather than the additive formula of Section~\ref{subsec:basicrrb}.
 Assign the selected set $X$ to open positions, so that $\CC_{\parallel}^o(\sigma)=X$.
 \choicestep \emph{Reserved positions.} Start with the set $K=\widehat{\mathrm H}\setminus X$ of incumbents not selected for open positions. Consider the remaining individuals, those who are neither incumbents nor in $X$, in priority order. Add an individual to $K$ if the enlarged set can still be matched to eligible reserved positions, as in footnote~\ref{fn:hall}. Continue until every individual has been considered, and assign the final set $K$ to eligible reserved positions.
\end{choicesteps} \smallskip
As in the basic rule, Step~1 uses Inequality~\eqref{eq:opentest} to ensure that every remaining incumbent can still be assigned a position. If some incumbents are not selected for open positions at the end of Step~1, they can all be assigned eligible reserved positions. This set is the starting point for Step~2, so all of these incumbents are retained. In Step~2, the rule selects another individual only if every member of the enlarged set can be assigned to an eligible reserved position (i.e., only if $\rho_q(K\cup\{i\})=|K|+1$). Category assignments remain flexible until all selections are made. Precedence RRB, presented below, must also allow all incumbents to be retained, but it assigns individuals to categories as it proceeds. Each assignment therefore reduces the capacity available for the remaining incumbents.

We use a \emph{fixed attribution convention}, independent of which selected individuals are incumbents and of their earlier categories, to determine their reserved-category assignments, and keep that convention unchanged when comparing full outcomes.\footnote{We specify the convention in Definition~\ref{def:attributionconvention} in Appendix~\ref{app:simpleresults}.}

An institution may instead specify an order in which reserved categories are processed, so that an individual eligible for several of them is considered for the earlier category first.\footnote{For UFC, see Section~\ref{app:ceara}. Turkey's state-university placement considers applicants first for general places, then for places reserved for school valedictorians, applicants affected by the 2023 earthquakes, and women aged 34 or older, in that order. See \"OSYM, \href{https://dokuman.osym.gov.tr/pdfdokuman/2024/YKS/TERCIH/tkilavuz_02082024.pdf}{\emph{2024 Higher Education Programs and Quotas Guide}}, p.~4.}
Giving a category precedence can leave positions in later categories for individuals who are ineligible for the earlier one. The institution then respects the selection made in an earlier category when considering claims to later positions, even if changing that selection would admit a higher-priority individual elsewhere.

Let $\triangleright$ be a strict precedence order, with the open category first. 
When institutional policy prescribes this precedence order, the account reports the requirement. Otherwise, the precedence order is a separate design lever. The full specification for precedence RRB is $(\sigma;\triangleright)$, where $\sigma$ contains the inputs specified in Section~\ref{subsec:flexrule}; the notation $\CC_{\triangleright}(\sigma)$ fixes the precedence order in the rule's index and suppresses it from those inputs.

\choiceheading{Precedence RRB choice rule, $\CC_{\triangleright}$.}{}
\begin{choicesteps}
\choicestep \emph{Open positions.} Apply Step~\ref{step:parallelopen} of the joint RRB choice rule, and assign the selected set $X$ to open positions, so that $\CC_{\triangleright}^o(\sigma)=X$.
\choicestep \emph{Reserved positions.} Process reserved categories in the prescribed precedence order $\triangleright$.
Within each category $c$, select individuals one at a time, up to its capacity.
At each selection, let $K$ be the set of incumbents not yet assigned during the procedure. Let $\bar q$ be the capacities that would remain after assigning one additional individual to category $c$,
consisting of the remaining capacity in $c$ plus the capacities of all later categories.

Among the unassigned individuals eligible for category $c$, select the highest-priority individual $i$ who satisfies
\begin{equation}\label{eq:sequentialtest}
\rho_{\bar q}(K\setminus\{i\})=|K\setminus\{i\}|.
\end{equation}
Assign her a position in category $c$. If she is an incumbent, remove her from $K$.
Continue within category $c$ until its capacity is filled or no such
individual exists. Leave any remaining capacity in $c$ unfilled and
proceed to the next category.
\end{choicesteps} \smallskip
In Step~1, Inequality~\eqref{eq:opentest} ensures that all remaining incumbents can still be assigned positions. At the end of Step~1, if there are any unassigned incumbents, they can all be assigned eligible reserved positions. However, selecting according to priority in each reserved category might compromise this possibility by using positions that an incumbent needs. Equality~\eqref{eq:sequentialtest} extends this requirement to reserved selection. In particular,
after assigning individual $i$ to a position in category $c$, all remaining incumbents must still be assignable to eligible positions within the remaining capacity of $c$ and the later categories.
The assignments that have already been made by the procedure stay fixed. The rule selects the eligible individual of highest priority who can be assigned while maintaining a feasible completion for all remaining incumbents. Equality~\eqref{eq:sequentialtest} can rule out assigning two types of individuals to the current category. The first type is a nonincumbent who would occupy a position that the remaining incumbents need. 
The second type is an incumbent who, when assigned to the current category, would prevent the remaining incumbents from all obtaining eligible positions.

For instance, suppose that, after assigning positions in the open category, there are two reserved categories with one position each and two incumbents who are still unassigned. The higher-priority incumbent is eligible for both, and the lower-priority incumbent is only eligible for the first category processed. Suppose there is also a nonincumbent, eligible only for the first category, who is of higher priority than both incumbents. If this individual is assigned to the first position, then only one position will be left for the two incumbents, so Equality~\eqref{eq:sequentialtest} rules it out. If the higher-priority incumbent is assigned to the first position, then the remaining incumbent will not have any position for which she is eligible, so Equality~\eqref{eq:sequentialtest} rules this out as well. The position goes to the lower-priority incumbent, and the higher-priority incumbent receives the position in the later category. Applying Equality~\eqref{eq:sequentialtest} throughout Step~2 ensures that every incumbent receives a position.

The two steps apply the same completion requirement, starting with the open category. We keep them separate to make the over-and-above structure and the common open selection of the RRB rules visible.

\subsubsection{The Cost of Protecting Incumbents}\label{par:nohrchoice}

For either RRB rule, choice depends on the inherited category-assignment profile only through its incumbent set. We therefore write $\sigma[K]$ for the problem obtained from $\sigma$ by replacing its protected set with $K\subseteq I$, holding the population, capacities, eligibility, priority, and any category precedence fixed. The members of $K$ must remain appointed, but their category assignments may change.

For $\CC\in\{\CC_{\parallel},\CC_{\triangleright}\}$, its ordinary benchmark is $\CC(\sigma[\emptyset])$. Applying the rule without protecting an incumbent's appointment leaves her in the population. With nonoverlapping reserves, $\CC_{\mathrm{RRB}}(\sigma[\emptyset])=\CC_{\mathrm{OA}}(I,q)$. With overlapping reserves, the ordinary joint and precedence outcomes can differ. Both RRB rules reproduce their corresponding benchmark whenever it already selects every incumbent. At each retention-feasible problem $\sigma$,
\begin{equation}\label{eq:benchmarkrecovery}
\CC(\sigma)=\CC(\sigma[\emptyset])\quad\Longleftrightarrow\quad
\widehat{\mathrm H}\subseteq\widehat\CC(\sigma[\emptyset]).
\end{equation}

An applicant selected by ordinary choice can lose her open position when incumbents are protected, yet still receive a reserved position. Alternatively, she can lose selection altogether. We distinguish these two costs, beginning with the changes that retention requires in the ordinary selection for open positions. Suppose $q^o\le |I|$, and let $O=\TopSet_{q^o}(I)$ be the individuals who receive open positions when no appointments are protected. Incumbents outside $O$ can be retained without changing this selection if they can be assigned to reserved categories. Any shortfall must instead be accommodated through open positions.

\begin{proposition}\label{prop:openadjustment}
At a retention-feasible problem with $q^o\le |I|$, the common selection module for open positions under joint and precedence RRB displaces the smallest possible number of members of $O=\TopSet_{q^o}(I)$ among all feasible sets of recipients filling the open positions and permitting retention of every incumbent. This number is
\[
|O\setminus\CC_{\parallel}^o(\sigma)|
=|O\setminus\CC_{\triangleright}^o(\sigma)|
=|\widehat{\mathrm H}\setminus O|-\rho_q(\widehat{\mathrm H}\setminus O).
\]
Moreover, for each $k\le q^o$, its $k$th highest-priority recipient of an open position has weakly higher priority than the $k$th highest-priority member of any other such selection.
\end{proposition}

Our next result concerns the set of recipients.

\begin{proposition}\label{prop:legacybound}
Fix $\CC\in\{\CC_{\parallel},\CC_{\triangleright}\}$. For any retention-feasible incumbent set $K$,
\[
\begin{aligned}
\widehat\CC(\sigma[K])\setminus\widehat\CC(\sigma[\emptyset])
 &=K\setminus\widehat\CC(\sigma[\emptyset]),\\
|\widehat\CC(\sigma[\emptyset])\setminus\widehat\CC(\sigma[K])|
 &\le |K\setminus\widehat\CC(\sigma[\emptyset])|.
\end{aligned}
\]
The inequality holds with equality whenever the two recipient sets have equal cardinality.
\end{proposition}
Every recipient selected by RRB but not by ordinary choice is an incumbent whose appointment must be protected. If there are $k$ such incumbents, protecting them displaces at most $k$ ordinary recipients.

\subsection{Characterizations of RRB Choice Rules}\label{subsec:rrbcharacterizations}

As we discussed in Section~\ref{para:allocationmodel}, when reserves are nonoverlapping, the O\&A choice rule is characterized by non-wastefulness, no justified envy, and maximal reserve protection.
The last axiom captures the institution's over-and-above policy. With inherited appointments, we not only have an additional axiom, incumbent retention, but also need to modify no justified envy and maximal reserve protection.
We do this in our broader model with overlapping reserves and study the joint implications of these axioms.

We modify how a priority claim is evaluated under no justified envy and maximal reserve protection. Without incumbent retention, non-wastefulness and no justified envy require that an unselected individual can neither fill a vacant position in an eligible category nor take such a position from a lower-priority recipient. With inherited appointments, we instead ask whether there is a feasible outcome that honors her claim while still retaining all incumbents. This changes the comparison in two ways. First, a claim that could be honored only by removing a protected appointment is not a valid claim. Second, we retain the incumbents in their appointments but not necessarily in their categories. A claim may be honored by moving an incumbent to another eligible category, subject to the priority and precedence restrictions specified below. The modified axioms allow for exactly the kinds of category changes that retain all incumbents.

In the case of overlapping reserves, we also need to determine how to compare priority claims across reserved categories. Under joint consideration, the reserved categories are on equal footing: 
we evaluate claims together and may reassign people across reserved categories. Under prescribed category precedence, a claim to a category must leave the selections in higher-precedence categories unchanged.
Under both approaches, open positions are considered before reserved ones, so the comparison for open positions is the same.

To see the difference, hold the recipients of open positions fixed and consider the reserved categories alone. Suppose an incumbent is eligible for two reserved categories, each with one position.
She holds the position in category $c$, and the other position is vacant.
A nonincumbent applicant with lower priority is eligible only for $c$. If the institution compares claims jointly, it can move the incumbent to the other category and give $c$ to this applicant; nobody loses an appointment.
If instead $c$ comes before the other category in the precedence order, the institution fills $c$ first and does not revisit it, so the nonincumbent applicant is not selected.

Given an outcome $\mathrm G$, let $\mathrm G^{\mathcal R}=\bigcup_{c\in\mathcal R}\mathrm G^c$ denote the set of recipients of reserved positions. We define the corresponding claims by asking which reassignments can honor them while maintaining all incumbents.

\begin{definition}\label{def:openclaims}
Fix a problem $\sigma$ and an outcome $\mathrm G$ retaining every incumbent. Each claim below is defined by the existence of a feasible outcome $\mathrm F$ retaining every incumbent and satisfying the stated conditions.
\begin{enumerate}[label=(\roman*)]
\item An individual $i\in I\setminus\mathrm G^o$ has a \emph{feasible open priority claim} if
\[
i\in\mathrm F^o,\qquad
\{j\in\mathrm G^o:j\mathrel{\pi}i\}\subseteq\mathrm F^o.
\]
\item An unselected individual $i\in I\setminus\widehat{\mathrm G}$ has a \emph{feasible joint reserved priority claim} if
\[
\mathrm F^o=\mathrm G^o,\qquad i\in\mathrm F^{\mathcal R},\quad \mbox{and} \quad 
\{j\in\mathrm G^{\mathcal R}:j\mathrel{\pi}i\}\subseteq\mathrm F^{\mathcal R}.
\]
\item Fix a precedence order $\triangleright$ with the open category first. An individual $i\in I$ has a \emph{feasible priority claim under precedence to category $v$}, where $i\notin\mathrm G^v$, if
\[
\mathrm F^w=\mathrm G^w\quad\text{for each }w\triangleright v,\qquad
i\in\mathrm F^v, \quad \mbox{and} \quad
\{j\in\mathrm G^v:j\mathrel{\pi}i\}\subseteq\mathrm F^v.
\]
She has a \emph{feasible priority claim under precedence} if she has such a claim to some category.
\end{enumerate}
\end{definition}
The requirements concerning higher-priority recipients prevent a claimant from displacing them from the positions over which her claim is evaluated.
In each case, $\mathrm F$ is an alternative outcome that honors $i$'s claim while preserving all protected appointments. What $\mathrm F$ has to keep fixed depends on the claim. With an open claim, any recipient of an open position who has higher priority than $i$ keeps her open position. Everyone else's category can change. For a joint reserved claim, the recipients of open positions stay fixed, and every higher-priority recipient of a reserved position remains selected, although her reserved category may change. For a claim under precedence to category $v$, the selections in all categories ahead of $v$ stay fixed, the higher-priority recipients of category $v$ stay there, and later categories may be reassigned. For $v=o$, this is the open claim in part~(i). Unlike part~(ii), part~(iii) applies to recipients of later categories as well as to unselected individuals; Definition~\ref{def:retainedprecedence} uses it for the former.

\begin{definition}\label{def:flexNJE}
A choice rule $\CC$ satisfies \emph{no justified envy subject to retention with joint comparisons of reserved categories} if, for each problem $\sigma$ and its outcome $\mathrm G=\CC(\sigma)$, no unselected individual has a feasible open priority claim or a feasible joint reserved priority claim.
\end{definition}

\begin{definition}\label{def:precedenceNJE}
A choice rule $\CC$ satisfies \emph{no justified envy subject to retention under precedence $\triangleright$} if, for each problem $\sigma$ and its outcome $\mathrm G=\CC(\sigma)$, no unselected individual has a feasible priority claim under precedence.
\end{definition}

With nonoverlapping reserves, the joint claims for reserved categories reduce to separate category comparisons, and Definitions~\ref{def:flexNJE} and~\ref{def:precedenceNJE} coincide. 
In this case we refer to the axiom simply as \emph{no justified envy subject to retention}.

Whereas no justified envy concerns unselected individuals, maximal reserve protection concerns a recipient of a reserved position who has a feasible claim to open consideration.
Honoring this claim permits her group's reserved position to benefit another eligible individual. The following condition extends maximal reserve protection to problems with inherited appointments.
\begin{definition}\label{def:flexVR}
A choice rule $\CC$ satisfies \emph{maximal reserve protection subject to retention} if, for each problem $\sigma$ and its outcome $\mathrm G=\CC(\sigma)$, no recipient of a reserved position has a feasible open priority claim.
\end{definition}
In particular, a reserved position can be used only when all open positions are filled. Otherwise, moving its recipient to a vacant open position would honor such a claim while retaining every incumbent.

On retaining outcomes, either version of no justified envy subject to retention implies non-wastefulness.\footnote{This is different from ordinary no justified envy in Section~\ref{para:allocationmodel}, which does not rule out a vacant position.}
To see why, suppose an individual is unselected and a position in one of her eligible categories is vacant. 
An outcome that awards her this position and leaves every other assignment unchanged is not only feasible but also retains all incumbents. Thus she has a feasible priority claim, and therefore the original outcome violates no justified envy subject to retention. Hence, 
an outcome that satisfies this axiom cannot have any such vacancy. For this reason, we do not impose non-wastefulness separately in the following characterizations.

\begin{theorem}\label{thm:flexcharacterization}
On the retention-feasible domain of this section, a choice rule satisfies incumbent retention, no justified envy subject to retention with joint comparisons of reserved categories, and maximal reserve protection subject to retention if and only if its set of recipients of open positions and its aggregate recipient set coincide with those of the joint RRB choice rule $\CC_{\parallel}$.
\end{theorem}
Theorem~\ref{thm:flexcharacterization} applies even when the inherited appointments were made under another procedure. 
The theorem's axioms pin down who receives open positions and who receives a position at all. In some cases, they do not pin down how those selected for reserved positions are distributed among overlapping categories. 
When reserves do not overlap, this innocuous multiplicity disappears.

\begin{corollary}\label{cor:basicrrbcharacterization}
On the retention-feasible domain with nonoverlapping reserves, a choice rule satisfies incumbent retention, no justified envy subject to retention, and maximal reserve protection subject to retention if and only if it is the RRB choice rule $\CC_{\mathrm{RRB}}$.
\end{corollary}
With no incumbents, retention imposes no restriction on claims. On this nonoverlapping domain, the axioms together are equivalent to those of Section~\ref{para:allocationmodel} and characterize the O\&A choice rule.

Our next axiom concerns prescribed precedence: once the recipients in a category have been selected, a claim to a later category cannot change them.
\begin{definition}\label{def:retainedprecedence}
A choice rule $\CC$ satisfies \emph{compliance with category precedence subject to retention} if, for each problem $\sigma$ and its outcome $\mathrm G=\CC(\sigma)$, no recipient of a position in category $v$ has a feasible priority claim under precedence to any category $w\triangleright v$.
\end{definition}
Compliance with category precedence requires that a recipient use an earlier eligible category whenever she can do so while preserving all incumbent appointments, the higher-priority recipients of that category, and the selections in preceding categories. For the open category, this is maximal reserve protection subject to retention.

\begin{theorem}\label{thm:precedencecharacterization}
Fix a precedence order $\triangleright$ with the open category first. On the retention-feasible domain of this section, a choice rule satisfies incumbent retention, no justified envy subject to retention under precedence $\triangleright$, and compliance with category precedence subject to retention if and only if it is the precedence RRB choice rule $\CC_{\triangleright}$.
\end{theorem}
Unlike Theorem~\ref{thm:flexcharacterization} for joint RRB, Theorem~\ref{thm:precedencecharacterization} characterizes the entire outcome of precedence RRB.

\subsection{Locking Incumbent Categories}\label{subsec:lockedchoice}
So far, we have assumed that the institution could change the categories of appointments it must retain.
If those categories also bind, the account must report this additional constraint and the institution must keep the assignments fixed.
Comparing this treatment with RRB separates the departures required to retain appointments from those caused by locking their categories.

With nonoverlapping reserves and a feasible inherited profile $\mathrm H=(\mathrm H^v)_v$, the \emph{retain-and-lock} choice rule $\CC_{\boxdot}$ keeps these assignments and applies the O\&A choice rule to the remaining population $I\setminus\widehat{\mathrm H}$ and capacities $(q^v-|\mathrm H^v|)_v$. It adds the resulting assignments to $\mathrm H$.
With overlapping reserves, the same construction gives joint and precedence variants by applying the corresponding ordinary rule to the remaining individuals and positions.
As with RRB, the two variants coincide when reserved categories do not overlap.

Starting from an empty outcome and nonoverlapping reserves, repeated $\CC_{\boxdot}$ at the aggregate capacities is the category-locked batch procedure applying the O\&A choice rule to each new batch. With one reserved category and the demand condition in Theorem~\ref{thm:compression}, that theorem shows that this procedure can lose nearly all of the boost that distinguishes over-and-above from minimum guarantee. Locking can also exclude individuals whom retention alone would admit. In the motivating example of Section~\ref{subsec:flexrule}, locking $k$ in the open category excludes $i$, whereas RRB admits $i$ by moving $k$ to category $c$.

Category locking can also prevent an admission after a departure, without any change in capacity. 
Suppose $i_1$ and $i_2$ are eligible for a single reserved category and occupy respectively one open and one reserved position. 
If $i_2$ leaves, moving $i_1$ to the reserved position allows an applicant who is only eligible for an open position to enter.
Locking $i_1$ in the open category leaves the reserved vacancy unusable for that applicant.\footnote{In \emph{Rajiv Mittal v. Maharshi Dayanand University}, a student admitted through a reserved category later qualified for an open vacancy. The university kept his category fixed. Moving him to the open category before offering the vacancy to someone else would have preserved his admission and freed a reserved place for another eligible applicant. Appendix~\ref{app:rajivdetails} discusses the case.}

\subsection{Continuation with Inherited Appointments}\label{subsec:flexcontinuation}

Having characterized RRB at a given problem, we now study its implementation over successive decisions.
After RRB has accommodated the original incumbents and selected additional recipients, the institution may also have to honor the new appointments.
We now ask if protecting these recipients entails any additional deviations from the outcome with only the original commitments.

Throughout Section~\ref{subsec:flexcontinuation}, eligibility, priority, and any category precedence remain fixed across dates. 
The next lemma, which applies at a fixed problem, shows that protecting selected recipients does not affect the outcome as long as the original incumbents remain protected.

\begin{lemma}\label{lem:selectedretention}
Fix $\CC\in\{\CC_{\parallel},\CC_{\triangleright}\}$ and a retention-feasible problem $\sigma$. For every set $K$,
\[
\widehat{\mathrm H}\subseteq K\subseteq\widehat\CC(\sigma)
\quad\Longrightarrow\quad \CC(\sigma[K])=\CC(\sigma).
\]
For joint RRB, full-outcome equality uses the same attribution convention.
\end{lemma}

\subsubsection{Capacity Expansion}\label{subsubsec:rrbexpansion}

In Section~\ref{sec:diagnosis}, we asked whether an institution can allocate positions with its choice rule as they are released and still reach the one-shot outcome.
We ask the same question here in a more general environment, in which some earlier appointments have to be honored.

Fix a retention-feasible initial problem $\sigma_0=(I,q_0,\mathrm H_0)$ and an RRB choice rule $\CC\in\{\CC_{\parallel},\CC_{\triangleright}\}$.
For a finite history $h_t=(\Delta q_1,\ldots,\Delta q_t)$ of nonnegative capacity releases, let
\[
q_t=q_0+\sum_{t'=1}^{t}\Delta q_{t'},\qquad
\sigma_t=(I,q_t,\mathrm H_0).
\]

We call $\CC(\sigma_t)$ the \emph{pooled outcome}.
That is, the rule is run on every position released through date $t$, with $\widehat{\mathrm H}_0$ as the only protected set.
For comparison, $\CC(\sigma_t[\emptyset])$ is the ordinary benchmark from Section~\ref{par:nohrchoice}, which has the same population and capacities and no incumbents protected.

With this initial problem fixed, a \emph{dynamic reserve procedure} assigns to each release history $h$ an outcome $\Phi(h)$ for $\bigl(I,q_0+q(h),\mathrm H_0\bigr)$ that retains every original incumbent.
It starts at $\Phi(\emptyset)=\mathrm G_0=\CC(\sigma_0)$. It satisfies \emph{recipient retention} if $\widehat\Phi(h)\subseteq\widehat\Phi(h')$ whenever $h\preceq h'$.
It satisfies \emph{resource-release invariance} if $\Phi(h)=\Phi(h')$ whenever $q(h)=q(h')$.

Cumulative implementation (Definition~\ref{def:cumulativeimplementation}) therefore gives the pooled outcome $\CC(\sigma_t)$ at each history $h_t$, with the original incumbents protected.

\begin{theorem}\label{thm:flexrelease}
Fix $\CC\in\{\CC_{\parallel},\CC_{\triangleright}\}$ on the domain of this section.
\begin{enumerate}[label=(\roman*)]
\item If $\sigma_0,\sigma_1$ differ only in capacities, with $q_1^v\ge q_0^v$ for each category and the incumbent set retention-feasible at $q_0$, then
\[
\widehat\CC(\sigma_0)\subseteq\widehat\CC(\sigma_1) \qquad \mbox{and} \qquad
\CC\!\left(\sigma_1[\widehat\CC(\sigma_0)]\right)=\CC(\sigma_1).
\]
\item From a retention-feasible initial problem, cumulative implementation is the unique dynamic reserve procedure on all finite release histories that satisfies resource-release invariance and recipient retention and whose one-shot outcomes, with the original incumbents protected, satisfy the axioms of Theorem~\ref{thm:flexcharacterization} for $\CC_{\parallel}$ and of Theorem~\ref{thm:precedencecharacterization} for $\CC_{\triangleright}$. With $\mathrm G_0=\CC(\sigma_0)$, at each date $t\ge1$,
\[
\mathrm G_t=\CC\!\left(\sigma_t[\widehat{\mathrm G}_{t-1}]\right)=\CC(\sigma_t).
\]
For $\CC_{\parallel}$, uniqueness concerns the set of recipients of open positions and the aggregate recipient set; the same attribution convention then determines the full outcome.
\end{enumerate}
\end{theorem}
Theorem~\ref{thm:flexrelease} is the counterpart of Theorem~\ref{thm:reattribution} for inherited appointments, with Theorems~\ref{thm:flexcharacterization} and~\ref{thm:precedencecharacterization} providing the one-shot choice. The institution can thus fill positions as they become available and protect its new appointees while obtaining the same outcome as RRB at the cumulative capacities with only the original incumbents protected. This holds even if ordinary choice would reject some of those original incumbents. Their appointments can still influence who is selected at later dates, but the commitments to the new appointees add no further departure. Combining Theorem~\ref{thm:flexrelease} with Proposition~\ref{prop:legacybound} identifies which original incumbents need this protection at each date.

\begin{corollary}\label{cor:dynamiclegacy}
In the setting of Theorem~\ref{thm:flexrelease}, let $\mathrm G_t$ be the outcome obtained by protecting earlier recipients at each date and let $\mathrm G_t^{\emptyset}=\CC(\sigma_t[\emptyset])$ be the corresponding outcome with no appointment guarantees. Then
\[
\widehat{\mathrm G}_t\setminus\widehat{\mathrm G}_t^{\emptyset}=\widehat{\mathrm H}_0\setminus\widehat{\mathrm G}_t^{\emptyset} \qquad \mbox{and} \qquad
|\widehat{\mathrm G}_t^{\emptyset}\setminus\widehat{\mathrm G}_t|\le |\widehat{\mathrm H}_0\setminus\widehat{\mathrm G}_t^{\emptyset}|,
\]
with equality whenever $|\widehat{\mathrm G}_t|=|\widehat{\mathrm G}_t^{\emptyset}|$. Moreover, $\widehat{\mathrm G}_t^{\emptyset}\subseteq\widehat{\mathrm G}_{t+1}^{\emptyset}$, so the set of original incumbents whom the ordinary benchmark would reject weakly shrinks over time.
\end{corollary}

Although the set of original incumbents rejected by ordinary choice can only shrink, the identities of the applicants displaced by their protection can change as capacities expand.

Prescribing category precedence does not require positions to become available in that order. Under precedence RRB, the prescribed order governs consideration at each decision, and because earlier category assignments may change, Theorem~\ref{thm:flexrelease} implies that different release orders give the same outcome at the same cumulative capacities.

\begin{example}\label{ex:legacyrelease}
At date 0, there is one open position and one reserved position at category $c$, occupied respectively by incumbents $i_1$ and $i_2$, both eligible for category $c$. The fixed population is $I=\{j_1,j_2,k,i_1,i_2\}$, with priority
\[
\pi:\quad j_1\quad j_2\quad k\quad i_1\quad i_2,
\]
where the nonincumbent applicant $k$ is eligible for category $c$, and the nonincumbent applicants $j_1,j_2$ are not. One further open position and one further reserved position at category $c$ are released, one at date 1 and the other at date 2.

If the open position is released at date 1, the date-1 selection assigns $\{j_1,i_1\}$ to the open category and $\{i_2\}$ to category $c$, because reserved capacity is one and so one incumbent must hold an open position. If the reserved position is released at date 1, the date-1 selection assigns $\{j_1\}$ to the open category and $\{i_1,i_2\}$ to category $c$. At date 2, either release order gives the outcome of RRB at the cumulative capacities, with $\{j_1,j_2\}$ assigned to the open category and $\{i_1,i_2\}$ to category $c$. All intermediate recipients remain appointed. On the open-first path, $\CC_{\parallel}$ itself assigns $i_1$ to the open category at date 1 and moves her to category $c$ at date 2. Thus reassignment from the open category to a reserved category is needed even for an appointment made by the intended rule. The O\&A choice rule at the final capacities would instead assign $\{j_1,j_2\}$ to the open category and $\{k,i_1\}$ to category $c$, rejecting $i_2$. Resource-release invariance therefore holds even when the original incumbent set binds and the benchmark outcome cannot be reproduced.
\end{example}

\subsubsection{Arrivals and Withdrawals}\label{subsubsec:rrbarrivals}

Both RRB choice rules can also be applied as individuals arrive. Let $I_t\subseteq I_{t+1}$ be the populations available at successive dates, with priority and eligibility fixed on $\mathcal I$ and capacities nondecreasing. Let $\mathrm H_0$ be a retention-feasible inherited profile on $I_0$, and let $\sigma_t=(I_t,q_t,\mathrm H_0)$. At each date, the account presents the enlarged population, the cumulative capacities, and every earlier recipient as an incumbent. Starting from $\mathrm G_0=\CC(\sigma_0)$, the continuation is
\[
\mathrm G_t=\CC\!\left(\sigma_t[\widehat{\mathrm G}_{t-1}]\right).
\]
The earlier outcome remains feasible, so all these appointments can be honored, and each $\sigma_t[\widehat{\mathrm G}_{t-1}]$ is a retention-feasible problem.

Theorems~\ref{thm:flexcharacterization} and~\ref{thm:precedencecharacterization} therefore apply at each date, and Proposition~\ref{prop:legacybound} bounds how far $\mathrm G_t$ departs from ordinary choice at $I_t$ and $q_t$. In two cases there is no departure. The first is when ordinary choice already selects every earlier recipient (Relation~\eqref{eq:benchmarkrecovery}). The second is when the initial outcome $\mathrm G_0$ is the ordinary one and every new arrival has lower priority than every earlier recipient (Lemma~\ref{lem:lowerpriorityarrivals}).

The situation is different when a person arrives who has higher priority than someone already selected. Ordinary choice at the expanded inputs may then reject that recipient. If it does, the continuation must deviate from ordinary choice to honor her appointment. The timing of releases then matters as well. Suppose a position is released early, and the person who receives it is ranked lower than someone arriving later. If the institution had waited, it might not have selected her, meaning there would be no appointment to honor. So retention may conflict with release invariance, even when changes of category are allowed.
Once arrivals cease, and eligibility, priority, and any category precedence remain fixed, we can apply Theorem~\ref{thm:flexrelease} from that date onward, taking the recipients already selected as the original incumbents. Further capacity expansion and protection of new appointments then reproduce RRB at the cumulative capacities with those incumbents protected.\footnote{We formalize these implications and the conflict between retention and release invariance with arrivals in Appendix~\ref{app:arrivalsretention}.}

Withdrawals are simpler. When a recipient leaves and her position remains available, either RRB rule retains all surviving recipients if the other inputs remain fixed. Reapplying the rule while protecting these recipients gives the same outcome as protecting only the surviving original incumbents.\footnote{We state and prove this consequence of the graph matching representation in Lemma~\ref{lem:rrbwithdrawals}.}

\subsection{Inherited Appointments with General Category Precedences}
\label{subsec:otherprecedences}

At each step, precedence RRB assigns an individual to the current category only if all remaining incumbents can still be assigned to eligible positions. 
In formulating the precedence RRB choice rule in Section~\ref{subsec:RRBvariants}, we presented these steps separately to highlight how they relate to over-and-above protection and to the other RRB rules. 
But the construction itself does not need to start with the open category.

Fix capacities, eligibility, a common priority, and a strict order of precedence over all categories. Process categories in this order, selecting the highest-priority eligible individual whose assignment leaves a feasible completion for the remaining incumbents. Any open positions that remain are included in that completion. We call this the \emph{retention-adjusted sequential choice rule}. It honors all inherited appointments and respects priority in the prescribed sequence as far as retention permits.\footnote{See Appendix~\ref{app:otherprecedences} for the construction of the retention-adjusted sequential choice rule and the proof that it is non-wasteful, reproduces ordinary sequential choice whenever that choice retains the incumbents, and gives the same outcome if more recipients of its own selection are protected.}

A word of caution is in order. That our retaining tool can be generalized to other precedences does not mean that it should be. In general, a reserve policy that is specified for problems without incumbents can be extended in more than one way to problems where incumbents must be retained. Which extension is suitable depends on the primary policy objective. 

Our construction retains incumbents while following the order in which categories are processed. 
Theorems~\ref{thm:flexcharacterization} and~\ref{thm:precedencecharacterization} and Corollary~\ref{cor:basicrrbcharacterization} show how RRB preserves over-and-above protection subject to retention.
For other policies, our construction may not preserve the policy objective in the same way, and other choice rules may be more suitable than the retention-adjusted sequential choice rule.

Consider the minimum guarantee choice rule with nonoverlapping reserves. If there are no incumbents, the rule fills reserved positions first and then open positions. With incumbents, one extension follows the same precedence through the retention-adjusted sequential rule. Another starts the process by admitting the incumbents and counts the eligible ones toward the minimum. Then it adds individuals by priority to meet any unmet minimum as far as possible, and after that it fills the remaining positions by priority. This is the guaranteed-enrollment choice rule of \citet{DIY2025}. 
If there are no incumbents and there are enough eligible applicants to fill the reserved positions, these two rules coincide with ordinary minimum guarantee. With inherited appointments, however, they can yield different outcomes.

For illustration, take the example of Section~\ref{subsec:flexrule} and process the reserved category first. Both rules retain the incumbent $k$. The retention-adjusted sequential rule gives the reserved position to $j$ and keeps $k$ in the open position, so $i$ is left out. Guaranteed enrollment counts $k$ toward the minimum and gives the other position to $i$. 
Under the retention-adjusted sequential rule, $i$ is left out so that the reserved position goes to $j$ by priority. But the minimum is already met, because the incumbent $k$ belongs to the protected group. So if the objective is minimum representation, guaranteed enrollment is the more suitable extension. In this example, it meets the minimum without excluding the highest-priority individual.\footnote{\citet[Sections~3.2 and~4.2.1]{CorreaEtAl2022} describe Chile's centralized school admissions system, which incorporates quotas as minimum guarantees and protects students seeking to transfer by giving them the highest priority in every category at their current school. Their current school is appended to their preference list, securing enrollment there if they cannot obtain a preferred assignment.}

Over-and-above protection calls for a different treatment of incumbents. 
Take a beneficiary whose priority is high enough for an open position. Under this policy, she gets the open position, and the reserved position is left for someone else in her group. An incumbent should be treated the same way when retention permits. RRB treats her this way. Counting her toward the reserve instead follows the minimum-representation objective and can cause the extra benefit intended by O\&A to be lost.
To see this, modify the example so that $i$ and $k$ are the eligible individuals and $i$ is the incumbent. Ordinary O\&A gives the open position to $i$ and the reserved position to $k$. Since it already retains $i$, RRB selects the same outcome. If $i$ counts toward the reserve, the other position goes to $j$ by priority, and $k$ is excluded. Although no adjustment was necessary to retain the incumbent, the boost is lost.

RRB maintains the extra benefit intended by over-and-above protection, whereas guaranteed enrollment treats incumbents as contributing to minimum representation. For other prescribed precedences, the appropriate retaining extension depends on the priority and protection principles that the institution seeks to maintain when honoring inherited appointments.\footnote{With insufficient eligible demand, a minimum guarantee must also be distinguished from a hard reserve, whose unused positions remain restricted.} This distinction matters in practice. In Section~\ref{app:cpnu}, we discuss a proposal for retaining incumbents in Brazil's federal recruitment, which follows an over-and-above policy. Like the guaranteed-enrollment choice rule, the proposal puts incumbents ahead of other candidates, but it does so under a different primary policy objective. We show that this may change the ordinary outcome even when no adjustment is needed to honor their appointments, and argue that precedence RRB is the more suitable retaining extension.

Other admission rules combine several forms of protection through a prescribed precedence. Dartford Grammar School, for example, considers a priority-area block followed by a block open to applicants at any address. Within each block, it considers pupil-premium places before score-based places. Its waiting list is reordered as applications come in. When category capacities are held fixed, the retention-adjusted sequential rule honors accepted admissions and follows this order as far as retention permits.\footnote{See the school's \href{https://www.kent.gov.uk/__data/assets/pdf_file/0016/234331/Dartford-Grammar-School-admissions-criteria-2027-2028.pdf}{\emph{Admissions Procedure for 2027 Entry}}, Sections~C, F, and~G. We consider score-based places after statutory-priority admissions and before transfers of unfilled places, with score thresholds included in eligibility. Priority-area applicants can also qualify for the block open to all addresses.}

\section{Multi-Institution Settings with Inherited Appointments}\label{sec:multiinstitution}

An individual who holds an appointment might want to move to another position, either within the same organization or elsewhere. In such a case, honoring her appointment does not require keeping her where she is. We now extend our model to heterogeneous positions. To do so, we use the term \emph{institution} for each program or position type with its own capacities and reserve policy. Individuals have preferences over these institutions, but positions within each institution are interchangeable. Integrating RRB with individual-proposing deferred acceptance \citep{GaleShapley1962} allows an incumbent to
keep her appointment or move to an institution that she prefers. Her category assignment remains subject to the receiving institution's reserve policy. Coordinated waiting-list design in admissions and public recruitment is an important application of this extension, which we develop in Section~\ref{sec:multiapplications}.

\subsection{The Current Problem with Multiple Institutions}\label{subsec:matchingproblem}

We consider the multi-institution problem at a given date and suppress the time index throughout this section. Let $I$ be a finite population and $\mathcal S$ a finite set of institutions. Each individual $i$ has a strict preference $\succ_i$ over $\mathcal S\cup\{\varnothing\}$, where $\varnothing$ denotes being unassigned; $\succeq_i$ is its weak counterpart. An institution is acceptable to $i$ if she prefers it to being unassigned. Individuals have preferences over the institutions they attend, not the categories through which they are counted there. Individuals may change their preferences between allocation dates.

Each institution $s$ has the reserve structure of Section~\ref{sec:rules}, with category capacities $q_s$, eligibility sets, and a baseline priority $\pi_s$. Let $q=(q_s)_{s\in\mathcal S}$. 
For the characterization below, each institution uses either joint or precedence RRB, with its category precedence fixed in the latter case. On the nonoverlapping domain both choice rules simplify to $\CC_{\mathrm{RRB}}$.

The account provides sets $H_s\subseteq I$ of individuals whose appointments at $s$ are protected. We assume that these sets are disjoint, so that no individual holds a protected appointment at more than one institution, and that each $H_s$ is retention-feasible at $q_s$. Let $H=(H_s)_{s\in\mathcal S}$. Denote individual $i$'s inherited institution by $\omega^H(i)$. That is, $\omega^H(i)=s$ for $i\in H_s$, and $\omega^H(i)=\varnothing$ if she has no protected appointment. 
Fix all inputs other than preferences.

For $K\subseteq J\subseteq I$, let $\sigma_s[J,K]$ denote institution $s$'s choice problem restricted to $J$, with the same capacities and with the appointments of $K$ protected. Let
\begin{equation}\label{eq:rrbmenuchoice}
\widehat{\CC}_s^H(J)=\widehat{\CC}_s\bigl(\sigma_s[J,H_s\cap J]\bigr)
\end{equation}
denote the recipient set selected by its RRB choice rule $\CC_s \in \{\CC_{\parallel}, \CC_{\triangleright}\}$. We use $\widehat{\CC}_s^{\emptyset}(J)$ for the ordinary counterpart with no protected appointments. Retention feasibility of $H_s$ ensures that the choice given in Relation~\eqref{eq:rrbmenuchoice} is defined on every set $J$. An incumbent is protected at her own institution whenever she applies there; she receives no such protection at another institution.

\begin{definition}\label{def:matching}
A \emph{matching} is a function $\mu:I\longrightarrow\mathcal S\cup\{\varnothing\}$. Let $\mu_s=\{i\in I:\mu(i)=s\}$.
\end{definition}

A matching specifies each individual's institution but not her category.
And yet our analysis and applications rely on category assignments. This may look like an oversight or an inconsistency in the model. It is not. 
Category assignments enter through the matching policy axioms in Definition~\ref{def:matchingpolicy} below, which require that the set of individuals who are matched to each institution admit a category assignment consistent with the institution's policy principles.

A matching could alternatively assign each individual to an institution--category pair. Our characterization holds in this case as well, as long as preferences are over institutions and the policy axioms apply to the recorded category assignments.
The institution assignments will be identified by Theorem~1 of \citet{DIY2025}, while the characterizations in Section~\ref{sec:rules} will yield the category assignments.
As for joint RRB--DA, we will have the same benign multiplicity as in Theorem~\ref{thm:flexcharacterization} unless we fix the attribution convention.
This multiplicity has no bearing on the matching in our setting, which specifies only institution assignments.

\begin{definition}\label{def:matchingfeasibility}
A matching $\mu$ is \emph{feasible} if, for every institution $s$, there is a category-assignment profile $\mathrm G_s$ feasible under its capacities and eligibility such that $\widehat{\mathrm G}_s=\mu_s$.
\end{definition}

\begin{definition}\label{def:matchingIR}
A matching $\mu$ is \emph{individually rational} if $\mu(i)\succeq_i\varnothing$ for every individual $i$.
\end{definition}

\begin{definition}\label{def:matchingretention}
A matching $\mu$ \emph{honors the inherited appointments} if $\mu(i)\succeq_i \omega^H(i)$ for every incumbent $i$.
\end{definition}

Let $\mathcal P_i$ be the set of all strict preferences over $\mathcal S\cup\{\varnothing\}$, let $\mathcal P=\prod_{i\in I}\mathcal P_i$, and let $\mathcal M$ be the set of feasible matchings. Let $\succ=(\succ_i)_{i\in I}$ denote a preference profile and $\succ_{-i}$ the preferences of individuals other than $i$.

\begin{definition}\label{def:directmechanism}
A \emph{direct mechanism} is a function $\varphi:\mathcal P\longrightarrow\mathcal M$ that assigns a feasible matching to each profile of reported preferences.
\end{definition}
We refer to a direct mechanism simply as a mechanism. Its input is the preference profile reported for the current problem; the institutional inputs are fixed as above.

\subsection{Integrating RRB with Individual-Proposing Deferred Acceptance}\label{subsec:rrbda}

For each institution $s$, first consider an arbitrary aggregate choice rule $\widehat{\CC}_s$ that selects from each set $J\subseteq I$ a subset admitting a feasible category-assignment profile at $s$. Given these choice rules and the reported preferences, the individual-proposing deferred acceptance algorithm \citep{GaleShapley1962,HatfieldMilgrom2005} proceeds as follows.

\choiceheading{Individual-proposing deferred acceptance algorithm.}{}
Initially, no individual is held by an institution and no proposal has been made. Each round $k=1,2,\ldots$ has two steps.
\begin{choicesteps}
\item \textbf{Step $k.1$:}\enspace Every individual who is not currently held and has an acceptable institution to which she has not yet proposed proposes to her most-preferred such institution, according to her reported preference.
\item \textbf{Step $k.2$:}\enspace For each institution $s$, let $J_s$ consist of the individuals it held from the preceding round and those who proposed to it in this round. Institution $s$ holds $\widehat{\CC}_s(J_s)$ and rejects the remaining individuals in $J_s$.
\end{choicesteps} \smallskip
If some individual not currently held has an acceptable institution to which she has not yet proposed, proceed to round $k+1$. Otherwise, assign each held individual to the institution holding her and leave every other individual unassigned. The algorithm terminates because each individual proposes to each institution at most once. 

Given aggregate choice rules $(\widehat{\CC}_s)_{s\in\mathcal S}$, the \emph{individual-proposing DA mechanism} selects, for each preference profile, the final matching of this algorithm. 
The \emph{RRB--DA mechanism}, denoted by $\varphi_{\mathrm{RRB}}(\,\cdot\,;q,H)$, is the individual-proposing DA mechanism with the aggregate choice rules $\widehat{\CC}_s^H$ in Relation~\eqref{eq:rrbmenuchoice}. 
We let $\varphi_{\mathrm{RRB}}(\succ;q,H)$ denote its matching at the reported preference profile $\succ$. When all institutions use joint RRB, we call it \emph{joint RRB--DA}; when they use precedence RRB with their prescribed category orders, we call it \emph{precedence RRB--DA}. Institutions can also use different variants according to their respective policies. With nonoverlapping reserves at every institution, both variants use the basic RRB choice rule.

We next define a property of matchings that plays a central role in the two-sided matching literature. For RRB--DA, it is applied with the choice rules $\widehat{\CC}_s^H$.

\begin{definition}\label{def:stability}
Given aggregate choice rules $(\widehat{\CC}_s)_{s\in\mathcal S}$, a matching $\mu$ is \emph{stable} if it is individually rational and, for every institution $s$,
\[
\widehat{\CC}_s(\mu_s)=\mu_s
\quad \mbox{and} \quad 
s\succ_i\mu(i)\quad\Longrightarrow\quad
 i\notin\widehat{\CC}_s(\mu_s\cup\{i\})
\quad\text{for every }i\in I.
\]
\end{definition}

To establish that RRB--DA produces a stable matching, is independent of the order in which proposals are processed, and satisfies strategy-proofness, defined below, we need to show that each institution's aggregate choice satisfies the following conditions on choice across sets of individuals.

\begin{definition}\label{def:menuproperties}
An aggregate choice rule $\widehat{\CC}$ satisfies
\begin{enumerate}[label=(\roman*)]
\item the \emph{substitutes condition} if $\widehat{\CC}(J')\cap J\subseteq\widehat{\CC}(J)$ whenever $J\subseteq J'\subseteq I$;
\item \emph{irrelevance of rejected individuals} if $\widehat{\CC}(J')=\widehat{\CC}(J)$ whenever $\widehat{\CC}(J)\subseteq J'\subseteq J\subseteq I$;
\item the \emph{law of aggregate demand} if $|\widehat{\CC}(J)|\leq|\widehat{\CC}(J')|$ whenever $J\subseteq J'\subseteq I$; and
\item \emph{path independence} if $\widehat{\CC}(J\cup J')=\widehat{\CC}(\widehat{\CC}(J)\cup J')$ for all $J,J'\subseteq I$.
\end{enumerate}
\end{definition}

Substitutes and irrelevance of rejected individuals ensure that DA produces an individual-optimal stable matching whose outcome does not depend on the order in which proposals are processed, and the law of aggregate demand additionally ensures strategy-proofness \citep{HatfieldMilgrom2005,AygunSonmez2013,HirataKasuya2014}. Our next lemma shows that the RRB choice rules satisfy all of them.

\begin{lemma}\label{lem:rrbmenus}
Fix the institutional inputs and a profile of protected sets $H=(H_s)_{s\in\mathcal S}$ such that each $H_s$ is retention-feasible at $q_s$. For every institution $s$, the aggregate choice rule $\widehat{\CC}_s^H$ satisfies substitutes, irrelevance of rejected individuals, the law of aggregate demand, and path independence.
\end{lemma}

An incumbent who proposes to her inherited institution under RRB--DA is never rejected there, since $\widehat{\CC}_s^H$ retains every member of $H_s$ on any set containing her. RRB--DA therefore honors inherited appointments, giving every incumbent an assignment she weakly prefers to her inherited institution.\footnote{With only open positions, each institution's RRB choice selects its incumbents first and fills the remaining positions by baseline priority, so RRB--DA coincides with the individual-proposing DA mechanism under priorities that rank each institution's incumbents above everyone else and use the baseline priority within each group.
Such ownership-adjusted priorities appear in several settings. \citet{GuillenKesten2012} show that MIT's undergraduate housing mechanism is equivalent to DA with existing tenants given top priority at their own rooms and characterize it by fairness, individual rationality, and incentive compatibility. The reassignment stage of the forward lottery deferred acceptance mechanism of \citet{FeigenbaumEtAl2020} has the same structure when the populations, protected assignments, and resulting priorities coincide, and \citet{CombeSchlegel2024} study reallocation through DA with ownership-adjusted priorities. With reserved categories, RRB--DA and priority adjustments differ; see Section~\ref{app:cpnu}.}

Institutions can also differ in the policies they implement. The retention-adjusted sequential rules of Section~\ref{subsec:otherprecedences}, including orders in which open does not come first, satisfy the same choice properties (Lemma~\ref{lem:sequentialmenus}). They can therefore be combined with RRB at other institutions in individual-proposing DA. A minimum-guarantee institution may instead use the guaranteed-enrollment choice rule of \citet[Proposition~2]{DIY2025}. 
The resulting mechanism remains strategy-proof, selects the individual-optimal stable matching for these institutional choice rules, and gives every incumbent her appointment or an institution she prefers.
We discuss this more general case in Section~\ref{app:mixedpolicies}; the characterization below concerns RRB.

\subsection{Characterization of the RRB--DA Mechanism}\label{subsec:matchingcharacterization}

In Section~\ref{sec:rules}, an institution allocates identical positions among a given population. With heterogeneous positions, individuals' preferences over institutions, together with their assignments under a given matching, determine which individuals an institution must consider. Given a matching $\mu$, let
\[
D_s(\mu)=\{i\in I:s\succeq_i\mu(i)\}
\]
be the set of individuals who weakly prefer institution $s$ to their assignment at $\mu$. We apply the institution's axioms from Theorem~\ref{thm:flexcharacterization} or~\ref{thm:precedencecharacterization}, according to its policy, to the problem $\sigma_s[D_s(\mu),H_s\cap D_s(\mu)]$. 
An incumbent of $s$ who is assigned under $\mu$ to an institution she prefers is not in $D_s(\mu)$, so retention at $s$ places no restriction on her; this is why a matching can honor her appointment by assigning her elsewhere.

\begin{definition}\label{def:matchingpolicy}
A matching $\mu$ satisfies institution $s$'s \emph{matching policy axioms} if there is a feasible category-assignment profile $\mathrm G_s$ with $\widehat{\mathrm G}_s=\mu_s$ that satisfies its applicable axioms below at $\sigma_s[D_s(\mu),H_s\cap D_s(\mu)]$:
\begin{enumerate}[label=(\roman*)]
\item for joint consideration, incumbent retention, no justified envy subject to retention with joint comparisons of reserved categories, and maximal reserve protection subject to retention;
\item for prescribed precedence, incumbent retention, no justified envy subject to retention under that precedence, and compliance with category precedence subject to retention.
\end{enumerate}
\end{definition}

Our final axiom is relational. It requires that no individual can benefit by misreporting her preferences.

\begin{definition}\label{def:matchingstrategyproofness}
A direct mechanism $\varphi$ is \emph{strategy-proof} if, for every preference profile $\succ\in\mathcal P$, individual $i\in I$, and alternative report $\succ_i'\in\mathcal P_i$,
\[
\varphi(\succ)(i)\succeq_i\varphi(\succ_i',\succ_{-i})(i).
\]
\end{definition}

We obtain our next characterization by invoking Theorem~1 of \citet{DIY2025}. Their theorem applies when each institution's choice rule is characterized by punctual axioms (solitary axioms in their terminology) and also satisfies path independence and the law of aggregate demand. On the full domain of strict preferences, individual-proposing DA based on these choice rules is then the unique mechanism satisfying individual rationality, strategy-proofness, and the extensions of those axioms to matchings. An extension evaluates each institution's assigned set as its choice from the individuals who weakly prefer it to their assignment, which is how Definition~\ref{def:matchingpolicy} is constructed.

In our setting, Theorems~\ref{thm:flexcharacterization} and~\ref{thm:precedencecharacterization} provide the punctual characterizations of each institution's aggregate choice, Lemma~\ref{lem:rrbmenus} provides path independence and the law of aggregate demand, and Definition~\ref{def:matchingpolicy} gives the matching extensions. The individual-optimality conclusion below is not part of their theorem; it follows from the corresponding DA result \citep[Theorem~4]{HatfieldMilgrom2005}, since the individually rational matchings satisfying the matching policy axioms coincide with the stable matchings for $(\widehat{\CC}_s^H)_{s\in\mathcal S}$.

\begin{theorem}\label{thm:rrbdacharacterization}
Fix the institutions' capacity vector $q$, eligibility, baseline priorities, retention-feasible protected set profile $H$, and their joint or prescribed-precedence comparisons. On the full domain of strict preferences over institutions and being unassigned, the RRB--DA mechanism $\varphi_{\mathrm{RRB}}(\,\cdot\,;q,H)$ is the unique mechanism satisfying individual rationality, strategy-proofness, and every institution's matching policy axioms. Its outcome honors each incumbent's appointment or assigns her an alternative she prefers. Moreover, it gives every individual her most-preferred assignment among all individually rational matchings satisfying the institutions' matching policy axioms. With nonoverlapping reserves, each institution's axioms reduce to those of Corollary~\ref{cor:basicrrbcharacterization}.
\end{theorem}

The characterization builds on work connecting priority-based allocation with strategy-proof matching mechanisms. \citet{AlcaldeBarbera1994} and \citet{BalinskiSonmez1999} present early characterizations in two-sided matching and student placement. For given institutional choice rules, \citet{HirataKasuya2017} and \citet{HatfieldKominersWestkamp2021} study when stability and strategy-proofness characterize a mechanism. \citet{SY2024} and \citet{GreenbergPathakSonmez2024} derive mechanisms from policy axioms for Indian reservations and US Army branching, respectively.

\section{Single-Institution Applications}\label{sec:casestudies}

Parallel to our analysis in Section~\ref{sec:rules}, in this section we consider applications with a common priority, open and reserved categories, and over-and-above protection with appointments that must be honored. 
Earlier selections may be protected as admissions proceed or employees are appointed; they may also survive a correction under court or administrative direction. RRB honors these commitments while maintaining the intended policy as far as retention permits. Sections~\ref{app:basicrrb}--\ref{app:ceara} show this for basic RRB under nonoverlapping reserves and for the choice between joint and precedence RRB when reserves overlap. 
Section~\ref{app:chilecontinuation} then considers how category reassignment and the protection of later selections affect continuing allocation. The fixed-capacity comparisons in these applications concern allocation before any redistribution of unfilled reserved positions.

\subsection{Waiting Lists with Nonoverlapping Reserves}\label{app:basicrrb}

In Chicago's selective-enrollment high school admissions, as in other systems with waiting lists, inherited rights arise naturally as part of the process. For its score-based admissions, the city assigns 30 percent of positions by admissions scores regardless of tier and then splits the remaining 70 percent evenly among four socioeconomic tiers, assigning them by the same scores within each tier.
Tiers are based on the city's census tracts, and each applicant belongs to exactly one tier, determined by home address, so the reserves are nonoverlapping. As studied by \citet{DurPathakSonmez2020}, these reserves are processed under the over-and-above policy.\footnote{Chicago Public Schools, \href{https://www.cps.edu/gocps/high-school/results/selective-enrollment/}{\emph{Selective Enrollment Selection Process}}. We consider score-and-tier places, excluding discretionary admissions. CPS \href{https://www.cps.edu/sites/cps-policy-rules/policies/600/602/602-2/}{Policy~602.2}, Section~V.1, gives the redistribution rules for unfilled tier places.}

A rolling waiting list opens after the initial offer-acceptance period. Eligible applicants who did not receive a selective-enrollment offer are placed automatically on the waiting list of every selective-enrollment program on their application for which they qualify; existing applicants can add programs, and students who did not apply in the fall can apply for the first time. The waiting lists are administered centrally through the GoCPS platform, with participation restrictions across programs: a student who holds an initial selective-enrollment offer is not placed on any other selective-enrollment waiting list, and one who receives a selective-enrollment waiting-list offer is removed from all other selective-enrollment waiting lists.\footnote{Chicago Public Schools, \href{https://www.cps.edu/gocps/high-school/results/rolling-waitlist/}{\emph{GoCPS Rolling Waitlist}}. Accepting a new offer means giving up the student's previously accepted offer.} 
We take these restrictions as given here and consider a single program's waiting list.
As students apply, the priority ranking of a waiting list is adjusted based on admissions scores, so a newcomer may be ranked higher than someone who received an offer before her. New offers are made only when the program has available positions, and an accepted admission is not rescinded when a higher-scoring student joins the list. Unlike regular admissions, the waiting-list process therefore cannot rely on the ordinary O\&A choice rule. Reapplied to the larger pool, that rule violates incumbent retention: it can reject a student whose admission must be honored. Once admissions are protected, over-and-above protection can be maintained only as far as these commitments allow, and how far depends on how the categories of admitted students are treated.

The account in this setting includes currently admitted students, whose positions are protected, and applicants who are still seeking admission, with their admission scores and tier eligibility. Capacities include both vacant positions and positions occupied by students. There is no information in the published instructions about what happens to an admitted student's category if an applicant with a higher score joins the list. 
If it is locked, her admission is protected, but the new applicant might miss out on admission unnecessarily. 
In contrast, RRB does not lock earlier category assignments. It retains every admission while allowing category assignments to change, so that the arrival can be admitted or a position in a tier becomes available for another student in that tier.

By Corollary~\ref{cor:basicrrbcharacterization}, basic RRB is the unique choice rule that satisfies incumbent retention and the retention-adjusted priority and protection axioms when accepted admissions are the protected appointments. In the setting with arrivals studied in Section~\ref{subsubsec:rrbarrivals}, it therefore provides a waiting-list design that maintains over-and-above protection subject to these commitments. RRB can be adopted program by program and requires no change to the current waiting-list structure. It can also serve as the first step of a broader reform. In Section~\ref{app:chicagocoordination}, we show that the same RRB rule, used as each program's choice rule in deferred acceptance, further allows a student to seek a preferred program without giving up her admission.

Dallas's score-based high-school magnet admissions have the same structure: open positions and nonoverlapping residential reserves, both filled by a common admissions score. The RRB design proposed above for Chicago therefore carries over. Unlike Chicago, Dallas allows a student to accept an offer while remaining on waiting lists for schools she ranks higher, so the coordinated proposal in Section~\ref{app:chicagocoordination} would build on an option that students there already have.\footnote{Institutional details for the Dallas setting are documented in Section~\ref{app:chicagocoordination}. Chicago's entry-grade selective-enrollment elementary programs are another parallel setting, with score-based open selection and socioeconomic-tier reserves; see CPS, \href{https://www.cps.edu/gocps/elementary-school/results/selection-process-selective-enrollment/}{\emph{Selective Enrollment Selection Process}}. There we consider score-and-tier places, holding separate IEP allocations fixed.}

\subsection{Overlapping Reserves with Joint Consideration}\label{app:jointrecruitment}

Amaz\^onia Azul Tecnologias de Defesa (AMAZUL), a Brazilian public company, reserved 20 percent of positions for Black applicants and 5 percent for people with disabilities in its 2022 recruitment. 
Since a Black applicant with a disability is eligible for both reserved categories, this is a case of overlapping reserves.
Members of both groups also compete for open positions, and a Black applicant who scores high enough to be selected in the open category does not count against the racial reserve.
Within each occupation, applicants share a common examination-score ranking.\footnote{See Sections~3.1--3.2 and~10 of AMAZUL's \href{https://selecon.org.br/wp-content/uploads/2022/01/EDITAL-001_2022_CONCURSO-PUBLICO-AMAZUL-CONSOLIDADO_RET01__02_03_25_02_22.pdf}{2022 recruitment notice}, and Article~3 of Brazil's \href{https://www.planalto.gov.br/ccivil_03/_ato2011-2014/2014/lei/l12990.htm}{federal racial-quota law}. The notice also allows unused reserved positions to go to other applicants or to be converted to open positions (items~3.1.13 and~3.2.10).}

Following an August 2024 settlement, AMAZUL adopted a dynamic list for the remainder of the recruitment.
When the open-category ranking reaches an employee originally appointed in the racial reserved category, she keeps her appointment but is counted in the open category. The next call gives the reserved position to another eligible applicant, after which the established call sequence resumes.\footnote{See items~2--5 of AMAZUL's August 2024 \href{https://selecon.org.br/wp-content/uploads/2022/01/Comunicado-Concurso_06082024.pdf}{notice for Concurso P\'ublico AMAZUL/Instituto Selecon~01/2022}.}

With overlapping reserves, it is also necessary for the institution to determine how claims across the two reserved categories should be compared.
The notice does not provide guidance on the order in which the reserved categories should be processed.
One natural approach would be to consider them jointly. Processing one reserve before the other would instead introduce a policy choice not specified in the notice.
Given the recipients of open positions, consider a disabled Black incumbent who can be placed in either reserved category. Counting her appointment in the disability category can leave a racial-reserve position for a Black applicant who is not disabled. Counting it in the racial reserve could instead admit a lower-priority disabled applicant who is not Black. Joint consideration honors the higher-priority applicant's claim whenever the reassignment retains every incumbent and every higher-priority recipient of a reserved position. Joint RRB implements these requirements, and Theorem~\ref{thm:flexcharacterization} characterizes it by incumbent retention, over-and-above protection, and joint comparison of claims across reserved categories.\footnote{AMAZUL's \href{https://selecon.org.br/wp-content/uploads/2022/01/Anexo-IX-ordem-de-convocacao-dos-aprovados-AMAZUL-2022.pdf}{Annex~IX} specifies a call sequence, without settling how claims across overlapping reserves are compared.}

A parallel application in school admissions comes from New South Wales, which has the same open-first structure and overlapping eligibility.
For its selective high schools, it has an Equity Placement Model where 80 percent of places are first offered to applicants according to their test performance (including qualifying members of the equity groups), followed by up to 20 percent offered to additional applicants from the equity groups. The equity groups concern socio-educational disadvantage, Aboriginal or Torres Strait Islander status, rural or remote schooling, and disability. 
An applicant may be eligible for several reserved categories. 
Places that are declined are offered again from each school's reserve list, while students who have accepted keep their places. The published procedures do not specify how an accepted student's place is counted as these later offers are made. Joint RRB would retain every accepted admission and let its category change, so that a student whose priority permits selection for a general place or through another of her equity groups
leaves her equity place for another eligible applicant. Claims across the equity reserves would be compared using the common test priority.\footnote{See the NSW Department of Education's \href{https://education.nsw.gov.au/schooling/parents-and-carers/choosing-a-school-setting/selective-high-schools/fair-access.html}{\emph{Fair access}} page and Sections~2.2--2.3 and~5.2 of its \href{https://education.nsw.gov.au/policy-library/policies/pd-2002-0006-06}{\emph{Selective high schools and opportunity class placement procedures}}. We consider Year~7 day places at single-sex selective schools, avoiding the additional gender allocations at coeducational schools. Eligibility includes the equity performance threshold. See Section~\ref{app:chicagocoordination} on protecting accepted offers.}

Brazil's Mais M\'edicos Especialistas program recruits specialist physicians for positions combining clinical service and training, providing a parallel application. It uses score rankings with over-and-above reserves and continues selection after earlier placements have been confirmed. RRB would honor these placements while allowing their categories to change. Basic RRB applies if recognized reserves do not overlap; with overlap, the choice between joint and precedence RRB depends on how claims across reserves are to be compared.\footnote{See items~5.5--5.6 and~9.2.5 of the Ministry of Health's \href{https://www.gov.br/saude/pt-br/acesso-a-informacao/participacao-social/chamamentos-publicos/2026/chamamento-publico-sgtes-ms-no-6-2026-pmm-e/edital}{2026 Mais M\'edicos Especialistas notice}. The Ministry's \href{https://www.gov.br/saude/pt-br/acesso-a-informacao/participacao-social/chamamentos-publicos/2026/chamamento-publico-sgtes-ms-no-1-2026-pmm-e}{call page} records successive selections.}

\subsection{Overlapping Reserves with Prescribed Precedence}\label{app:ceara}

Brazil's Sistema de Sele\c{c}\~ao Unificada (SISU) is the Ministry of Education's system for admission to participating public institutions, based on scores on the national secondary school examination, ENEM. Federal universities reserve places for applicants who attended public schools, with additional reserves based on income, disability, and ethnic-racial criteria. Since the reform introduced in SISU in 2024, federal law requires over-and-above protection: applicants are considered first for open positions and only then for the reserved categories.\footnote{See Article~3, paragraph~2, of Lei~12.711/2012, as amended by \href{https://www.planalto.gov.br/ccivil_03/_ato2023-2026/2023/lei/l14723.htm}{Lei~14.723 of November~13, 2023}.}

Reserved categories at the Federal University of Cear\'a (UFC) are defined by combinations of attributes, so an applicant may qualify for more than one category.
Let $\mathrm P$ denote completion of secondary education in public schools, $\mathrm L$ low income, $\mathrm D$ disability, $\mathrm Q$ quilombola status, and $\mathrm E$ self-declaration as Black, pardo, or Indigenous. Let $\mathcal A=\{\mathrm P,\mathrm L,\mathrm D,\mathrm Q,\mathrm E\}$ be the set of attributes and $\alpha(i)\subseteq\mathcal A$ be the set of attributes of applicant $i$.
Each reserved category $v$ described below is denoted by its set of required attributes, so that $I^v=\{i\in I:v\subseteq\alpha(i)\}$. Additional attributes do not disqualify an applicant from a category.\footnote{Under SISU's earlier rules, applicants chose a single category. \citet{AygunBo2021} show how this could penalize applicants with additional attributes and encourage them not to declare those attributes. They propose consideration for every qualifying category, with precedence set by the policymaker. \citet{BoSenkevics2023} develop the policy proposal.}
In this section, we denote the open category by $\emptyset$, since it is available to all applicants regardless of their attributes. For example, an applicant with attributes $\{\mathrm P,\mathrm L,\mathrm E\}$ is qualified for $\emptyset$, $\{\mathrm P\}$, $\{\mathrm P,\mathrm E\}$, $\{\mathrm P,\mathrm L\}$, and $\{\mathrm P,\mathrm L,\mathrm E\}$.
UFC specifies the following precedence order of categories, $\triangleright_{\mathrm{UFC}}$:\footnote{See UFC's \href{https://sisu.ufc.br/pt/entenda-a-classificacao-e-selecao-de-candidatos-nas-convocacoes-de-suplentes-do-sisu-2024/}{explanation of waiting-list selection} (July~2024) and the Ministry of Education's \href{https://sisu.mec.gov.br/static/pdf/PORTARIA\%20NORMATIVA\%20N\%C2\%BA\%2021,\%20DE\%205\%20DE\%20NOVEMBRO\%20DE\%202012\%20(Alterada\%20pela\%20Portaria\%20n\%C2\%BA\%202.027-2023).pdf}{SISU rules}, Article~20, as revised in~2023.}
{\small
\abovedisplayskip=4pt plus 2pt minus 1pt
\belowdisplayskip=4pt plus 2pt minus 1pt
\thinmuskip=1mu \medmuskip=2mu \thickmuskip=2mu
\[
\emptyset\triangleright_{\mathrm{UFC}}\{\mathrm P\}\triangleright_{\mathrm{UFC}}\{\mathrm P,\mathrm D\}\triangleright_{\mathrm{UFC}}\{\mathrm P,\mathrm Q\}\triangleright_{\mathrm{UFC}}\{\mathrm P,\mathrm E\}\triangleright_{\mathrm{UFC}}\{\mathrm P,\mathrm L\}\triangleright_{\mathrm{UFC}}\{\mathrm P,\mathrm L,\mathrm D\}\triangleright_{\mathrm{UFC}}\{\mathrm P,\mathrm L,\mathrm Q\}\triangleright_{\mathrm{UFC}}\{\mathrm P,\mathrm L,\mathrm E\}.
\]\par}

After the nationally coordinated regular round, UFC uses its waiting list to send invitations. In March 2024, adjustments by the Ministry required UFC to correct its waiting list, adding those who had previously requested inclusion but were not included and revising the classifications of other listed candidates. Earlier invitations had meanwhile been sent out. UFC indicated that it would honor those invitations, and their recipients would continue on with the remainder of the admission process.\footnote{UFC's \href{https://sisu.ufc.br/pt/lista-de-espera-nota-explicativa-sobre-aplicacao-da-nova-lei-de-cotas/}{March~8, 2024 correction notice} preserves the earlier invitations. They remain subject to the enrollment and eligibility checks in UFC's \href{https://prograd.ufc.br/wp-content/uploads/2024/02/edital-03-2024-lista-de-espera-e-suplentes-sisu-2024.pdf}{2024 admissions notice}.} Those recipients correspond to the incumbents in our model. A formerly omitted candidate enters the current choice problem despite having applied before. Recognizing that someone qualifies for a category other than the one she was previously recognized for alters her eligibility, not the population.

Applying this model requires fixing the program and campus, class schedule (for example, daytime or evening), and admissions cycle. Examination scores with the applicable tie-breaking rule give the priority order, and the corrected list gives the population and recognized qualifications. The account includes the recipients of protected invitations, with capacities counting both occupied places and places available for further admissions. The prescribed precedence is $\triangleright_{\mathrm{UFC}}$.\footnote{The Ministry's \href{https://ufr.edu.br/ingresso/wp-content/uploads/2024/01/PORTARIA-No-2.027-DE-16-DE-NOVEMBRO-DE-2023.pdf}{2023 SISU rules} allow unused reserved places to be redistributed (Article~28).}

Using ordinary choice on the corrected list would follow the prescribed open-first precedence, but could violate incumbent retention by rescinding an earlier invitation. The retain-and-lock choice rule satisfies incumbent retention if the inherited category-assignment profile is feasible. However, locking can prevent a priority or category-precedence claim from being honored even when category reassignment would accommodate the claim without withdrawing any earlier invitation. The institution must honor these invitations while maintaining its score priority and prescribed category precedence as far as retention permits. 
By Theorem~\ref{thm:precedencecharacterization}, precedence RRB is the only choice rule that does so on the retention-feasible domain.
The prescribed precedence order at UFC motivates choosing this variant over joint RRB.

The invitations protected in UFC's correction become commitments for continuing admissions, which we consider in Section~\ref{app:chilecontinuation}.

\subsection{Category Reassignment during Continuing Allocation}\label{app:chilecontinuation}

Reconsidering earlier category assignments while preserving appointments is central to our analysis. This practice already appears in several institutions. 
Universidad Adolfo Ib\'a\~nez in Chile reassigns categories within a program while its admissions continue. In each participating science and engineering program, there are regular (open) positions plus positions reserved for women (+MC) and recipients of an academic-excellence scholarship (BEA). Each program uses the same weighted admissions score in every category, and the national admissions procedure considers regular positions first, followed by +MC positions, and finally BEA positions. A woman who holds the scholarship may be eligible for both reserved categories.\footnote{See the Chilean Ministry of Education's \href{https://acceso.mineduc.cl/wp-content/uploads/2025/12/Oferta-Definitiva-2026-FINAL31_12_25-1_compressed-1.pdf}{\emph{Oferta Definitiva de Carreras, Vacantes y Ponderaciones: Proceso de Admisi\'on 2026}}. The abbreviations stand for M\'as Mujeres Cient\'ificas and Beca de Excelencia Acad\'emica. We do not consider admission routes that use different priorities.}

The university regulation states that when the program's regular waiting list reaches a student admitted via +MC or BEA, she keeps her admission but is counted in the regular category, leaving the reserved position available for another eligible applicant.\footnote{See Articles~6--7 of Universidad Adolfo Ib\'a\~nez's \href{https://uai.cdn7pm.net/documentos/reglamento-de-admision-pregrado-2025.pdf}{\emph{Reglamento de Admisi\'on de los Estudiantes de Pregrado}}, dated February~14, 2025. These articles give the eligibility rules, and item~3 of each article gives the category reassignment. Under Section~4.1.4 of the Ministry's \href{https://www.bcn.cl/leychile/navegar?idNorma=1184028\&idParte=10378024\&idVersion=2026-06-05}{Resolution~5250 Exenta}, consolidated June~5, 2026, only these groups may hold the additional positions.}

We consider a single program here and coordinated continuation in Section~\ref{app:chilecoordination}.
To formulate our proposal for this setting, let the account include admitted students who have protected positions together with those currently being considered, along with their score priority, eligibility, and the prescribed precedence.
Capacities count occupied positions as well as available vacancies. A student who permanently leaves the program no longer has a protected appointment there, and her departure makes an existing position available without increasing capacity. We propose precedence RRB, which takes into account the remaining commitments and maintains priority and category precedence as far as retention permits. The university's reserved-to-regular reassignment is one of the moves allowed by the rule.

The same approach could be used at the Federal Institute of Cear\'a (IFCE), which fills vacancies through successive waiting-list calls. RRB would preserve confirmed enrollments while allowing their category assignments to change.\footnote{See Sections~9--10 of IFCE's \href{https://portal.ifce.edu.br/documents/17570/Edital_1-2026-DI-PROEN-REITORIA-IFCE.pdf}{\emph{SiSU 2026 general admissions notice}} and pp.~2--3 of its \href{https://portal.ifce.edu.br/documents/18696/Fortaleza_SiSU2026_Lista_de_espera.pdf}{\emph{SiSU 2026: Lista de Espera, Campus Fortaleza}}.}

Category reassignment is also practiced in India, between the open category and vertical reserves. The Central Seat Allocation Board (CSAB) allows for it even if the admitted student does not participate in its special rounds.
This is illustrated by an example in the 2025 CSAB brochure, where an admission in the OBC-NCL category is counted in the open category for the same program. The student retains her admission, and the reserved position becomes available to another eligible applicant.\footnote{See clauses~26--27, 48, and~50 of CSAB's \href{https://cdnbbsr.s3waas.gov.in/s305a70454516ecd9194c293b0e415777f/uploads/2025/05/202508181225350702.pdf}{\emph{Information Brochure, Special Rounds 2025}}, and the example on p.~14. We consider open and vertical-reserve positions within a fixed program, ranked by the JEE(Main) Common Rank List.}

For this reassignment, the account must include the admitted student's priority and eligibility even though she is not seeking a new admission. The current population therefore cannot be identified with the applicants participating in the new round.\footnote{See also JoSAA's provision for an SC admission to move to the open category at the same institute and program, even after the student has selected ``Freeze'': \href{https://cdnbbsr.s3waas.gov.in/s313111c20aee51aeb480ecbd988cd8cc9/uploads/2026/06/20260601632081678.pdf}{\emph{Business Rules}, 2026--27}, clauses~34--35, and \href{https://cdnbbsr.s3waas.gov.in/s313111c20aee51aeb480ecbd988cd8cc9/uploads/2026/06/202606061208276078.pdf}{\emph{Frequently Asked Questions}, 2026}, Question~24.}

\emph{Pramod Kumar Singh v.\ State of Uttar Pradesh} shows how large this practice can be. In a continuing police recruitment, 1,252 existing appointments were counted in the open category. No appointment was withdrawn, and 1,252 reserved positions became available to other applicants.\footnote{\label{fn:pramod}See paragraphs~2(G)--(H) and~6--8 of the Supreme Court of India's 2021 judgment in \href{https://api.sci.gov.in/supremecourt/2020/7940/7940_2020_34_1501_27024_Judgement_16-Mar-2021.pdf}{\emph{Pramod Kumar Singh v.\ State of Uttar Pradesh}}. Appendix~\ref{subsec:otherretention} distinguishes these reattributions from the 3,295 new appointments made in the same exercise.}

In all of these settings, a selection made at one stage has to be honored at later stages. Whether positions reopen because recipients withdraw (Lemma~\ref{lem:rrbwithdrawals}) or category capacities increase (Theorem~\ref{thm:flexrelease}), RRB with the later selections protected gives the same outcome as RRB protecting only the original incumbents, provided the other inputs are unchanged.

\section{Multi-Institution Applications}\label{sec:multiapplications}

Section~\ref{sec:casestudies} focuses on each institution individually. An RRB choice rule ensures that any incumbent at an institution can keep her appointment. A single-institution choice rule does not itself coordinate moves across institutions. RRB--DA provides this coordination while allowing an incumbent to retain her current appointment if no preferred assignment is available (Section~\ref{sec:multiinstitution}). In the applications below, we propose reforms based on this mechanism for school and university admissions, medical residency, and public recruitment. In each proposal, an institution's account includes its incumbents and their positions, so category assignments can be reconsidered as people move. Two of these settings, Chicago's waiting lists and Chilean university admissions, build on our proposals in Sections~\ref{app:basicrrb} and~\ref{app:chilecontinuation}, respectively. Our first proposal concerns protected participation in waiting lists. We then compare retention through RRB with raising incumbent priorities, and examine how later selections can be protected as capacities increase. Our last proposal allows institutions with distinct reserve policies to take part in the same coordinated mechanism.

\subsection{Coordinated Waiting Lists}\label{app:chicagocoordination}

In Section~\ref{app:basicrrb}, we proposed RRB for a single program's waiting list in Chicago Public Schools. Under current GoCPS rules, however, a student who holds a selective-enrollment offer is not placed on any other selective-enrollment waiting list and must decline the offer to join one.\footnote{Chicago Public Schools, \href{https://www.cps.edu/gocps/high-school/results/rolling-waitlist/}{\emph{GoCPS Rolling Waitlist}}. Our coordinated proposal allows applicants to rank all acceptable participating schools.} 
This practice may deter students from applying to their higher-ranked schools when additional positions open later on, which in turn may result in welfare loss.
The same problem motivated the \emph{you request my house--I get your turn} (YRMH--IGYT) mechanism of \citet{AbdulkadirogluSonmez1999}. Under serial dictatorship with squatting rights, an existing tenant can keep her house by staying out, but must give up its protection to seek another. Their mechanism allows her to participate while guaranteeing her existing house or one she prefers.

The design below provides the same assurance for an accepted admission. It removes the restriction by coordinating the waiting lists through the individual-proposing deferred acceptance algorithm, with RRB governing how each school honors appointments under its reserve policy.

Under this design, a student who has been admitted can rank schools she prefers, while retaining her current position if she does not move. Each school uses the account described in Section~\ref{app:basicrrb}, with its own admitted students protected and the participating applicants considered through their proposals. If a student wishes to remain in her current position, she ranks her current school and no others. She remains in the account, but her category can change while her school does not. If she wishes to move, she lists the schools she prefers, placing her current school last. The mechanism guarantees her a seat at her current school; it does not guarantee the category she was originally admitted under.

Unlike in the single-institution design of Section~\ref{app:basicrrb}, 
the schools do not run RRB separately. The assignment is made centrally by RRB--DA, in which each school's acceptances are determined by its RRB choice rule. When a student moves, there are two ways her move can affect the allocation. First, her vacated position becomes available for someone else. Second, her receiving school may shift some of its incumbents among categories to make room for her. A procedure that removes accepted students from the account and fills only the vacancies cannot do the second.

Theorem~\ref{thm:rrbdacharacterization} identifies basic RRB--DA as the unique mechanism satisfying individual rationality, strategy-proofness, and the multi-institution counterparts of incumbent retention, no justified envy subject to retention, and maximal reserve protection subject to retention at each school. 
The theorem also provides a welfare justification for this design. Among all individually rational matchings satisfying these policy principles, RRB--DA gives every student her most-preferred assignment and therefore Pareto dominates every other such matching. Unlike the current system, 
no student needs to relinquish her current admission to seek an improvement. 
Arrivals can be included in a new matching problem, with the earlier accepted admissions protected. 

The magnet schools of the Dallas Independent School District already permit an accepted admission to serve as a fallback while an applicant waits for a school she ranks higher. We consider the district's score-based high-school magnet programs, which assign 30 percent of positions districtwide and the rest by residential feeder pattern, using the same admissions score. Our proposal would thus coordinate waiting-list selection and category reassignment across these schools while preserving a participation option that the district already offers. Basic RRB--DA provides such a design, with accepted admissions protected at their schools.\footnote{Dallas Independent School District, \href{https://www.dallasisd.org/departments/magnets-and-special-programs/magnet-selectionprocess}{\emph{Magnet Selection Process}} and \href{https://www.dallasisd.org/about-our-schools/discover-dallas-isd/how-k-12-students-are-assigned-to-schools}{\emph{How K--12 Students Are Assigned to Schools}}. We consider qualified in-district applicants to score-based high-school programs; we abstract from the sibling provisions. Later first-come-first-served selection and the current limit on magnet applications are separate from the unrestricted-preference mechanism characterized here.}

In New South Wales, a student can hold an accepted selective-school offer and wait for a higher choice, but only until the reserve decision date. After that date, she has to give up the offer to stay on a higher-choice waiting list.\footnote{See ``Reserve decision date'' on the NSW Department of Education's \href{https://education.nsw.gov.au/schooling/parents-and-carers/choosing-a-school-setting/selective-high-schools/outcomes}{\emph{Outcomes}} page. Before this date, a later higher-choice offer may be declined without losing the accepted place.}
Joint RRB--DA, based on the reserve structure in Section~\ref{app:jointrecruitment}, would coordinate category reassignment across schools and allow this protected participation to continue until the closure of allocations. 

The same participation dilemma as in Chicago arises in Brazil's ENARE residency allocation. Within a medical specialty, an applicant can choose an institution in each selection opportunity, but once she is selected in one opportunity she is excluded from later ones. Earlier selections are preserved. A reform that adopts RRB--DA would remove this dilemma. An applicant who has been selected could take part in later opportunities and would end up with her current residency or one she prefers. Each program would keep its original score priority, and earlier placements would remain protected.\footnote{See Sections~7.1 and~13.7--13.9 of \href{https://enare2026.conhecimento.fgv.br/docs/36fa57ca8c688040ad69c657f67c4a5d.pdf}{\emph{ENARE 2026/2027, Resid\^encia M\'edica---Acesso Direto}}. Applicants selected for open positions do not count against reserves. Which RRB variant applies depends on how each program compares claims across its reserved categories.}

\subsection{Retention through RRB versus Priority Bonuses}
\label{app:cpnu}

Brazil's Concurso P\'ublico Nacional Unificado (CPNU) coordinates applications to public-service jobs. The initial allocation in the first CPNU used individual-proposing deferred acceptance with job-specific examination scores \citep{BoCoutinhoAlves2025}. The prescribed category precedence, denoted by $\triangleright_{\mathrm{CPNU}}$, is as follows at each job where the relevant categories have positive capacity:
\[
o\ \triangleright_{\mathrm{CPNU}}\ \mathrm D\
\triangleright_{\mathrm{CPNU}}\ \mathrm B\
\triangleright_{\mathrm{CPNU}}\ \mathrm I.
\]
Here $\mathrm D$, $\mathrm B$, and $\mathrm I$ denote reserved categories for individuals with disabilities, Black individuals, and Indigenous individuals. The protected groups can overlap. The recruitment notice requires that Black and Indigenous applicants selected for open positions not count against their reserves. The treatment of the disability reserve and the complete precedence above are specified in the published design.\footnote{See items~3.4.6--3.4.7.1 of the 2024 Block~1 \href{https://www.gov.br/gestao/pt-br/concursonacional/editais/edital_bloco1_versaoretificada01out2024.pdf}{recruitment notice} for Black applicants and items~3.6.6--3.6.7 for Indigenous applicants. Applicants with disabilities appear in both the disability and the open lists (item~3.1.2.4). \citet[pp.~142--143]{BoCoutinhoAlves2025} describe the selection sequence.}

\citet{BoCoutinhoAlves2025} also propose a coordinated waiting-list procedure in which current appointees can apply for preferred jobs without risking the loss of their appointments. Incumbents who participate are given priority over all other candidates at their current job.

To compare their design with one based on precedence RRB--DA, we treat reserved categories as hard reserves at fixed capacities.\footnote{In \citet{BoCoutinhoAlves2025}, the proposed coordinated waiting-list procedure allows nonbeneficiaries to fill unused reserved positions. One soft-reserve variant of our construction maintains the category ordering and gives beneficiaries priority within each reserve, based on their ranking in the exam within each group. The permitted fallback assignments to nonbeneficiaries are counted when we check whether all remaining incumbents can still be assigned. We prove in Appendix~\ref{app:softsequential} that this variant reproduces ordinary choice whenever ordinary choice already retains the incumbents, and that it satisfies the choice rule properties required for integration with individual-proposing deferred acceptance.}

Our proposal keeps the same individual-proposing deferred acceptance procedure, participating population, and positions, but replaces priority bonuses with precedence RRB under the original examination ranking and prescribed category precedence. 
Both designs aim to make sure an incumbent keeps her current appointment unless she receives one she prefers.
They do so differently. Their design increases the priority of an incumbent at her current job above all others, then uses the ordinary choice rule. We leave the priority as it is and incorporate retention in the choice rule. Precedence RRB chooses according to the original examination ranking, subject to all incumbents being retained.
We argue that the priority bonus has a drawback. It can change who else a job chooses, even when no change to priorities is required to retain the incumbent.

Consider a situation with one job with one open and one reserved position. Let $i$ be the incumbent and $j$ and $k$ be two applicants. Only $i$ and $k$ are part of the group protected by the reserve. Suppose the baseline priority is $j\mathrel\pi i\mathrel\pi k$. If we use ordinary O\&A, $j$ will be selected for the open position and $i$ for the reserved position. This already retains $i$, so incumbent retention raises no issue under the ordinary choice rule, and RRB will choose the same outcome. Now assume that $i$ is given the highest priority and ordinary O\&A is used. This will place $i$ in the open position and $k$ in the reserved position, excluding $j$, even though she has the highest baseline priority.

We have discussed in Section~\ref{subsec:otherprecedences} why the way incumbents are retained should match the primary policy objective. A priority bonus does more than retain incumbents; it ranks them above all other candidates at their current jobs. If that ranking is indeed what one wants, then a priority bonus is a coherent choice. It is not a neutral means of retaining incumbents using the original examination ranking.

The contrast with priority bonuses motivates two properties that hold for RRB--DA under any RRB variant, and so apply to the other over-and-above applications as well. 

Fix all current inputs and the preference profile $\succ$. Let $\mu^H=\varphi_{\mathrm{RRB}}(\succ;q,H)$ be the outcome of RRB--DA with protected sets $H$, and let $\mu^{\emptyset}=\varphi_{\mathrm{RRB}}(\succ;q,\emptyset)$ be the outcome when no appointment is protected. The latter is the outcome of individual-proposing DA under the institutions' ordinary choice rules, which we call ordinary DA. The first property is that RRB--DA reproduces ordinary DA whenever no adjustment is needed to honor inherited appointments.

\begin{proposition}\label{prop:dabenchmark}
The matchings $\mu^H$ and $\mu^{\emptyset}$ coincide if and only if ordinary DA already honors every inherited appointment:
\[
\mu^H=\mu^{\emptyset}
\quad\Longleftrightarrow\quad
\mu^{\emptyset}(i)\succeq_i \omega^H(i)
\quad\text{for every incumbent }i.
\]
\end{proposition}

If ordinary DA assigns each incumbent either her position or something she prefers, then RRB--DA has no effect on assignments.
Increasing the priorities of incumbents does not have this property, as the example above illustrates. 
If, on the other hand, ordinary DA fails to honor an appointment, correcting it can cascade to other institutions through subsequent applications.
The single-institution displacement bound of Proposition~\ref{prop:legacybound} does not bound this chain.

The second property is that RRB--DA harms as few individuals as its policy principles allow.
Let $\mathcal M^H$ denote the set of individually rational matchings satisfying every institution's retention-adjusted matching policy axioms. For any matching $\nu$, let
\[
L(\nu;\mu^{\emptyset})
=\{i:\mu^{\emptyset}(i)\succ_i\nu(i)\}
\]
be the set of individuals made worse off relative to ordinary DA. The individual-optimality conclusion of Theorem~\ref{thm:rrbdacharacterization} gives the following consequence.

\begin{corollary}\label{cor:daharm}
For every matching $\nu\in\mathcal M^H$,
$L(\mu^H;\mu^{\emptyset})\subseteq L(\nu;\mu^{\emptyset})$.
\end{corollary}

The comparison is with matchings satisfying the retention-adjusted policy principles, not all feasible matchings honoring appointments. RRB--DA also does not minimize the number of changed assignments, a count that includes voluntary moves to better positions. A priority bonus can sometimes produce fewer losses while violating a feasible claim under the baseline priority.\footnote{See Appendix~\ref{app:disruptionboundary} for an example.}

For CPNU, suppose the objective is to protect appointments while otherwise keeping the original examination ranking and the over-and-above policy. Then these comparisons favor honoring appointments through the choice rule. A priority bonus would add the ranking of incumbents above all other candidates at their current jobs as a further policy commitment.

\subsection{Protected Upgrades as Capacities Grow}\label{app:chilecoordination}

In Chilean university admissions, a student selected at one program can remain on waiting lists for programs she ranked higher. Subsequent enrollment elsewhere can lead her to withdraw from the earlier program.\footnote{Ministry of Education, \href{https://www.bcn.cl/leychile/navegar?idNorma=1184028\&idParte=10378024\&idVersion=2026-06-05}{Resolution~5250 Exenta}, consolidated June~5, 2026, Sections~4.1.6--4.1.7, on remaining on higher-preference waiting lists and withdrawal after enrollment elsewhere.} Section~\ref{app:chilecontinuation} describes a second adjustment: at Universidad Adolfo Ib\'a\~nez, an admitted student's move from a reserved to the regular category frees a reserved position without changing her program.

Both changes are compatible with coordinated continuation using the accounts of Section~\ref{app:chilecontinuation}. Each program applies precedence RRB, considering regular positions first, followed by +MC and BEA positions in the applicable prescribed order.
Individuals rank programs according to their preferences, and individual-proposing DA coordinates their applications. Our proposal involves categories that share the program's weighted-score priority.\footnote{We consider regular, +MC, and BEA places; PACE uses a different selection score. For the historical university-proposing procedure and flexible quotas for score ties, see \citet{RiosEtAl2021}.} 

Another setting where precedence RRB--DA can be applied is the admissions system of the University of S\~ao Paulo (FUVEST).
According to the 2027 rules, the applicants are considered in a fixed order of categories: open places, then public-school reserves, then reserves for Black, pardo, and Indigenous public-school applicants. All are evaluated according to the same final score. Applicants in the last category are also eligible for the public-school reserve. Students can keep an admission and compete for a more preferred course in regular calls. At the final allocation stage, our proposal enables the applicant to rank all participating courses that she is eligible for, without the current four-course limitation.
With courses as institutions, precedence RRB--DA meets these guarantees, preserves the prescribed reserve order, and allows reassignment between categories.\footnote{See Articles~8, 12, 15--17, and~23 of the University of S\~ao Paulo's \href{https://www.fuvest.br/wp-content/uploads/fuvest2027-resolucao-DOE.pdf}{\emph{FUVEST 2027 Admissions Rules}}. Enrollment status M permits protected upgrading during regular calls, but not the subsequent waiting-list stage.}

These university applications also motivate a continuation question that applies to RRB--DA across institutions more generally. Suppose a coordinated allocation is finished, and we protect all the assignments made during that process. What if, afterward, the capacity of each institution's categories were to weakly increase? Will rerunning RRB--DA with the new assignments protected take us further away from the allocation that protects only the original commitments? In our next result, we show that it will not, provided that nothing besides capacities changes.

Let $q_0\leq q_1\leq\cdots\leq q_T$ be category-capacity vectors, with inequalities applying separately to every institution and category. Fix the population, preferences, eligibility, baseline priorities, and category precedences. Let $H^0=(H_s^0)_{s\in\mathcal S}$ be the profile of original protected sets, with each $H_s^0$ retention-feasible at institution $s$'s initial capacities. Start with $\mu_0=\varphi_{\mathrm{RRB}}(\succ;q_0,H^0)$ and, for $t\geq1$, define
\[
H_s^t=(\mu_{t-1})_s,
\qquad
\mu_t=\varphi_{\mathrm{RRB}}(\succ;q_t,H^t).
\]

Thus, as capacities weakly increase, each run of the RRB--DA mechanism protects the most recent assignments, including those made to individuals who originally had no appointment. 

\begin{theorem}\label{thm:dacontinuation}
Under the preceding assumptions, for every date $t$,
\[
\mu_t=\varphi_{\mathrm{RRB}}(\succ;q_t,H^0),
\qquad
\mu_t(i)\succeq_i\mu_{t-1}(i)\quad\text{for all }i\text{ and }t\geq1.
\]
Consequently, protecting the subsequent assignments adds no departure from pooled RRB--DA with the original commitments, and the final matching is independent of the schedule of capacity releases.
\end{theorem}

\subsection{Coordination beyond Over-and-Above Protection}\label{app:mixedpolicies}

My School DC, the common application system for public schools in Washington, DC, shows that a coordinated system need not impose the same reserve policy on all schools. Participating schools can give an Equitable Access preference to applicants who are homeless, in foster care, receiving public assistance, or overage for their grade in high school. They implement it through either a priority preference or designated seats. With the former, eligible applicants move up in the school's priority order. With the latter, eligible applicants are considered first for a set number of restricted places and then for general places if not chosen. Designated seats that go unfilled remain available for eligible applicants who apply later.\footnote{My School DC, \href{https://www.myschooldc.org/node/75261}{\emph{Applying to Equitable Access Schools}}. The stated eligibility criteria identify the protected group; having several qualifying attributes does not create separate reserve categories.}

Students can already be enrolled in one school while on waiting lists for other schools they rank higher. If a later offer is accepted, it replaces the current enrollment.\footnote{My School DC, \href{https://www.myschooldc.org/node/98}{\emph{Waitlists}}. Applicants arriving after the lottery can enter above earlier applicants when their priority preferences warrant it.} Our proposal would coordinate these later allocations, including both enrolled students and applicants seeking places in each school's account. This keeps the accepted school as a fallback while reconsidering category assignments according to each school's policy.

Which retaining rule is chosen makes a difference. In a school that only uses a priority preference, the preference enters the baseline ranking, and RRB with no reserved category accommodates incumbents. A school committed to its designated-first sequence can use the retention-adjusted sequential rule from Section~\ref{subsec:otherprecedences}. If a school chooses minimum representation as its goal, it could instead count eligible incumbents toward that minimum via the guaranteed-enrollment rule of \citet{DIY2025}.

Once these decisions are taken, individual-proposing DA can coordinate these rules together, including RRB at any institution adopting over-and-above protection. Section~\ref{subsec:rrbda} gives the strategy-proofness and individual-optimality conclusions for these institutional choices.\footnote{Sibling priorities are held fixed here; priority changes arising from reassigning siblings are not covered.}

In New Zealand, the University of Otago's health-professional admissions provide a parallel. Applicants can accept a place and remain on waiting lists of other programs. The Radiation Therapy and Dental Technology programs each stipulate a common ranking across admission categories and an equity-first sequence, in which Priority pathways are filled before the General pathway. At specified pathway capacities and recognized eligibility, their retention-adjusted sequential rules could be coordinated by DA. The General pathway here takes only applicants not applying through a Priority pathway, so it is represented by its actual eligibility set.\footnote{See Appendix~\ref{app:otagodetails} for the published rules of these two programs.}

\par\medskip
\begingroup\small\linespread{.95}\selectfont\setlength{\bibsep}{0pt plus .5pt}

\bibliographystyle{econ-aea}
\bibliography{ledger-account-choice-Nov2026-ArXiv}

\endgroup

\par\bigskip\appendix
\titleformat{\section}{\normalfont\Large\bfseries}{Appendix~\thesection:}{.5em}{}

\section{Proofs for the Single-Institution Setting}\label{app:principalproofs}
We first prove the implementation results. We then characterize RRB choice and prove its displacement and continuation properties.\footnote{Additional results are in Supplemental Sections~\ref{app:releaseextensions}--\ref{app:retentionextensions}.}
\setcounter{equation}{0}
\renewcommand{\theequation}{\thesection.\arabic{equation}}

\subsection{Proof of Theorem~\ref{thm:compression}}\label{app:compressionproof} Fix a history $h$ as in Theorem~\ref{thm:compression}, and let $h_t$ be its history through date $t$.  Define, for each $k\in\mathbb Z_+$ and $J \subseteq I$,
\[
g(k,J)=|\TopSet_k^\pi(J)\medcap I^\varnothing| \qquad \text{with}
\quad \TopSet_0^\pi(J)=\OO.
\]
For each fixed $J$, $g(\cdot,J)$ is nondecreasing with increments zero or one. For equal-sized recipient sets $A$ and $B$, their \emph{replacement distance} is $|A\setminus B|=|B\setminus A|$.

\begin{claim}\label{cl:compressionprefix} A viable category-locked batch procedure fills every position in each batch along $h$. After $n$ individuals have been selected in total, the recipients are the highest-priority $n-k$ members of $I^c$ and the highest-priority $k$ members of $I^\varnothing$, where $0\le k\le g(n,I)$. Two such recipient sets under two viable category-locked batch procedures with counts $k\le k'$ selected in $I^\varnothing$ have replacement distance $k'-k$.
\end{claim}
\begin{proof}
Suppose that after $n$ selections, an unselected individual $j$ has higher priority than a selected individual $i$ with the same category-$c$ eligibility. Then $j$ was available when $i$ was selected and eligible for her position, contradicting no justified envy. Thus, the recipients are the highest-priority individuals within each of $I^c$ and $I^\varnothing$, and their set has the form
\begin{equation}\label{eq:prefixform}
\TopSet_{n-k}^\pi(I^c)\medcup\TopSet_k^\pi(I^\varnothing)
\end{equation}
for some integer $k$ with $0\le k\le g(n,I)$.
For the upper bound, suppose $k>g(n,I)$. Then the first $n$ individuals contain fewer than $k$ members of $I^\varnothing$, so at least one selected member of $I^\varnothing$ is outside the first $n$, and at least one reserve-eligible individual among the first $n$ is unselected. When the former was selected for an open position, the latter was available and had higher priority, violating no justified envy.
Two such $n$-person sets with $I^\varnothing$ counts $k\le k'$ have replacement distance $k'-k$.

Every position in each batch is filled. After $n$ selections at least $|I^c|-n\ge q^{\Sigma}(h)-n$ category-$c$-eligible individuals remain, enough to fill all positions still to be allocated. Leaving a currently available position empty would violate non-wastefulness.

\end{proof}

\begin{claim}\label{cl:onebatch}
Consider a residual population $J$ at a date $t$ and suppose $|J\medcap I^c|\ge \Delta q_t^o+\Delta q_t^c$. The number $\ell$ of members of $I^\varnothing$ selected in the current batch by any viable category-locked batch procedure satisfies
\[
g(\Delta q_t^o,J)
\le \ell
\le
\min\{\Delta q_t^o,g(\Delta q_t^o+\Delta q_t^c,J)\}.
\]
The O\&A and minimum guarantee choice rules attain the lower and upper bounds, respectively.
\end{claim}
\begin{proof}
For the lower bound, every member of $I^\varnothing$ among the first $\Delta q_t^o$ residual individuals must be selected. Otherwise, an unselected individual eligible for an open position has higher priority than one of its recipients. The upper bound $\ell\le \Delta q_t^o$ follows because reserve-ineligible individuals can occupy only the $\Delta q_t^o$ open positions. For the other upper bound, suppose $\ell>g(\Delta q_t^o+\Delta q_t^c,J)$. Since all $\Delta q_t^o+\Delta q_t^c$ positions are filled, exactly $\Delta q_t^o+\Delta q_t^c-\ell$ selected individuals are eligible for $c$. The first $\Delta q_t^o+\Delta q_t^c$ residual individuals contain $\Delta q_t^o+\Delta q_t^c-g(\Delta q_t^o+\Delta q_t^c,J)>\Delta q_t^o+\Delta q_t^c-\ell$ such individuals, so at least one of them is unselected. At least one of the $\ell$ selected individuals without eligibility for reserved categories lies below these first $\Delta q_t^o+\Delta q_t^c$ individuals and must occupy an open position. The omitted individual eligible for category $c$ has higher priority and is eligible for that open position, contradicting no justified envy. Hence, $\ell\le g(\Delta q_t^o+\Delta q_t^c,J)$.

The O\&A choice rule attains the lower bound because it fills the $\Delta q_t^o$ open positions first. The minimum guarantee choice rule attains the upper bound. If the first $\Delta q_t^o+\Delta q_t^c$ residual individuals contain at least $\Delta q_t^c$ members of $I^c$, the minimum guarantee choice rule removes the first $\Delta q_t^c$ such individuals from these first $\Delta q_t^o+\Delta q_t^c$ individuals and its open step selects the remaining $\Delta q_t^o$ people. If there are fewer, selecting the first $\Delta q_t^c$ members of $I^c$ leaves the first $\Delta q_t^o$ members of $I^\varnothing$ as the highest residual candidates for open positions. Both cases give the upper expression.

\end{proof}

\begin{claim}\label{cl:residualprefix}
Let $L \subseteq I$ be the earlier recipient set and $J=I\setminus L$. Suppose $L$ has the priority structure in Claim~\ref{cl:compressionprefix}, i.e.,
\[
L=\operatorname{Top}_{n-k}^{\pi}(I^c)
\cup
\operatorname{Top}_{k}^{\pi}(I^\varnothing),
\]
has size $n$, and contains $k\le g(n,I)$ members of $I^\varnothing$. For any integer $m$ such that $0\le m\le |I|-n$, we have
\begin{equation}\label{eq:residualprefix}
k+g(m,J)
=
\min\{k+m,g(n+m,I)\}.
\end{equation}
\end{claim}
\begin{proof}
For $m=0$, the identity follows from $k\le g(n,I)$. Suppose $m>0$. If $g(n+m,I)\le k+m$, the first $n+m$ original individuals contain at least $n-k$ reserve-eligible individuals and at least $k$ reserve-ineligible individuals. Therefore, they contain all earlier recipients. Deleting those recipients leaves exactly $g(n+m,I)-k$ reserve-ineligible individuals among the first $m$ residual individuals. If $g(n+m,I)>k+m$, the first $n+m$ original individuals contain fewer than $n-k$ reserve-eligible individuals, all of whom were selected earlier, and more than $k+m$ reserve-ineligible individuals, only $k$ of whom were selected earlier. Therefore, their first $m$ residual individuals are all reserve-ineligible. Both cases give Relation~\eqref{eq:residualprefix}.\qedhere
\end{proof}

\begin{proof}[\bfseries Proof of Theorem~\ref{thm:compression}]
Let
\[
k_t=|\widehat\Psi(h_t)\medcap I^\varnothing|,
\quad
k'_t=|\widehat\Psi_{\mathrm{MG}}(h_t)\medcap I^\varnothing|,
\quad
k''_t=|\widehat\Psi_{\mathrm{OA}}(h_t)\medcap I^\varnothing|.
\]
At date $t$, let $n=q^{\Sigma}(h_{t-1})$. Applying Claims~\ref{cl:onebatch}--\ref{cl:residualprefix} to each procedure's own residual population gives
\begin{align}
 k''_t&=\min\{k''_{t-1}+\Delta q_t^o,g(n+\Delta q_t^o,I)\},\label{eq:oarecurrence}\\
 k'_t&=\min\{k'_{t-1}+\Delta q_t^o,g(n+\Delta q_t^o+\Delta q_t^c,I)\},\label{eq:mgrecurrence}\\
 \min\{k_{t-1}+\Delta q_t^o,g(n+\Delta q_t^o,I)\}
 &\le k_t\le
 \min\{k_{t-1}+\Delta q_t^o,g(n+\Delta q_t^o+\Delta q_t^c,I)\}.
 \notag
\end{align}
 Both bounds are nondecreasing in the number of reserve-ineligible individuals selected before date \(t\). Since all three counts start at zero, induction gives
\[
k''_t\le k_t\le k'_t
\]
for every viable category-locked batch procedure. By Claim~\ref{cl:compressionprefix}, each procedure selects the highest-priority individuals within the reserve-eligible and reserve-ineligible populations. Thus, the number selected from each population determines exactly who is selected.

Let $\Delta_t=k'_t-k''_t$. If
\[
k''_{t-1}+\Delta q_t^o\le g(n+\Delta q_t^o,I),
\]
then Relations~\eqref{eq:oarecurrence}--\eqref{eq:mgrecurrence} imply $\Delta_t\le\Delta_{t-1}$. Otherwise $k''_t=g(n+\Delta q_t^o,I)$ and
\[
\begin{aligned}
\Delta_t&\le k'_{t-1}+\Delta q_t^o-g(n+\Delta q_t^o,I)\le\Delta q_t^o,\\
\Delta_t&\le g(n+\Delta q_t^o+\Delta q_t^c,I)-g(n+\Delta q_t^o,I)\le\Delta q_t^c.
\end{aligned}
\]
The first bound uses $k'_{t-1}\le g(n,I)\le g(n+\Delta q_t^o,I)$; the second uses the unit increments of $g$. Hence,
\[
\Delta_t\le\max\{\Delta_{t-1},\min\{\Delta q_t^o,\Delta q_t^c\}\}\le b(h_t).
\]
By Relation~\eqref{eq:prefixform}, $\Delta_t=\bigl|\widehat\Psi_{\mathrm{OA}}(h_t)\setminus \widehat\Psi_{\mathrm{MG}}(h_t)\bigr|$, as required. Any two recipient sets generated by viable category-locked batch procedures have reserve-ineligible counts between $k''_t$ and $k'_t$, so their replacement distance is also at most $b(h_t)$.

To attain the bound, take one batch with $\Delta q_1^o=\Delta q_1^c\ge 1$ and priority consisting of $\Delta q_1^o$ members of $I^c$, then $\Delta q_1^o$ members of $I^\varnothing$, then another $\Delta q_1^o$ members of $I^c$. The minimum guarantee choice rule selects the first two blocks; the O\&A choice rule selects the two category-$c$-eligible blocks. Their replacement distance is $\Delta q_1^o=b(h_1)$. The proof uses only the realized residual outcomes, so allowing the viable category-locked batch procedure to depend on past history does not change the conclusion. \end{proof}

\subsection{Proof of Proposition~\ref{prop:lockingimpossibility}}\label{app:lockingimpossibility}
\begin{proof}[\bfseries Proof of Proposition~\ref{prop:lockingimpossibility}]
Take $I=\mathcal I=\{r_1,u,r_2\}$, $\mathcal I^c=\{r_1,r_2\}$, and priority $r_1\mathrel{\pi}u\mathrel{\pi}r_2$. Consider histories with $q^{\Sigma}(h)\le2$. They satisfy Assumption~\ref{ass:abundantdemand} and $q^{\Sigma}(h)<|I|$. Non-wastefulness and no justified envy imply $r_1$ is assigned to the open category after the first batch of history $h=((1,0),(0,1))$ is released. After the second batch, category locking keeps $r_1$ in the open category, and by non-wastefulness, $r_2$ must be assigned to the reserved category. On the other hand, after the first batch of history $h'=((0,1),(1,0))$ is released, the same axioms assign $r_1$ to the reserved category. After the second batch, category locking keeps her there, and non-wastefulness and no justified envy imply $u$ is assigned to the open category. %
The two histories $h$ and $h'$ have the same endpoint total capacity vector $(1,1)$ but lead to different recipients, contradicting resource-release invariance. %
\end{proof}

\subsection{Proof of Theorem~\ref{thm:reattribution}}\label{app:batchinginvariance}
\begin{proof}[\bfseries Proof of Theorem~\ref{thm:reattribution}]
For any dynamic reserve procedure $\Phi$, define its one-shot choice rule by
\[
\CC_\Phi(q)=\Phi((q)),
\]
where $(q)$ releases all capacities in one batch. Resource-release invariance requires $\Phi(h)=\CC_\Phi(q(h))$ for each history $h$. Conversely, under this equality, histories with the same aggregate capacities give the same outcome.

Suppose also that $\Phi$ satisfies non-wastefulness, no justified envy, and recipient retention. Applying the first two axioms to one-shot histories shows that $\CC_\Phi$ satisfies them.

Fix $q\le q'$. Resource-release invariance gives
\[
\Phi((q))=\CC_\Phi(q),
\qquad
\Phi((q,q'-q))=\CC_\Phi(q').
\]
Therefore, recipient retention implies
\[
\widehat{\CC}_\Phi(q)\subseteq \widehat{\CC}_\Phi(q'),
\]
so $\CC_\Phi$ satisfies capacity monotonicity. The representing choice rule is unique because it must agree with $\Phi$ at every one-shot history.

Conversely, suppose $\CC_\Phi$ satisfies non-wastefulness, no justified envy, and capacity monotonicity. By part~(i), $\Phi(h)=\CC_\Phi(q(h))$. The first two axioms hold at each history. If $h\preceq h'$, then $q(h)\le q(h')$, and capacity monotonicity implies
\[
\widehat\Phi(h)=\widehat{\CC}_\Phi(q(h))\subseteq\widehat{\CC}_\Phi(q(h'))=\widehat\Phi(h').
\]
Thus, $\Phi$ retains recipients. It also satisfies resource-release invariance, since its outcome depends only on aggregate capacities.

\end{proof}

We prove the following result regarding properties of the O\&A and minimum guarantee choice rules. %

\begin{appendixlemma}\label{lem:benchmarkcapacity}
With fixed population and priority and disjoint reserved eligibility, the O\&A and minimum guarantee choice rules are capacity monotone. Consequently, their cumulative implementations retain earlier recipients.
\end{appendixlemma}
\begin{proof}
It suffices to consider one additional position. Under O\&A, an additional reserved position leaves open recipients unchanged and selects the next remaining eligible individual in that category, if one exists. An additional open position goes to the highest-priority individual not already assigned open, if one exists. If she held a reserved position, it goes to the highest-priority remaining eligible individual, when one exists. All other assignments are unchanged, so every earlier recipient remains selected.

Under minimum guarantee, an additional open position selects the next remaining individual after reserved positions are filled, if one exists. An additional category-$c$ position has no effect if all eligible individuals already receive positions in $c$. Otherwise, let $i$ be the next eligible individual assigned to $c$. If $i$ was unselected, open recipients are unchanged. If she held an open position, every other open recipient remains selected, and the next remaining individual receives the vacated position, if anyone remains. Disjoint eligibility leaves other reserved selections unchanged. Thus, both rules are capacity monotone, and their cumulative implementations retain recipients.
\end{proof}

\subsection{Proofs of Theorems~\ref{thm:flexcharacterization}--\ref{thm:precedencecharacterization}, Corollary~\ref{cor:basicrrbcharacterization}, and Proposition~\ref{prop:openadjustment}}\label{app:simpleresults}
We first establish the matroid properties used to characterize RRB. Both rules choose open recipients by priority subject to retention. Joint RRB then chooses reserved recipients from the reserved-position transversal matroid; precedence RRB processes the reserved categories in order. For joint RRB, we use the following attribution convention.

\begin{appendixdefinition}\label{def:attributionconvention}
For fixed population $I$, categories $\mathcal V$, eligibility $(I^v)_v$, capacities $q$, and priority $\pi$, let $O,R\subseteq I$ be disjoint sets with a feasible assignment of $O$ to open positions and $R$ to reserved positions. Define
\[
\mathscr A(O,R)=\left\{\mathrm G:\ \mathrm G\text{ is feasible},\quad
\mathrm G^o=O,\quad \bigcup_{c\in\mathcal R}\mathrm G^c=R\right\}.
\]
A \emph{fixed attribution convention} is a function $\tau$ assigning an element of $\mathscr A(O,R)$ to each such tuple $(I,\mathcal V,(I^v)_v,q,\pi,O,R)$. It does not depend on the inherited profile or on which selected individuals are incumbents. Comparisons of full outcomes use the same function $\tau$.
\end{appendixdefinition}
This convention resolves ties between feasible category assignments of already selected individuals. It does not break ties in individual priority, which is strict.

For distinct sets $X,Y\subseteq I$, let $X\succ_\pi Y$ mean that the highest-priority individual who belongs to one set but not the other belongs to $X$. Every nonempty family of subsets of $I$ has a unique highest-priority element under this order.

\begin{appendixlemma}\label{lem:priorityclaim}
Let $\mathscr F$ be a nonempty family of subsets of a finite set $E$ with strict order $\pi$, and let $L\in\mathscr F$. Then $L$ is the highest-priority element of $\mathscr F$ under $\succ_\pi$ if and only if there are no $i\in E\setminus L$ and $X\in\mathscr F$ such that
\[
i\in X,\qquad \{j\in L:j\mathrel{\pi}i\}\subseteq X.
\]
\end{appendixlemma}
\begin{proof}
Suppose $L$ is not highest-priority, and let $X$ be highest-priority. The highest-priority individual in their symmetric difference belongs to $X$, as do all higher-priority members of $L$. Conversely, suppose $i$ and $X$ satisfy the displayed conditions. Every member of $L\setminus X$ has lower priority than $i\in X\setminus L$. Thus, $X\succ_\pi L$, a contradiction.
\end{proof}

We are ready to introduce our machinery for the RRB characterization. The matroid terminology and basic facts below follow \citet[Chapter~1]{Oxley2011}. A family $\mathcal J$ of subsets of a ground set $E$ satisfies
\emph{heredity} if every subset of a member of $\mathcal J$
also belongs to $\mathcal J$; that is,
$X\in\mathcal J,\quad Y\subseteq X$ implies
 $Y\in\mathcal J$. A \emph{matroid} $\mathbf{M}=(E,\mathcal J)$ consists of a finite ground
set $E$ and a family $\mathcal J$ of \emph{independent sets}
satisfying three conditions: $\emptyset\in\mathcal J$; $\mathcal J$ satisfies heredity; and, whenever
$X,Y\in\mathcal J$ satisfy $|X|<|Y|$, some $i\in Y\setminus X$
satisfies $X\cup\{i\}\in\mathcal J$. The third condition is called the \emph{augmentation property}. The \emph{rank function} of $\mathbf{M}$ is
\[
\rho_{\mathbf{M}}(X)=\max\{|J|:J\subseteq X,\ J\in\mathcal J\},
\qquad X\subseteq E.
\]
Every independent set extends to a maximal independent set,
called a \emph{basis}. All bases have cardinality $\rho_{\mathbf{M}}(E)$.
A set $X$ is \emph{spanning} if $\rho_{\mathbf{M}}(X)=\rho_{\mathbf{M}}(E)$.

A \emph{transversal matroid} is defined by a bipartite graph:
its ground set is one side of the graph, and a subset is
independent if its members can be assigned to distinct adjacent nodes
on the other side. In our setting, the ground set is the current
population $I$. Represent each reserved category $c\in\mathcal R$
by $q^c$ distinct positions, and join individual $i$ to every
position in category $c$ whenever $i\in I^c$.
Let $\mathcal J_q$ be the family of subsets of $I$ whose members
can all be assigned to distinct eligible reserved positions.
The resulting \emph{reserved-position transversal matroid}
is $\mathbf{M}_q=(I,\mathcal J_q)$, with rank function
\[
\rho_q(L)=\rho_{\mathbf{M}_q}(L)
=\max\{|J|:J\subseteq L,\ J\in\mathcal J_q\},
\qquad L\subseteq I.
\]
Thus,
\[
\mathcal J_q=\{L\subseteq I:\rho_q(L)=|L|\}.
\]
That is, $L\in\mathcal J_q$ if and only if all members of $L$ can
simultaneously be assigned to eligible reserved positions, with at most
$q^c$ of them assigned to category $c$ for each $c\in\mathcal R$.

Fix a priority order $\pi$ on $E$. Define
$P_k=\operatorname{Top}_k^\pi(E)$ for each $k=0,\ldots,|E|$.
The \emph{priority-greedy algorithm} starts with the empty set, which is an independent set.
It processes each individual in order $\pi$ and adds her
if the enlarged set remains an independent set. We call its final selection the \emph{priority-greedy basis}.

For $X,Y\subseteq E$, we say that $X$ \emph{Gale dominates} $Y$
if $|X\cap P_k|\ge |Y\cap P_k|$ for every $k$.
Equivalently, $|X|\ge |Y|$ and, when both sets are listed in
priority order, the $\ell$th individual in $X$ has at least as high
priority as the $\ell$th individual in $Y$ for every $\ell\le |Y|$.
This comparison is stronger than the highest-priority
comparison under $\succ_\pi$. The following lemma is standard; see \citet{Gale1968}.

\begin{appendixlemma}[Priority-greedy selection, \citealp{Gale1968}]
\label{lem:greedyprefix}
Let $\mathbf{M}=(E,\mathcal J)$ be a matroid with rank function $\rho_{\mathbf{M}}$,
and let $\pi$ be a strict priority order on $E$.
The priority-greedy algorithm selects a basis $B$ satisfying
\[
|B\cap P_k|=\rho_{\mathbf{M}}(P_k),
\qquad k=0,\ldots,|E|.
\]
Consequently, $B$ Gale dominates every independent set of $\mathbf{M}$
and is the unique highest-priority basis under $\succ_\pi$.
\end{appendixlemma}

\begin{appendixlemma}[Priority-greedy selection with required individuals]
\label{lem:protectedcontraction}
Let $\mathbf{M}=(E,\mathcal J)$ be a matroid with rank function $\rho_{\mathbf{M}}$,
let $\pi$ be a strict priority order on $E$, and let
$K\in\mathcal J$ be a set of individuals who must be selected.

Start with $K$ as the initial independent set. Process individuals
in $E\setminus K$, one at a time, in the priority order of $\pi$,
adding each individual if the enlarged set remains an
independent set of $\mathbf{M}$. The final selection, which we refer to as  $L_K$, is a basis of $\mathbf{M}$
containing $K$. It Gale dominates every independent set
containing $K$ and is the highest-priority basis among
those containing $K$.

Equivalently, define the \emph{contraction}, denoted by $\mathbf{M}/K$, on the ground set
$E\setminus K$ with the following collection of independent sets: for each $Z\subseteq E\setminus K$, $Z$ is independent in the contraction if and only if $K\cup Z$ is independent in $\mathbf{M}$.
Then, $\mathbf{M}/K$ is a matroid with the rank function
\[
\rho_{\mathbf{M}/K}(Z)=\rho_{\mathbf{M}}(K\cup Z)-|K|.
\]
If $Z_K$ is its priority-greedy basis under the restriction
of $\pi$ to $E\setminus K$, then $L_K=K\cup Z_K$.
\end{appendixlemma}

\begin{proof}
The empty set is independent in $\mathbf{M}/K$, and heredity
follows from heredity in $\mathbf{M}$. If $Z_1,Z_2$ are independent
in $\mathbf{M}/K$ and $|Z_1|<|Z_2|$, augmentation of $K\cup Z_1$
by an individual in $Z_2\setminus Z_1$ proves augmentation
in $\mathbf{M}/K$. Thus, $\mathbf{M}/K$ is a matroid.

For the rank function, extend $K$ to a maximal independent
subset of $K\cup Z$. Its members outside $K$ form a
largest independent subset of $Z$ in $\mathbf{M}/K$, with
cardinality $\rho_{\mathbf{M}}(K\cup Z)-|K|$. Therefore,
\[
\rho_{\mathbf{M}/K}(Z)=\rho_{\mathbf{M}}(K\cup Z)-|K|.
\]

Each inclusion test in priority-greedy selection starting
with $K$ is identical to the corresponding test in $\mathbf{M}/K$.
Thus, $L_K=K\cup Z_K$, and
\[
|L_K|
=|K|+\rho_{\mathbf{M}/K}(E\setminus K)
=\rho_{\mathbf{M}}(E),
\]
so $L_K$ is a basis of $\mathbf{M}$.

For any independent set $Y$ containing $K$,
$Y\setminus K$ is independent in $\mathbf{M}/K$.
Applying Lemma~\ref{lem:greedyprefix} to $\mathbf{M}/K$, with
the priority order inherited from $\pi$, shows that
$Z_K$ Gale dominates $Y\setminus K$. For every $k$, adding $K$ to both sets increases each
set's number of members among the $k$ highest-priority
individuals by $|K\cap P_k|$. Thus, the inequalities
defining Gale dominance remain valid. Hence, $L_K$
Gale dominates every independent set containing $K$
and is the unique highest-priority basis among those
containing $K$.
\end{proof}

Fix a choice problem
$
\sigma=\bigl(I,\mathcal V,(I^v)_{v\in\mathcal V},q,\pi,\mathrm H\bigr)$.
Let $\mathscr X^o(\sigma)$ be the family of sets $X\subseteq I$ of size $\min\{q^o,|I|\}$ for which every incumbent outside $X$ can be assigned to an eligible reserved position. For a feasible outcome $\mathrm G$ retaining every incumbent, define
\[
\mathscr X^{\mathcal R}(\mathrm G,\sigma)=
\{Y\subseteq I\setminus\mathrm G^o:
\widehat{\mathrm H}\setminus\mathrm G^o\subseteq Y,\quad
\rho_q(Y)=|Y|=\rho_q(I\setminus\mathrm G^o)\}.
\]
The latter family consists of the largest sets of recipients of reserved positions that retain all incumbents not assigned an open position, holding $\mathrm G^o$ fixed.

We use the following standard characterization of matroid bases \citep[Section~1.2]{Oxley2011}. A nonempty family $\mathscr B$ of subsets of a finite
set $E$, all of the same cardinality, is the family of bases
of a matroid on $E$ if and only if it satisfies the
\emph{basis exchange property}: for any $X,Y\in\mathscr B$
and any $i\in X\setminus Y$, there exists
$j\in Y\setminus X$ such that
\[
(X\setminus\{i\})\cup\{j\}\in\mathscr B.
\]
Since every member of $\mathscr X^o(\sigma)$ has cardinality
$\min\{q^o,|I|\}$, showing that this family is nonempty and
satisfies basis exchange establishes that it is the family
of bases of a matroid on $I$.

\begin{appendixlemma}\label{lem:feasibleopen}
On the retention-feasible domain, $\mathscr X^o$ is nonempty and forms the family of bases of a matroid on $I$. For $q^o\le |I|$, a set $X\subseteq I$ with $|X|\le q^o$ extends to an element of $\mathscr X^o$ if and only if
\[
|\widehat{\mathrm H}\setminus X|-\rho_q(\widehat{\mathrm H}\setminus X)\le q^o-|X|.
\]
The common selection module for open positions in Step~\ref{step:basicopen} of either RRB choice rule is greedy on this family and selects its highest-priority element under $\succ_\pi$.
\end{appendixlemma}
\begin{proof}
The reserved-position transversal matroid $\mathbf{M}_q$ has rank function $\rho_q$. Lemma~\ref{lem:retentionfeasibility} gives an outcome retaining every incumbent. Fill open positions up to $\min\{q^o,|I|\}$ by adding an unselected individual or moving someone from a reserved position to an open position. This proves nonemptiness of $\mathscr X^o$. By its definition, a set filling the open positions belongs to this family exactly when the remaining incumbents can be assigned to distinct eligible reserved positions.

For necessity, suppose $X$ is contained in some member
of $\mathscr X^o(\sigma)$. At most
$\rho_q(\widehat{\mathrm H}\setminus X)$ incumbents
outside $X$ can receive reserved positions. Therefore,
at least
\[
|\widehat{\mathrm H}\setminus X|
-\rho_q(\widehat{\mathrm H}\setminus X)
\]
of these incumbents must receive open positions outside
$X$. Since only $q^o-|X|$ such positions are available,
the inequality is necessary.

For sufficiency, suppose the inequality holds. Choose
a set $B\subseteq\widehat{\mathrm H}\setminus X$ of
cardinality $\rho_q(\widehat{\mathrm H}\setminus X)$
whose members can all be assigned to eligible reserved
positions. Make those assignments, and assign everyone
in $X$ and $(\widehat{\mathrm H}\setminus X)\setminus B$
to open positions. The inequality ensures that these
assignments use at most $q^o$ open positions.
Fill any remaining open positions with other individuals,
moving them from reserved positions if necessary.
There are enough distinct individuals because
$q^o\le |I|$. Every incumbent remains appointed, and
the resulting set of open recipients contains $X$
and belongs to $\mathscr X^o(\sigma)$.

Applying this condition to $X\cup\{i\}$ gives
Inequality~\eqref{eq:opentest} in the definition of Step 1 of the RRB choice rules.

To show that $\mathscr X^o(\sigma)$ is the family of
bases of a matroid, we verify the basis exchange property.
Suppose first that $q^o\le |I|$.

Let $\mathbf{N}$ be the restriction of $\mathbf{M}_q$
to $\widehat{\mathrm H}$, i.e., we consider only subsets of $\widehat{\mathrm H}$, with each subset remaining independent or dependent exactly as in the original matroid.
Thus, a subset of $\widehat{\mathrm H}$ is independent
in $\mathbf{N}$ precisely when all its members can
simultaneously receive eligible reserved positions. Moreover,
the bases of $\mathbf{N}$ are the largest
such incumbent subsets.

Consider the family $
\mathscr B
=
\big \{
\widehat{\mathrm H}\setminus B:
B\text{ is a basis of }\mathbf{N}
\big \}$.
This family is nonempty, and every member has cardinality
$
d=|\widehat{\mathrm H}|-\rho_q(\widehat{\mathrm H})$.
We first verify that $\mathscr B$ satisfies basis exchange.
Let $B_1,B_2$ be bases of $\mathbf{N}$, and let
$i\in B_2\setminus B_1$. The set
$(B_1\cap B_2)\cup\{i\}$ is independent because it is
contained in $B_2$. Extend it to a maximal independent
subset of $B_1\cup\{i\}$. Since $B_1$ is a basis, this
extension has $|B_1|$ members and hence equals
$(B_1\setminus\{j\})\cup\{i\}$ for some
$j\in B_1\setminus B_2$. Taking complements in
$\widehat{\mathrm H}$ gives
\[
\left(
(\widehat{\mathrm H}\setminus B_1)\setminus\{i\}
\right)\cup\{j\}
\in\mathscr B.
\]
This is exactly the basis exchange property for
$\mathscr B$. Consequently, $\mathscr B$ is the family
of bases of a matroid on $I$, which we denote by
$\mathbf{D}$.

On the ground set $\widehat{\mathrm H}$, this construction
is called the \emph{dual} of $\mathbf{N}$: its bases are
the complements of the bases of $\mathbf{N}$.
The individuals in $I\setminus\widehat{\mathrm H}$
belong to no basis of $\mathbf{D}$ and hence to no
independent set of $\mathbf{D}$. Thus, individuals outside
$\widehat{\mathrm H}$ contribute no rank:
\[
\rho_{\mathbf{D}}(X)
=
\rho_{\mathbf{D}}(X\cap\widehat{\mathrm H})
\qquad\text{for every }X\subseteq I.
\]

We next derive the rank of $\mathbf{D}$.
For $X\subseteq I$, every independent set of $\mathbf{D}$
is contained in one of its bases, so
\[
\rho_{\mathbf{D}}(X)
=
|X\medcap\widehat{\mathrm H}|
-
\min_{\substack{B\text{ is a basis}\\
                 \text{of }\mathbf{N}}}
|X\cap B|.
\]
Every basis $B$ of $\mathbf{N}$ has
$\rho_q(\widehat{\mathrm H})$ members, at most
$\rho_q(\widehat{\mathrm H}\setminus X)$ of whom can
lie outside $X$. This bound is attained by extending
a maximal independent subset of
$\widehat{\mathrm H}\setminus X$ to a basis of
$\mathbf{N}$. Therefore,
\[
\rho_{\mathbf{D}}(X)
=
|X\medcap\widehat{\mathrm H}|
-\rho_q(\widehat{\mathrm H})
+\rho_q(\widehat{\mathrm H}\setminus X).
\]
In particular, $\rho_{\mathbf{D}}(I)=d$, and
\[
d-\rho_{\mathbf{D}}(X)
=
|\widehat{\mathrm H}\setminus X|
-\rho_q(\widehat{\mathrm H}\setminus X).
\]
Hence, $X$ spans $\mathbf{D}$ if and only if all
incumbents outside $X$ can simultaneously receive eligible
reserved positions. By the definition of $\mathscr X^o(\sigma)$,
\[
\mathscr X^o(\sigma)
=
\left\{
X\subseteq I:
|X|=q^o,\quad \rho_{\mathbf{D}}(X)=d
\right\}.
\]
Thus, $\mathscr X^o(\sigma)$ consists precisely of the
$q^o$-element spanning sets of $\mathbf{D}$.
Although nonincumbents contribute no rank in $\mathbf{D}$,
they may belong to these spanning sets and receive
open positions. Therefore, they make no contribution
to satisfying incumbent retention.

We can now verify basis exchange for
$\mathscr X^o(\sigma)$. Take
$X,Y\in\mathscr X^o(\sigma)$ and $i\in X\setminus Y$,
and let $Z=X\setminus\{i\}$.
If $Z$ spans $\mathbf{D}$, then
$Z\cup\{j\}$ remains spanning for any
$j\in Y\setminus X$. Since it has $q^o$ members,
it belongs to $\mathscr X^o(\sigma)$.

Otherwise, removing the single individual $i$ reduces
rank by exactly one, so $\rho_{\mathbf{D}}(Z)=d-1$.
Choose an independent set $J\subseteq Z$ of size $d-1$.
Since $Y$ spans $\mathbf{D}$, it contains an independent
set of size $d$. Therefore, the augmentation property
gives some $j\in Y\setminus J$ such that
$J\cup\{j\}$ is independent.
This individual cannot belong to $Z$, because $J$
is a largest independent subset of $Z$.
Also, $j\ne i$ because $i\notin Y$.
Thus, $j\in Y\setminus X$, and $Z\cup\{j\}$ spans
$\mathbf{D}$. It has $q^o$ members. Therefore, it
belongs to $\mathscr X^o(\sigma)$.
This proves basis exchange.

Nonemptiness was established above, so
$\mathscr X^o(\sigma)$ is the family of bases of a
matroid on $I$. Its independent sets are exactly
the sets that extend to a member of
$\mathscr X^o(\sigma)$. The common selection module
for open positions in Step 1 of the RRB choice rules examines individuals in priority
order and uses the extension test to determine whether
each proposed addition remains independent in this
matroid. Therefore, it is the priority-greedy algorithm.
Lemma~\ref{lem:greedyprefix} gives its highest-priority
selection and its Gale-domination comparisons.

Finally, if $q^o>|I|$, all individuals can receive open
positions. Therefore, every proposed addition admits
a retaining completion and satisfies Inequality~\eqref{eq:opentest}.
In this case, $\mathscr X^o(\sigma)=\{I\}$, which is
the family of bases of the matroid on $I$ in which
every subset is independent.\qedhere

\end{proof}

\begin{appendixlemma}\label{lem:claimwitnesses}
For a feasible outcome $\mathrm G$ retaining every incumbent, the open claim in Definition~\ref{def:openclaims}(i) is equivalent to the existence of a set $X\in\mathscr X^o(\sigma)$ containing the claimant and every higher-priority member of $\mathrm G^o$. The joint reserved claim in Definition~\ref{def:openclaims}(ii) is equivalent to the existence of a set $Y\in\mathscr X^{\mathcal R}(\mathrm G,\sigma)$ containing the claimant and every higher-priority member of $\mathrm G^{\mathcal R}$.
\end{appendixlemma}
\begin{proof}
A set $X\in\mathscr X^o(\sigma)$ gives a retaining outcome with open recipients $X$. Conversely, let $\mathrm F$ be an outcome witnessing an open claim. Add unselected individuals or move reserved recipients to open positions until there are $\min\{q^o,|I|\}$ individuals in the open category. No recipient of $\mathrm F$ loses her appointment, and every member of $\mathrm F^o$ remains in the open category. The resulting set belongs to $\mathscr X^o(\sigma)$ and contains the claimant and all required higher-priority recipients.

Similarly, a set $Y\in\mathscr X^{\mathcal R}(\mathrm G,\sigma)$ gives a retaining outcome with open recipients $\mathrm G^o$. Conversely, extend the reserved selection with a joint reserved claim to a basis of the restriction of $\mathbf{M}_q$ to $I\setminus\mathrm G^o$. The basis still contains the remaining incumbents, the claimant, and all required higher-priority recipients, so it belongs to $\mathscr X^{\mathcal R}(\mathrm G,\sigma)$.
\end{proof}

\begin{proof}[\bfseries Proof of Theorem~\ref{thm:flexcharacterization}]
Suppose a rule satisfies the axioms. Incumbent retention and no justified envy subject to retention imply non-wastefulness, as observed in Section~\ref{subsec:rrbcharacterizations}. Let $L$ be its open selection. If $|L|<\min\{q^o,|I|\}$, some individual is unselected or assigned reserved. The former violates non-wastefulness; the latter violates maximal reserve protection subject to retention. Thus, $L\in\mathscr X^o$. No justified envy excludes open claims by unselected individuals; maximal reserve protection subject to retention excludes those by reserved recipients. By Lemmas~\ref{lem:claimwitnesses} and~\ref{lem:priorityclaim}, $L$ is the highest-priority element of $\mathscr X^o$.

Fix $L$ and let $K=\widehat{\mathrm H}\setminus L$. Retention makes $K$ independent in the restriction of $\mathbf{M}_q$ to $I\setminus L$. By Lemma~\ref{lem:protectedcontraction}, the remaining selections are independent sets of its contraction by $K$. If the reserved selection were not a basis, extending it would give an unselected individual a joint reserved claim, contrary to no justified envy. Lemmas~\ref{lem:claimwitnesses} and~\ref{lem:priorityclaim} then identify it as the highest-priority basis containing $K$: the greedy basis of the contraction together with $K$.

Conversely, Lemma~\ref{lem:priorityclaim} rules out open and reserved claims for these two maximizing selections. Feasibility and retention follow by construction. A vacant open position cannot coexist with a recipient of a reserved position or an unselected individual. A vacant reserved position with an eligible unselected individual would permit adding that individual to the set of recipients of reserved positions, contradicting its maximality. Thus, non-wastefulness also holds. These comparisons do not depend on how the recipients of
reserved positions are assigned among reserved categories.
Thus, the set of recipients of open positions and the set
of all recipients are uniquely determined.
\end{proof}

\begin{proof}[\bfseries Proof of Corollary~\ref{cor:basicrrbcharacterization}]
Theorem~\ref{thm:flexcharacterization} determines the recipients of open positions and the aggregate recipient set. Each remaining recipient has only one eligible reserved category, and joint RRB coincides with $\CC_{\mathrm{RRB}}$ on this domain.
\end{proof}

\begin{proof}[\bfseries Proof of Theorem~\ref{thm:precedencecharacterization}]

Fix a retention-feasible problem $\sigma$. We first prove the following claim.

\begin{claim}\label{cl:highestpriority}Under the ranking that ranks outcomes first by category
precedence and then by $\succ_\pi$ at the first category where they
differ, precedence RRB selects the unique highest-ranked feasible
outcome retaining every incumbent.
\end{claim}

\begin{proof}
Consider a category $v$.
Suppose the recipient sets in earlier categories are fixed at those selected by precedence RRB. We refer to a selection of individuals $X$
for category $v$ as \textit{feasible} if it can be extended to a feasible outcome
that retains every incumbent and agrees with the fixed earlier selections.

We first show that precedence RRB chooses the highest-priority feasible selection under $\succ_\pi$ in category $v$.
After fixing the next assignment for an individual $i$, the feasibility conditions in Inequality~\eqref{eq:opentest} in Step 1, if $v=o$, and
Equation~\eqref{eq:sequentialtest} in Step 2, if $v \in \mathcal R$, of precedence RRB hold exactly when all remaining incumbents can
be assigned to eligible positions
remaining in category $v$ or later categories. A rejected assignment cannot become feasible after additional assignments are fixed: any outcome satisfying the additional assignments would also have been available before they were fixed. Since positions within a category are identical, the procedure considers each
unassigned eligible individual in order of $\pi$ and assigns her
whenever her assignment
allows for retention of every remaining incumbent. A position is left empty only if
no eligible individual can occupy it in such an outcome.

Therefore, after all categories are processed using the precedence order,
the procedure ends with a feasible outcome that retains every incumbent.

Consider a category $v$ again. Let $L$ be the rule's selection in category $v$. Suppose a feasible selection
$X$ satisfies $X \succ_\pi L$. Then the highest-priority individual $i$ in   $ (X\setminus L) \cup (L \setminus X)$ is in $(X\setminus L)$.
Thus, every individual selected in category $v$ for $L$ before $i$ precedes her in $\pi$ and belongs to $X$. Since $X$ satisfies the capacity constraint for $v$ and also contains $i$, the category cannot already be filled when it is $i$'s turn. Thus, we showed that when $i$ is processed, there exists an outcome with the feasible category-$v$
selection $X$ (i.e., retaining every incumbent
while preserving the assignments already made).
Thus, the procedure would assign $ i$, leading to a contradiction.

This in turn proves the claim by iterating the above finding over all categories in the precedence order.
\end{proof}

Now let $\mathrm G$ be the outcome of a rule satisfying the axioms for problem $\sigma$.
Incumbent retention ensures that $\mathrm G$ retains every incumbent.
Fix a category $v$ and the rule's earlier-category selections.
A person who is assigned to an earlier category cannot have a feasible
claim to $v$, because her assignment must remain fixed.
No justified envy subject to retention rules out claims by unselected
individuals. Compliance with category precedence subject to retention
rules out claims by individuals assigned to later categories. These cases hold for
every individual outside $\mathrm G^v$. Therefore, by Lemma~\ref{lem:priorityclaim}
 $\mathrm G^v$ is the highest-priority feasible
selection conditional on the earlier-category selections.
Starting with the open category and proceeding in precedence order,
these selections coincide with those of precedence RRB, by Claim~\ref{cl:highestpriority}.

Conversely, precedence RRB is feasible and retains every incumbent,
as established in the proof of Claim~\ref{cl:highestpriority}. In each category, its highest-priority feasible
selection admits no priority claim by Lemma~\ref{lem:priorityclaim}.
This gives both no justified envy and compliance with category
precedence, subject to retention. If an eligible unselected individual
faced a vacant position, assigning her to it while leaving all other
assignments unchanged would give her a feasible priority claim.
Thus, non-wastefulness also holds.
\end{proof}

\begin{proof}[\bfseries Proof of Proposition~\ref{prop:openadjustment}]
Lemmas~\ref{lem:feasibleopen} and~\ref{lem:greedyprefix} give componentwise priority dominance over every other element of $\mathscr X^o$. Let $K=\widehat{\mathrm H}\setminus O$ and $d=|K|-\rho_q(K)$. Every feasible set filling the open positions must include at least $d$ members of $K$, since the rest of $K$ needs to be fit into reserved positions. It follows that at least $d$ members of $O$ have to pass on open positions.

Let $B_K$ be a basis of the restriction of $\mathbf{M}_q$ to $K$ and extend it to a basis $B_H$ of its restriction to $\widehat{\mathrm H}$. Then $B_H\cap K=B_K$. Add $K\setminus B_K$ to $O$ and remove $d$ people from $(O\setminus\widehat{\mathrm H})\medcup(B_H\cap O)$. This set has size $d+q^o-|\widehat{\mathrm H}|+\rho_q(\widehat{\mathrm H})\ge d$ by retention feasibility.  Since every remaining incumbent outside of the open category is part of $B_H$, this construction is feasible and removes $d$ members of $O$.
Since RRB's open selection Gale dominates every element of $\mathscr X^o$,
it keeps at least $q^o-d$ members of $O$. This is exactly the same number of members of $O$ kept in the set just constructed.
Thus, it displaces at most $d$ members of $O$.
The two reverse inequalities imply that it displaces exactly $d$.
\end{proof}

\subsection{Proofs of Proposition~\ref{prop:legacybound} and Lemma~\ref{lem:selectedretention}}\label{app:matchingcomparisons}

We introduce the following concept for the proofs in this subsection.
A \emph{graph matching} pairs individuals with distinct eligible positions,
with no more than one position per individual. Let a graph matching be denoted by its list (or set) of individual--position pairs. We assign a distinct label to each position so that we may compare graph matchings for different capacity vectors. Increasing capacity makes additional
labeled positions available. Observe that every feasible outcome can be represented by a graph matching (possibly many).

We first prove some preparatory lemmas.

\begin{appendixlemma}\label{lem:fixedrepresentation}
Fix the population, eligibility, priority $\pi$, any category precedence, if there is any,
and a finite collection of capacity vectors. For each of the two
RRB rules, there is a strict ranking of graph matchings that represents the rule
in the sense explained below. In particular, each rule's ranking is
fixed across all capacity vectors in this collection and all incumbent
sets. The representing ranking may differ between joint RRB and precedence RRB.

For any capacity vector in the collection and any retention-feasible
incumbent set $K$, let $\beta$ be the highest-ranked graph matching under the
ranking for the rule under consideration, among those that use only
available positions and assign a position to every member of $K$. Then:
\begin{enumerate}[label=(\roman*)]
\item For joint RRB, $\beta$ has the same open recipient set and
aggregate recipient set as the rule's outcome.
\item For precedence RRB, $\beta$ has the same recipient set in
every category as the rule's outcome.
\end{enumerate}
\end{appendixlemma}

\begin{proof} First we fix a linear tie-breaking order $\rightthreetimes$ of all eligible individual--position pairs.

Take two graph matchings: For joint RRB, first compare the sets of open recipients by $\succ_\pi$. If they are equal, compare the sets of reserved recipients by $\succ_\pi$.
For precedence RRB, process categories in precedence order and compare the sets of recipients by $\succ_\pi$ when they first become different. In other words, for each comparison, rank one graph matching over the other based on the highest-priority person assigned in one but not the other, following their respective orders of processing across categories for either rule, as explained.

If two graph matchings are still tied, rank higher the graph matching containing the first pair under tie-breaker $\rightthreetimes$ that is not assigned in the other graph matching. This last comparison distinguishes all pairs of different graph matchings without altering the previous comparisons involving the recipient sets. The number of graph matchings is finite because the set of individuals and the collection of capacity vectors are finite. Thus, every nonempty feasible family has a unique highest-ranked member.

Under the joint comparison, an open vacancy can be filled by adding
an unselected individual or moving a reserved recipient to open,
whenever fewer than $\min\{q^o,|I|\}$ individuals are assigned open.
This raises the open recipient set under $\succ_\pi$ and retains
every incumbent. Hence, the highest-ranked graph matching fills the open
positions to that number. It chooses the highest-priority feasible
open recipient set and, conditional on that set, the highest-priority
feasible reserved recipient set retaining the remaining incumbents.
These are precisely the selections identified in the proof of
Theorem~\ref{thm:flexcharacterization}. For precedence RRB, the proof
of Theorem~\ref{thm:precedencecharacterization} establishes the same
conclusion for the category-by-category comparison.

All comparisons, including the final order of individual--position
pairs, are fixed independently of which positions are available and
which individuals are incumbents. Therefore, those changes affect only
the feasible graph matchings. For joint RRB, apply the prescribed attribution
convention to the resulting recipient sets to obtain the category
outcome.
\end{proof}

Consider either RRB rule and let $\beta$ be the highest-ranked feasible graph matching retaining all incumbents under its ranking in Lemma~\ref{lem:fixedrepresentation}. We call $\beta$ the \textit{representative graph matching for the RRB rule} at the given problem. For joint RRB, the attribution convention determines the reserved-category assignments separately.

\begin{appendixlemma}\label{lem:reserveassignment}
Consider either RRB rule and
the fixed ranking constructed for it in Lemma~\ref{lem:fixedrepresentation}. Let $\beta$
be the highest-ranked feasible graph matching retaining every member of a set of individuals $K$,
and let $L$ be its recipient set in a retention-feasible problem with incumbent set $K$. The
following statements hold for its representative graph matching:
\begin{enumerate}[label=(\roman*)]
\item If we add a position to the problem, every recipient in $L$ will remain assigned by the rule.
\item If we remove a position unused by $\beta$ from the problem, its representative graph matching remains $\beta$.
\item If we change only the incumbent set from $K$ to $K'$ such that $K\subseteq K'\subseteq L$, its representative graph matching remains $\beta$.
\item If we change only the incumbent set from $K$ to a retention-feasible set $K' \supseteq K$, the new recipients outside $L$ are exactly $K'\setminus L$, and at most $|K'\setminus L|$ old recipients are unassigned under the new representative graph matching.
\end{enumerate}
\end{appendixlemma}

\begin{proof}
Fix an RRB rule and the fixed ranking constructed for it in Lemma~\ref{lem:fixedrepresentation}.
For each modification considered in the lemma, we denote the
representative graph matching at the original capacities and incumbent set by $\beta$ and the representative graph matching at the modified capacities
or incumbent set by $\beta'$.

First consider the individual--position pairs common to $\beta$ and
$\beta'$. These assignments will not change. We set them aside. Represent every remaining pair by an edge connecting its individual to its position, labeling the edge $\beta$ or $\beta'$ depending on which graph matching contains the pair. This operation forms a labeled undirected graph.

Each individual and each position belongs to at most one pair in
each graph matching. At most two remaining edges meet at an individual or position. Where two edges meet, one is in $\beta$, and the other is in $\beta'$.

This graph has two types of \emph{components}, which are defined as maximal connected subgraphs.
Starting at a node with only one edge and following the connected edges yields a \emph{chain}, which is a sequence of distinct individuals
and positions with two separate end nodes. If a connected group
has no end node, then every node in it has two edges. Since there are finitely many nodes, following the edges must return to the starting node and yield a \emph{cycle}. A cycle has the same starting node as the ending node, and all other nodes are distinct. In both chains and cycles, consecutive edges alternate between belonging to $\beta$ and $\beta'$. Different chains and cycles share
no individual or position. Therefore, each identified one is a component of the graph.

A chain can have two end nodes consisting of individuals, two end nodes consisting of positions, or one end node of each type. End nodes are not assigned in one of the two graph matchings. In a cycle, all individuals are assigned in both graph matchings but switch their positions.

For a component $C$, let $\beta_C$ and $\beta'_C$ denote its pairs
from $\beta$ and $\beta'$, respectively. \emph{Switching $C$ in $\beta$}
means replacing $\beta_C$ by $\beta'_C$, resulting in the graph matching
\[
\beta_{\oplus C}=(\beta\setminus\beta_C)\cup\beta'_C.
\]
Switching it in $\beta'$ makes the reverse replacement, resulting in the graph matching
\[
\beta'_{\oplus C}=(\beta'\setminus\beta'_C)\cup\beta_C.
\]
These are indeed graph matchings, because neither switch creates an assignment conflict, as different
components share no individuals or positions.

Every individual in a chain who is not an end node and every individual in a cycle is assigned to different positions in the two graph matchings. After the switch, she remains assigned, although her position changes, and her assignment category changes only if those two positions belong to different categories. On the other hand, an individual who is an end node of a chain is assigned in only one graph matching, so she may become assigned or unassigned when the component is switched. A position that is an end node in a chain is used by only one graph matching; switching changes whether that position is used or not.

Every pair that is introduced after a switch consists of a position and an individual eligible for this position, because this pair belongs to one of
the two graph matchings, which satisfy eligibility in their respective problems. Consequently, a
switch is \emph{feasible} in a problem whenever all positions
used after the switch are available in the problem in question, and every required incumbent remains assigned.

We next establish the ranking comparison used below. Fix a
component $C$ of the graph.
We compare $\beta$ with $\beta_{\oplus C}$, and separately compare $\beta'$ with $\beta'_{\oplus C}$.

The assignments outside $C$ can differ between the two problems,
but they remain unchanged within each comparison. Furthermore,
none of the individuals in $C$ is assigned outside $C$ in either
graph matching. Hence, at each graph matching comparison ranking fixed in Lemma~\ref{lem:fixedrepresentation}, the individuals
whose membership differs between $\beta_C$ and $\beta'_C$ determine the comparison between $\beta$ and $\beta_{\oplus C}$ (and also separately between $\beta'$ and $\beta'_{\oplus C}$).

For example, suppose the first differing comparison concerns
open recipients, $\beta_C$ assigns $i$ to open and $\beta'_C$ assigns
$j$ to open, and $i\mathrel{\pi}j$, where $i$ is the highest-priority differing person among the open recipients. If the set of open category recipients preceding $i$ under $\pi$ is $J_1$ in $\beta$, then the ranking between $\beta$ and $\beta_{\oplus C}$ is determined by $\succ_\pi$ for
$J_1\cup\{i\}$ and $J_1\cup\{j\}$. Similarly, if the set of open category recipients preceding $i$ under $\pi$ is $J_2$ in $\beta'$, the ranking between $J_2\cup\{i\}$ and $J_2\cup\{j\}$ through $\succ_\pi$ determines the ranking between the graph matchings $\beta'_{\oplus C}$ and $\beta'$, respectively.
We have
\[
J_1\cup\{i\}\succ_\pi J_1\cup\{j\}
\qquad \& \qquad
J_2\cup\{i\}\succ_\pi J_2\cup\{j\}.
\]
Thus, $\beta$ is ranked over $\beta_{\oplus C}$ in the original problem while $\beta'_{\oplus C}$ is ranked over $\beta'$ in the modified problem.

We apply the same reasoning to every recipient-set comparison in the fixed ranking of the rule under consideration: (i) open recipients first and all reserved recipients second together under the joint RRB, and (ii) using the category order of the precedence RRB, considering each category's selection in order.
If all recipient-set comparisons tie, the fixed tie-breaker order $\rightthreetimes$
of eligible individual--position pairs decides between
the two.

Consequently, the ranking prefers the same assignment subgraph, $\beta_C$ or $\beta'_{C}$, in both comparisons. Since the switches themselves are in opposite directions,
one replaces $\beta_C$ with $\beta'_{C}$, and the other replaces $\beta'_{C}$ with $\beta_C$.
If both switches were feasible, whichever representative graph matching
uses the less-preferred subgraph could be improved by switching
to the preferred subgraph. This contradicts its being
highest-ranked. We conclude that no differing component can
be switched feasibly in both problems.

\begin{claim}\label{cl:matchingcapacity}
Adding a position leaves every individual assigned in $\beta$ assigned in $\beta'$.
\end{claim}

\begin{proof}
Modify the problem by adding a single position $p$. Every member of $K$ is assigned
in both $\beta$ and $\beta'$. If her pair is common, it is unchanged
by any switch. Otherwise, she has one pair in each graph matching
and is an internal individual of a chain or an individual in
a cycle. Thus, she remains assigned after either switch.

Consider a component not containing $p$. All its positions are
available in both problems, and either switch retains every
member of $K$. Therefore, both switches would be feasible,
contradicting the comparison established before the claim. Hence, a
differing component has to include $p$.

If $\beta'$ leaves $p$ unused, there is no differing component as $p$ did not exist in the original problem, and
so $\beta'=\beta$. Otherwise, $p$ is in a pair with a single edge, belonging to
$\beta'$, because it was unavailable under $\beta$, i.e., it is
an end node of a chain.

If the other end node is an individual $j$, the chain has an odd
number of edges: it starts at a position and ends at an
individual. Since the chain's edges alternate between the graph matchings and the first edge belongs
to $\beta'$, the last edge also belongs to $\beta'$. Then $j$
is assigned in $\beta'$ but not in $\beta$. All other individuals in the chain are
internal and remain assigned.

If the other end node is a position, every individual in the
chain is internal and remains assigned. The switch changes
which positions are used but leaves every recipient assigned.

These are all possible end node types, so adding $p$ leaves
all old recipients assigned.
\end{proof}

Modifying the problem by deleting a position that is not used by $\beta$ restricts the set of feasible
graph matchings. However, $\beta$ remains feasible. Since no feasible graph matching is
above $\beta$ in the fixed ranking before the deletion, none can rank above $\beta$ afterward, either.
This proves part~(ii).

Similarly, if $K\subseteq K'\subseteq L$ and $K'$ is the set of incumbents in the modified problem, this restricts the set of feasible graph matchings but
leaves $\beta$ feasible. Thus, $\beta$ remains the representative graph matching in the modified problem as well, proving
part~(iii).

\begin{claim}\label{cl:matchingprotection}
If we modify the problem by expanding the incumbent set from $K$ to a retention-feasible
$K'$, new recipients in $\beta'$ are exactly $K'\setminus L$,
and at most $|K'\setminus L|$ old recipients are unassigned in $\beta'$.
\end{claim}

\begin{proof}
Suppose, contrary to the claim, an individual $i\notin L$ becomes a new recipient in
$\beta'$ but does not belong to $K'$. She has a pair in $\beta'$
and none in $\beta$, so she is an end node of a chain.

Switching this chain in $\beta'$ causes only $i$ to become unassigned, while every internal
individual remains assigned. If the other end node is an
individual, she becomes assigned after the switch. On the other hand, if it is a
position, no other individual becomes assigned or unassigned. Since
$i\notin K'$, the switch retains every member of $K'$.

Switching the same chain in $\beta$ also retains $K$: Every
member of $K \subseteq K'$ is assigned in both graph matchings and, hence,
cannot be an individual end node of the chain. As
capacities do not change, all positions
used by either switch are available in both problems. Then both switches would be
feasible, contradicting the component comparison established before the claim.

We showed that every new recipient in $\beta'$ belongs to $K'\setminus L$.
Conversely, as $\beta'$ retains $K'$, every individual in $K'\setminus L$ becomes a
recipient. This proves that the set of new recipients in $\beta'$ is exactly $K'\setminus L$.

Now let $j\in L$ be an old recipient who is not assigned under
$\beta'$. She is not in $K'$ and is an end node of a chain
with a pair of $\beta$. Switching this chain in $\beta$ makes
only $j$ unassigned. Since $j\notin K'$ and $K\subseteq K'$, this switch
retains every member of $K$.

If the other end node is a position, switching the chain in $\beta'$ would assign $j$
without making any person unassigned: all other individuals in the
chain are internal and remain assigned. Thus, it would
retain $K'$. Since capacities are fixed, both switches would
be feasible in their respective problems, contradicting the component comparison. Hence, this particular chain cannot have a position end node.

We showed that the other end node must be an individual $i$.
Because the chain starts at the individual $j$ with a pair
of $\beta$, alternation implies that it ends at $i$ with a
pair in $\beta'$. Thus, $i$ is assigned in $\beta'$ but not
in $\beta$, so $i\notin L$. Switching this chain in $\beta'$
makes only $i$ unassigned. If $i\notin K'$, that switch would retain
$K'$, again making both switches feasible in their respective problems. Therefore,
$i\in K'\setminus L$.

Thus, every old recipient who becomes unassigned in $\beta'$ is connected along her chain
to a recipient in $K'\setminus L$. A chain has only
two end nodes, and different components do not share any individuals.
Hence, distinct old recipients who become unassigned in $\beta'$ are connected to distinct individuals in $K'\setminus L$. Therefore, at most $|K'\setminus L|$ old recipients
become unassigned in $\beta'$. This proves part~(iv).
\end{proof}

The two claims and the other arguments complete the proofs of the four assertions.
\end{proof}

\begin{proof}[\bfseries Proof of Proposition~\ref{prop:legacybound}]
Use the fixed ranking in Lemma~\ref{lem:fixedrepresentation}, first
with no required incumbents and then with incumbent set $K$.
The last assertion of Lemma~\ref{lem:reserveassignment} gives both
comparisons. When the two recipient sets have equal cardinality,
the number of individuals who become assigned equals the number who become unassigned. If the ordinary
benchmark already selects every member of $K$, its representative graph matching remains
feasible and no competing graph matching ranks higher after retention is
required. Conversely, equality with that benchmark requires it to
select every member of $K$.
For precedence RRB, this establishes equality of the full category
outcomes. For joint RRB, it establishes equality of the open recipient
sets and the aggregate reserved recipient sets. Applying the same fixed
attribution convention to these sets gives the same assignment among
reserved categories, since the convention does not depend on which
selected individuals are incumbents. Thus, the full outcomes coincide
for joint RRB as well, proving Relation~\eqref{eq:benchmarkrecovery}.
\end{proof}

\begin{proof}[\bfseries Proof of Lemma~\ref{lem:selectedretention}]
Use the fixed ranking in Lemma~\ref{lem:fixedrepresentation}.
Requiring additional current recipients to remain selected removes
some feasible graph matchings but leaves the representative graph matching feasible.
Therefore, it remains highest-ranked, as in part~(iii) of
Lemma~\ref{lem:reserveassignment}. For joint RRB, the open and
aggregate recipient sets remain the same, and the fixed attribution
convention gives the same category outcome.
\end{proof}

\subsection{Proofs of Theorem~\ref{thm:flexrelease} and Corollary~\ref{cor:dynamiclegacy}}\label{app:retainedcontinuation}
\begin{proof}[\bfseries Proof of Theorem~\ref{thm:flexrelease}]
For each rule, use its fixed graph matching ranking from Lemma~\ref{lem:fixedrepresentation} and represent increased category capacities by additions of position identities. Lemma~\ref{lem:reserveassignment} preserves every old recipient for each rule. By Lemma~\ref{lem:selectedretention}, protecting these recipients at the new capacities leaves the outcome unchanged, proving part~(i). For joint RRB the sets of recipients of open positions and the aggregate recipient sets coincide; using the same attribution convention gives the stated profile equality.

Induction proves the cumulative recursion in part~(ii). Every release history can be pooled into a one-shot release on this domain, as in Theorem~\ref{thm:reattribution}(i). Therefore, resource-release invariance identifies a dynamic reserve procedure with its
one-shot outcome at the same cumulative capacities, holding $q_0$ and $\widehat{\mathrm H}_0$ fixed, and Theorem~\ref{thm:flexcharacterization} or~\ref{thm:precedencecharacterization} identifies the corresponding selection. Conversely, cumulative implementation is invariant and, by part~(i), retains all preceding recipients.
The precise uniqueness is that of the one-shot characterization, concerning the set of recipients of open positions and the aggregate recipient set for joint RRB and the full category profile for the precedence rule.
\end{proof}

\begin{proof}[\bfseries Proof of Corollary~\ref{cor:dynamiclegacy}]
Theorem~\ref{thm:flexrelease} identifies the outcome at each date with cumulative implementation protecting only $\widehat{\mathrm H}_0$. Proposition~\ref{prop:legacybound} gives the two comparisons at each date. Apply Theorem~\ref{thm:flexrelease}(i) to the same sequence with no original incumbents to obtain the benchmark inclusion. Taking its complement within $\widehat{\mathrm H}_0$ proves the final claim.
\end{proof}

\section{Proofs for the Multi-Institution Setting}\label{app:multiinstitutionproofs}
\setcounter{equation}{0}
\renewcommand{\theequation}{\thesection.\arabic{equation}}

\subsection{Proofs of Lemma~\ref{lem:rrbmenus} and Theorem~\ref{thm:rrbdacharacterization}}

\begin{proof}[\bfseries Proof of Lemma~\ref{lem:rrbmenus}]
Fix an institution $s$. A graph matching in this proof consists of eligible individual--position pairs at $s$, with at most one pair for each individual and each position. Label the positions and fix the ranking of graph matchings in Lemma~\ref{lem:fixedrepresentation}.

We use this same ranking for every subset $J\subseteq I$. This is possible as individuals outside $J$ do not affect the ranking between two graph matchings that assign positions only to individuals in $J$: At each recipient-set comparison, if the two sets are different, then the highest-priority individual who belongs to one set but not the other is in $J$.
If all recipient-set comparisons tie, then every pair that belongs to one graph matching but not the other involves an individual in $J$ as well. Thus, we obtain the same ranking between these two graph matchings when we restrict the priority order to $J$ and the tie-breaker $\rightthreetimes$ to the pairs involving individuals in $J$.

Since $H_s$ is retention-feasible, there is a feasible graph matching that assigns a position to every member of $H_s$. Keeping only its pairs involving individuals in $H_s\cap J$ shows that this set can be retained for every $J\subseteq I$. By Lemma~\ref{lem:fixedrepresentation}, the unique highest-ranked feasible graph matching on $J$ retaining $H_s\cap J$ has recipient set $\widehat{\CC}_s^H(J)$.

We use the graph and the comparison of graph matchings in the proof
of Lemma~\ref{lem:reserveassignment}. For two graph matchings $\beta$
and $\beta'$, remove their common pairs. Represent each remaining
pair by an edge joining its individual to its labeled position,
and label the edge by the graph matching containing it. Ignore vertices with no remaining edge. Each vertex has at most two edges, and
where two edges meet, one belongs to each graph matching. Thus, each
component is a chain or a cycle with alternating edges. Different
components share no individual or position.

For a component $C$, let $\beta_C$ and $\beta'_C$ be its pairs
from the two graph matchings. Switching $C$ gives
\[
\beta_{\oplus C}=(\beta\setminus\beta_C)\cup\beta'_C,
\qquad
\beta'_{\oplus C}=(\beta'\setminus\beta'_C)\cup\beta_C.
\]
As shown in the proof of Lemma~\ref{lem:reserveassignment}, the
ranking prefers the same assignment subgraph, $\beta_C$ or $\beta'_C$, in both
comparisons. Indeed, pairs outside $C$ remain unchanged within each comparison. At the first differing recipient-set comparison, the same highest-priority differing individual
determines which assignment subgraph is preferred.
If all recipient sets agree, the same first differing pair in the fixed tie-breaking order determines the comparison. Hence, if both switches are feasible and retain the protected individuals in their respective problems, one of them improves the representative graph matching, a contradiction.

\begin{claim}\label{claim:menuoneaddition}
When one individual is added to a set, the new representative graph matching
either has the same recipient set, adds only that individual, or
replaces one earlier recipient by that individual.
\end{claim}

\begin{proof}
Let $\beta$ and $\beta'$ be the representative graph matchings on $J$ and
$J\cup\{i\}$, retaining $H_s\cap J$ and $H_s\cap(J\cup\{i\})$,
respectively, where $i\notin J$. If $\beta'$ leaves $i$ unassigned,
then $i\notin H_s$. Thus, the protected sets are the same, and
each graph matching is feasible in the other problem. Since each is
highest-ranked in its problem, uniqueness gives $\beta=\beta'$.

Suppose $\beta'$ assigns $i$. Consider a component of the graph
that does not contain $i$. Every individual in this component
belongs to $J$, and every position is available in both problems. Each member of $H_s\cap J$ is assigned under both graph matchings.
If she belongs to the component, she has one edge from each graph matching and remains assigned after either switch. Individual
$i$, who may also be protected, is outside this component.
Thus, both switches are feasible and preserve the required
protected appointments. The preceding ranking comparison rules
out such a component.

Individual $i$ has one edge, belonging to $\beta'$. Hence, the
graph consists of a single chain with end node $i$. Every internal
individual has one edge from each graph matching and remains assigned. If the other end node is a position, its sole edge belongs to
$\beta'$, so it was unused under $\beta$. In this case, $i$ is
added without displacing anyone.
If the other end node is an
individual, her sole edge belongs to $\beta$. She is the only
earlier recipient who becomes unassigned.
\end{proof}

By Claim~\ref{claim:menuoneaddition},
\[
\widehat{\CC}_s^H(J\cup\{i\})\cap J
\subseteq\widehat{\CC}_s^H(J),
\qquad
|\widehat{\CC}_s^H(J)|
\leq|\widehat{\CC}_s^H(J\cup\{i\})|.
\]
Adding individuals one at a time proves substitutes and the law
of aggregate demand.

For irrelevance of rejected individuals, suppose
$\widehat{\CC}_s^H(J)\subseteq J'\subseteq J$.
Since $H_s\cap J\subseteq\widehat{\CC}_s^H(J)$, we have
$H_s\cap J'=H_s\cap J$. The representative graph matching on $J$ remains
feasible on $J'$ and retains the same protected set. Every
feasible graph matching on $J'$ retaining that set was also feasible
on $J$. Hence, the same graph matching remains uniquely highest-ranked,
and $\widehat{\CC}_s^H(J')=\widehat{\CC}_s^H(J)$.

To prove path independence, fix $J,J'\subseteq I$. By substitutes, every individual in $J$ who is selected from $J\cup J'$ is also selected from $J$. Hence,
\[
\widehat{\CC}_s^H(J\cup J')\subseteq\widehat{\CC}_s^H(J)\cup J'\subseteq J\cup J'.
\]
By irrelevance of rejected individuals,
\[
\widehat{\CC}_s^H(J\cup J')=\widehat{\CC}_s^H\bigl(\widehat{\CC}_s^H(J)\cup J'\bigr).\qedhere
\]
\end{proof}

\begin{proof}[\bfseries Proof of Theorem~\ref{thm:rrbdacharacterization}]
Fix an institution $s$ and a set $J$. Restricting a feasible assignment of $H_s$ to $H_s\cap J$ shows that this set is retention feasible. Theorem~\ref{thm:flexcharacterization} or
Theorem~\ref{thm:precedencecharacterization}, depending on the institution's policy, identifies the recipient set $\widehat{\CC}_s^H(J)$ from its axioms. In the joint case, the theorem also identifies the open recipient set. The possible differences in reserved-category assignments do not affect the recipient set.

These axioms apply to a single choice problem, with the institutional inputs and protected sets fixed. Lemma~\ref{lem:rrbmenus} gives path independence and the law of aggregate demand.
Moreover, preferences range over all strict orders on institutions and being unassigned, and Definition~\ref{def:matchingpolicy} applies the institution's axioms to $D_s(\mu)$. Therefore, \citet[Theorem~1]{DIY2025} implies that RRB--DA is the unique mechanism satisfying individual rationality, strategy-proofness, and the matching policy axioms.

By the institution-level characterizations, a matching satisfies the matching policy axioms if and only if
\begin{equation}\label{eq:matchingfixedpoint}
\mu_s=\widehat{\CC}_s^H(D_s(\mu))\qquad\text{for every }s
\end{equation}
holds. We next show that, among individually rational matchings, Relation~\eqref{eq:matchingfixedpoint} is equivalent to stability for these aggregate choice rules. Suppose the relation holds. Every individual in $D_s(\mu)\setminus\mu_s$ is rejected from $D_s(\mu)$.
By irrelevance of rejected individuals, removing all of them leaves the chosen set $\mu_s$ unchanged, so $s$ accepts its assigned set. For any individual $i$ who prefers $s$ to $\mu(i)$, removing all of them except $i$ also leaves choice unchanged. Thus, $i$ is rejected from $\mu_s\cup\{i\}$, and the matching is stable.

Conversely, suppose the matching is stable, so each institution accepts its assigned set and rejects every individual who prefers it when she is added to that set.
By substitutes, every individual in $D_s(\mu)\setminus\mu_s$ is also rejected from $D_s(\mu)$. Irrelevance of rejected individuals then gives
\[
\widehat{\CC}_s^H(D_s(\mu))=\widehat{\CC}_s^H(\mu_s)=\mu_s.
\]
Therefore, the individual-optimal stable matching is also individually optimal among the individually rational matchings satisfying the matching policy axioms. Individual optimality and strategy-proofness of DA follow from \citet[Theorems~4 and~11]{HatfieldMilgrom2005}, with irrelevance of rejected individuals as in \citet{AygunSonmez2013}.

Take an incumbent $i\in H_s$, and let $\mu$ be the DA outcome. If $s\succ_i\mu(i)$, individual rationality implies that $s$ is acceptable. Hence, $i$ proposed to $s$ and was rejected, contradicting incumbent retention.
Thus, $\mu(i)\succeq_i s=\omega^H(i)$. For nonoverlapping reserves, Corollary~\ref{cor:basicrrbcharacterization} gives the same institution-level characterization for both RRB rules.
\end{proof}

\subsection{Proofs of Proposition~\ref{prop:dabenchmark} and Corollary~\ref{cor:daharm}}

\begin{proof}[\bfseries Proof of Proposition~\ref{prop:dabenchmark}]
If $\mu^H=\mu^{\emptyset}$, Theorem~\ref{thm:rrbdacharacterization} implies that $\mu^{\emptyset}$ honors every inherited appointment. Conversely, suppose $\mu^{\emptyset}(i)\succeq_i\omega^H(i)$ for every incumbent $i$. Consider ordinary individual-proposing DA. By path independence, each institution can choose from all the individuals who have proposed to it so far.

Take $i\in H_s$ who proposes to $s$ in this run. If $s$ rejects $i$, every subsequent proposal is to an institution below $s$, giving $\omega^H(i)=s\succ_i\mu^{\emptyset}(i)$, contrary to the hypothesis. Therefore, $H_s\cap J\subseteq\widehat{\CC}_s^{\emptyset}(J)$ at every proposal set $J$ reached at $s$.
By Relation~\eqref{eq:benchmarkrecovery}, $\widehat{\CC}_s^H(J)=\widehat{\CC}_s^{\emptyset}(J)$ on these proposal sets. The first proposals coincide. Equal proposal sets give equal selections, rejections, and next proposals. By induction, $\mu^H=\mu^{\emptyset}$.
\end{proof}

\begin{proof}[\bfseries Proof of Corollary~\ref{cor:daharm}]
Let $\nu\in\mathcal M^H$. By Theorem~\ref{thm:rrbdacharacterization}, $\mu^H(i)\succeq_i\nu(i)$ for every $i$. If $i\in L(\mu^H;\mu^{\emptyset})$, then $\mu^{\emptyset}(i)\succ_i\mu^H(i)\succeq_i\nu(i)$, so $i\in L(\nu;\mu^{\emptyset})$.
\end{proof}

\subsubsection{Why the Policy Class Matters}\label{app:disruptionboundary}

Consider three institutions $a,b,c$ and four individuals $i,j,k,\ell$. Institution $a$ has one open and one reserved position, with only $i,k$ eligible for the reserved category; $b$ and $c$ each have one open position. All three institutions have baseline priority $j\mathrel\pi k\mathrel\pi\ell\mathrel\pi i$. The acceptable institutions, in preference order, are
\[
i:\ a\succ_i b,\qquad
j:\ a\succ_j b,\qquad
k:\ a\succ_k c,\qquad
\ell:\ c.
\]
Individual $i$ has a protected appointment at institution $a$. Let $\nu$ be the DA outcome with ordinary O\&A after moving $i$ first in $a$'s priority. In the order $(i,j,k,\ell)$, the ordinary DA, RRB--DA, and modified-priority assignments are
\[
\mu^{\emptyset}=(b,a,a,c),\qquad
\mu^H=(a,a,c,\varnothing),\qquad
\nu=(a,b,a,c).
\]
At institution $a$, the open-category recipient is $j$ in the first two matchings and $i$ in $\nu$; the other recipient is assigned the reserved position. Compared with $\mu^{\emptyset}$, the numbers of individuals whose assignments change are $3$ vs. $2$, who are worse off are $2$ vs. $1$, and who become unassigned are $1$ vs. $0$ in $\mu^H$ vs. $\nu$, respectively, while both matchings protect $i$'s appointment.

However, $j$ prefers $a$ to $\nu(j)=b$ and has higher baseline priority than both $i$ and $k$. Assigning $j$ the open category and $i$ the reserved category at institution $a$ is feasible; it also retains $i$.
Therefore, $\nu$ violates no justified envy subject to retention; hence,  $\nu \notin \mathcal M^H$. Without this restriction, all three counts can be lower than under RRB--DA, even though the reserves in this example are nonoverlapping.

\subsection{Proof of Theorem~\ref{thm:dacontinuation}}

In this subsection, only capacities and protected appointments may change. Let $\widehat{\CC}_s^{q,H}(J)$ denote institution $s$'s aggregate choice from $J$ at capacities $q$, with $H_s\cap J$ protected.

\begin{appendixlemma}\label{lem:damonotonicity}
If $q'\geq q$ category by category and $H$ is retention feasible at $q$, then
\[
\varphi_{\mathrm{RRB}}(\succ;q',H)(i)\succeq_i\varphi_{\mathrm{RRB}}(\succ;q,H)(i)
\qquad\text{for every }i.
\]
\end{appendixlemma}

\begin{proof}
For every set $J$, Theorem~\ref{thm:flexrelease}(i), applied with the protected set $H_s\cap J$, gives
\[
\widehat{\CC}_s^{q,H}(J)\subseteq \widehat{\CC}_s^{q',H}(J).
\]
Let $R_s^q(J)=J\setminus \widehat{\CC}_s^{q,H}(J)$ be the set of rejected individuals. By substitutes, $R_s^q(J)\subseteq R_s^q(J')$ whenever $J\subseteq J'$. The capacity comparison above gives $R_s^{q'}(J)\subseteq R_s^q(J)$.

Consider the following version of DA. Start with no rejections. At each round, an individual applies to every acceptable institution to which she has not yet applied and for which all institutions she strictly prefers have rejected her in earlier rounds. Each institution chooses from all its proposals. Stop when no new proposal is made. Let $J$ be institution $s$'s earlier proposers and $J'$ its new proposers. By path independence, choosing from $J\cup J'$ or $\widehat{\CC}_s^{q,H}(J)\cup J'$ gives the same recipients. By substitutes, no earlier rejection is reversed. Thus, this version has the same outcome as DA. It terminates because each individual proposes to each institution at most once.

We show by induction that at every round the rejected pairs under $q'$ are a subset of those under $q$. This holds before the first round. Suppose it holds after a given round. Let $J_s'$ and $J_s$ be the sets of all individuals who have proposed to $s$ by the next round under $q'$ and $q$, respectively. Any individual in $J_s'$ has been rejected by every institution she prefers to $s$ in earlier rounds under $q'$, and hence under $q$. Thus, she has also proposed to $s$ under $q$ by this round, so $J_s'\subseteq J_s$. Therefore,
\[
R_s^{q'}(J_s')\subseteq R_s^{q'}(J_s)\subseteq R_s^q(J_s).
\]
The inclusion continues to hold after either run terminates. If an individual assigned to $s$ under $q$ were assigned to a worse alternative under $q'$, she would have to be rejected by $s$ under $q'$ but not under $q$, a contradiction. An individual unassigned under $q$ cannot be worse off under $q'$, by individual rationality.
\end{proof}

\begin{appendixlemma}\label{lem:daselectedprotection}
Let $\mu=\varphi_{\mathrm{RRB}}(\succ;q,H)$ and $K=(K_s)_{s\in\mathcal S}$, where $K_s=\mu_s$. If $q'\geq q$, then
\[
\varphi_{\mathrm{RRB}}(\succ;q',K)=\varphi_{\mathrm{RRB}}(\succ;q',H).
\]
\end{appendixlemma}

\begin{proof}
Let $\mu'=\varphi_{\mathrm{RRB}}(\succ;q',H)$. By Lemma~\ref{lem:damonotonicity}, $\mu'(i)\succeq_i\mu(i)$ for every $i$. Consider the DA run giving $\mu'$, with each institution choosing from all its proposals. If $i\in K_s$ proposes to $s$ and is rejected, then $s=\mu(i)\succ_i\mu'(i)$, a contradiction. Hence, every member of $K_s$ who proposes to $s$ remains selected there.

Let $i\in H_s$ propose to $s$ in this run. The appointment guarantee at $q$ gives $\mu(i)\succeq_i s$. If $\mu(i)\succ_i s$, then $\mu'(i)\succeq_i\mu(i)\succ_i s$, so $i$ never proposes to $s$, leading to a contradiction. Therefore, $\mu(i)=s$ and $i\in K_s$.

It follows that at every proposal set $J$ at $s$,
\[
H_s\cap J\ \subseteq\ K_s\cap J\ \subseteq\ \widehat{\CC}_s^{q',H}(J).
\]
By Lemma~\ref{lem:selectedretention},
\[
\widehat{\CC}_s^{q',K}(J)=\widehat{\CC}_s^{q',H}(J)
\]
for every proposal set $J$ reached at $s$ in the run with $H$.

Now consider DA at the same capacities $q'$ and preferences, with protected profile $K$. In the first round, individuals make the same proposals in both runs. The displayed equality gives the same choices and hence the same rejections at each institution. Next, suppose that the two runs have the same rejections through a given round. As preferences are also the same, individuals make the same proposals in the next round of both runs. Then, each institution $s$ chooses from the same proposal set $J$ in both runs, and this set is reached at $s$ in the run with $H$. Thus, we can use the displayed equality for $J$ as well, i.e., $s$ makes the same choices and the same rejections in both runs. This in turn proves that the two runs coincide by iterating the argument over the rounds. Therefore,
\[
\varphi_{\mathrm{RRB}}(\succ;q',K)=\mu'.
\]
Observe that this argument uses $H_s\cap J\subseteq K_s\cap J$ only for proposal sets reached in the run with $H$.
\end{proof}

\begin{proof}[\bfseries Proof of Theorem~\ref{thm:dacontinuation}]
By construction, $\mu_0=\varphi_{\mathrm{RRB}}(\succ;q_0,H^0)$. Suppose $\mu_{t-1}=\varphi_{\mathrm{RRB}}(\succ;q_{t-1},H^0)$. The sets assigned under $\mu_{t-1}$ are retention feasible at $q_{t-1}$ and hence at $q_t$. Applying Lemma~\ref{lem:daselectedprotection} with $K=(K_s)_{s\in\mathcal S}$ and $K_s=(\mu_{t-1})_s$ gives
\[
\mu_t=\varphi_{\mathrm{RRB}}(\succ;q_t,K)=\varphi_{\mathrm{RRB}}(\succ;q_t,H^0).
\]
By Lemma~\ref{lem:damonotonicity}, $\mu_t(i)\succeq_i\mu_{t-1}(i)$ for every $i$. Induction gives both conclusions at every date. At the final date, $\mu_T=\varphi_{\mathrm{RRB}}(\succ;q_T,H^0)$ depends only on the final capacities and the fixed initial inputs. Thus, it is independent of the schedule of releases.
\end{proof}

\titleformat{\section}{\normalfont\Large\bfseries}{\thesection}{1em}{}

\clearpage

\setcounter{page}{1}\setcounter{section}{0}\setcounter{equation}{0}
\renewcommand{\thepage}{S.\arabic{page}}
\renewcommand{\thesection}{S.\arabic{section}}
\renewcommand{\theequation}{S.\arabic{equation}}
\setcounter{footnote}{0}
\renewcommand{\thefootnote}{S.\arabic{footnote}}
\setcounter{proposition}{0}
\renewcommand{\theproposition}{S.\arabic{proposition}}
\setcounter{lemma}{0}
\renewcommand{\thelemma}{S.\arabic{lemma}}
\setcounter{corollary}{0}
\renewcommand{\thecorollary}{S.\arabic{corollary}}
\setcounter{definition}{0}
\renewcommand{\thedefinition}{S.\arabic{definition}}
\setcounter{example}{0}
\renewcommand{\theexample}{S.\arabic{example}}
\setcounter{claim}{0}
\renewcommand{\theclaim}{S.\arabic{claim}}
\pdfbookmark[0]{Supplemental Appendix}{supplement-start}

\begin{center}{\Large\bfseries Supplemental Appendix}\par\medskip
{\large Reserve Systems in Dynamic Settings:\\Account Design and Inherited Rights}\par\smallskip
Tayfun S\"onmez and M. Utku \"Unver\par
\end{center}

\section{Implementation across Releases}\label{app:releaseextensions}
These results support Section~\ref{sec:diagnosis}: they compare repeated and cumulative choice, establish the Parcoursup equivalence, and show how the number of reserved categories and arrivals affect the compression bound.

\subsection{Repeated and Cumulative Choice}\label{app:mgimplementation}

Theorem~\ref{thm:compression} compares repeated O\&A with repeated minimum guarantee. It does not compare either rule with its own cumulative implementation. 
That is what we do in the next result. 
Compared with cumulative implementation, repetition can only increase the number of reserve-eligible recipients under minimum guarantee. Under O\&A, it can only decrease that number.

\begin{proposition}\label{prop:mgasymmetry}
Under the assumptions of Theorem~\ref{thm:compression}, at each date repeated application of the minimum guarantee choice rule selects weakly more reserve-eligible individuals than its cumulative implementation, while repeated application of the O\&A choice rule selects weakly fewer than its cumulative implementation. The repeated and cumulative minimum guarantee recipient sets coincide at date $t$ if the minimum binds in each residual batch through $t$, or if it binds in none of them.
\end{proposition}
Here the minimum binds when the highest-priority individuals up to the batch's total capacity include fewer reserve-eligible individuals than the batch's reserved capacity. The equality concerns recipient sets, not necessarily category assignments.

\begin{proof}
Use $g$, $k'_t$, and $k''_t$ from the proof of Theorem~\ref{thm:compression}, and let $o_t=q^o(h_t)$ and $n_t=q^\Sigma(h_t)$. Applying Claim~\ref{cl:onebatch} to one batch with the cumulative capacities shows that cumulative minimum guarantee selects $\min\{o_t,g(n_t,I)\}$ reserve-ineligible individuals. Under repetition, each such individual occupies an open position, so $k'_t\le o_t$. Also, Relation~\eqref{eq:mgrecurrence} gives $k'_t\le g(n_t,I)$. Hence, repeated minimum guarantee selects weakly fewer reserve-ineligible individuals. Both implementations fill all $n_t$ positions. As a result, repetition selects weakly more reserve-eligible individuals.

Cumulative O\&A selects $g(o_t,I)$ reserve-ineligible individuals. We prove $k''_t\ge g(o_t,I)$ by induction, starting from date zero. Given $k''_{t-1}\ge g(o_{t-1},I)$, the fact that $g$ increases by at most one at each rank gives
\[
k''_{t-1}+\Delta q_t^o\ge g(o_{t-1},I)+\Delta q_t^o\ge g(o_t,I).
\]
Also, $n_{t-1}\ge o_{t-1}$ gives $g(n_{t-1}+\Delta q_t^o,I)\ge g(o_t,I)$. Therefore, both terms in the minimum in Relation~\eqref{eq:oarecurrence} are at least $g(o_t,I)$, proving the induction step. Repeated O\&A consequently selects weakly fewer reserve-eligible individuals.

Finally, suppose the minimum binds in every residual batch through date $t$. In each batch, we consider the highest-priority individuals up to its total capacity. The number of reserve-eligible individuals in this group is below the reserved capacity. Thus, the reserved step selects all reserve-eligible individuals in this group and fills the remaining reserved positions with lower-priority reserve-eligible individuals.
 At least as many reserve-ineligible individuals as open positions remain in the initial segment, so the open step selects only reserve-ineligible individuals. Summing across batches gives $k'_t=o_t$. Together with $k'_t\le g(n_t,I)$, this yields $k'_t=\min\{o_t,g(n_t,I)\}$. Both implementations select the highest-priority individuals within each group. As a result, the equality of these counts also implies the equivalence of their recipient sets.

Instead, suppose the minimum binds in none of the residual batches through $t$.  Then, in each batch, the highest-priority individuals in the residual population up to the batch's total capacity contain enough reserve-eligible individuals to fill its reserved positions. Thus, minimum guarantee selects precisely this initial segment by priority in the residual population. By induction across batches, the union of residual recipients after date $t$ is the first $n_t$ individuals in the original priority order. Every batch contributes at least its reserved capacity in reserve-eligible recipients, so this union contains at least $q^c(h_t)$ such individuals. It consequently satisfies the cumulative minimum, and cumulative minimum guarantee selects this same set.
\end{proof}

Equality of minimum guarantee recipient sets across the schedules in Example~\ref{ex:200} does not extend to all populations, even with one reserved category. Let $I=\mathcal I=\{r_1,r_2,r_3,r_4,u_1,u_2\}$, $\mathcal I^c=\{r_1,r_2,r_3,r_4\}$, and
\[
\pi:\quad r_1\quad r_2\quad u_1\quad u_2\quad r_3\quad r_4.
\]
Release one open and one category-$c$ position at each of two dates, so $h=((1,1),(1,1))$. Repeated application of the minimum guarantee choice rule assigns $r_1$ to category $c$ and $r_2$ to the open category at date 1, then $r_3$ to category $c$ and $u_1$ to the open category at date 2. Cumulative implementation at date 2 instead assigns $r_1,r_2$ to category $c$ and $u_1,u_2$ to the open category. Hence,
\[
\widehat\Psi_{\mathrm{MG}}(h)=\{r_1,r_2,r_3,u_1\},
\qquad
\widehat\Phi^{\CC_{\mathrm{MG}}}(h)=\{r_1,r_2,u_1,u_2\}.
\]
This history satisfies Assumption~\ref{ass:abundantdemand}, with $|\mathcal I^c|=q^{\Sigma}(h)=4<|I|=6$. Cumulative implementation retains both first-date recipients but reassigns $r_2$ from the open category to category $c$. Thus, category reassignment can also matter for the minimum guarantee policy implemented through allocation categories. \citet{AbizadaBo2021} study this failure of separate batch guarantees and correct it by counting preceding hires.
\subsection{Parcoursup: Calling Orders with Historical Counting}\label{app:callingresult}

We establish the fixed-list equivalence stated in Section~\ref{app:parcoursup}, with the minimum for grant recipients as the only additional restriction.

Fix a finite population $I$, a strict priority $\pi$, and a set $I^c\subseteq I$ of applicants receiving secondary-school grants. Let $a_0=0$ and $a_k-a_{k-1}\in\{0,1\}$ for $k=1,\ldots,|I|$, and let $m_k=\min\{a_k,|I^c|\}$. Both $m_k$ and $k-m_k$ are nondecreasing in $k$. Parcoursup's single-minimum construction takes $a_k=\lceil\lambda k\rceil$.

Construct the calling order one position at a time. At position $k$, if, for positions $1$ to $k-1$, fewer than $m_k$ grant recipients have been placed, we select the highest-priority remaining grant recipient; otherwise, we select the highest-priority remaining applicant.

To compare the calls for individuals with the calls for batches, consider the following retained rule. At capacity $k$, the account presents the initial applicants, the minimum $m_k$, and a protected set $K$ of applicants that have already been put into the calling order. This is different from enrollment or pending offers; it is about who was placed in the calling order earlier. For $K\subseteq I$ with $|K|\le k$, the single-minimum retained rule $\mathrm{C}_{\mathrm{GE}}^{\mathrm{call}}(I,k,m_k,K)$ retains the set $K$, adds grant recipients in priority order until either $m_k$ is reached or capacity is filled, and then fills any remaining positions in priority order. Note that the earlier grant recipients in $K$ already count toward the minimum; if they were also subtracted from $m_k$, their opportunities would be counted twice.

\begin{corollary}\label{cor:calling}
Among complete calling orders satisfying the attainable minimum for grant recipients $m_k$ among their first $k$ applicants for each $k$, the calling procedure gives the unique order that places the higher-priority applicant first at the first position where it differs from any other such order. Its first $k$ recipients are
\begin{equation}\label{eq:callingspecialization}
L_k=\widehat\CC_{\mathrm{MG}}\big(I,(k-m_k,m_k)\big).
\end{equation}
For $\ell\le k$, applying the single-minimum retained rule at capacity $k$ while protecting $L_\ell$ gives $L_k$. Consequently, processing an increasing sequence of cumulative calling capacities gives the same recipient sets at common capacities.
\end{corollary}
\begin{proof}
We first verify that the procedure meets every minimum and can construct a complete order. Suppose its first $k-1$ positions meet their minima. If these positions contain fewer than $m_k$ grant recipients, they contain exactly $m_k-1$, because the previous minimum was met and $m_k-m_{k-1}\leq 1$. A remaining grant recipient exists because $m_k\leq|I^c|$, and placing her next meets the new minimum. Otherwise, we can place any remaining applicant next. We continue in this way until every applicant is placed.

To prove uniqueness, consider another complete calling order that satisfies all minima. Let $k$ be the first position at which this complete order differs from the order we constructed. Then the first $k-1$ applicants are the same in both orders. If this common set of applicants contains fewer than $m_k$ grant recipients, both orders must select a grant recipient next. Our procedure selects the highest-priority remaining grant recipient. If the initial common set of applicants already meets the minimum $m_k$, our procedure selects the highest-priority remaining applicant. The two orders select different applicants in position $k$. Thus, in either case, our procedure selects the higher-priority of the two applicants in this position. This proves uniqueness.

Next, compare the minimum guarantee selections at capacities $k-1$ and $k$. By Lemma~\ref{lem:benchmarkcapacity}, every earlier recipient remains selected, and exactly one applicant is added. If $m_k=m_{k-1}$, the additional position is open and goes to the highest-priority remaining applicant. If $m_k=m_{k-1}+1$, the additional position is reserved. Let $i$ be the $m_k$-th highest-priority grant recipient. Since minimum guarantee selects grant recipients in their priority order, $i$ is already selected if and only if the earlier selection contains at least $m_k$ grant recipients. If she is not already selected, she receives the additional reserved position, and the open selection is unchanged. Otherwise, she moves from an open position to the additional reserved position, and the highest-priority remaining applicant receives her open position. Thus, at every $k$, the additional recipient is exactly the next applicant in the calling order. This proves Relation~\eqref{eq:callingspecialization}.

Finally, consider the single-minimum retained rule at capacity $k$, with $L_\ell$ protected, and let $a_c=|L_\ell\cap I^c|$. The calling procedure selects grant recipients in their priority order, so $L_\ell$ contains the first $a_c$ grant recipients. If $a_c<m_k$, the retained rule adds the next $m_k-a_c$ grant recipients. These applicants belong to $L_k$, since minimum guarantee at capacity $k$ selects the first $m_k$ grant recipients. If $a_c\geq m_k$, no additions are needed to meet the minimum. Since $L_\ell\subseteq L_k$, everyone retained or added belongs to $L_k$. Hence, this step meets the minimum without exceeding capacity.

Next, the rule fills the remaining positions with the highest-priority remaining applicants. Every member of $L_k$ who remains to be selected was chosen in the open step of the minimum guarantee rule at capacity $k$. Indeed, every member of its reserved selection either belongs to $L_\ell$ or was added in the preceding step. By the definition of the open step, each member of $L_k$ who has not yet been selected has higher priority than every applicant outside $L_k$. Hence, the retained rule selects these applicants and gives $L_k$. If $\ell=k$, the same conclusion holds because every member of $L_k$ is already retained. Then, we can apply this argument at each subsequent cumulative capacity. This proves the continuation claim.
\end{proof}

For the proportional minimum, the cumulative recipient sets also coincide with those of the sequential adjusted minority reserves rule of \citet[Theorem~2 and Corollary~1]{AbizadaBo2021}.

\subsection{Several Nonoverlapping Reserved Categories}\label{app:multigroup}
Let $\mathcal R=\{c_1,\ldots,c_m\}$ be a set of reserved categories with pairwise disjoint eligible populations $(I^c)_{c\in\mathcal R}$. Let $I^\varnothing$ be the population ineligible for all reserved categories, and fix $I=I^\varnothing\cup\left(\bigcup_{c\in\mathcal R}I^c \right)\subseteq\mathcal I$. At date $t'$, the capacity vector released is
\[
\Delta q_{t'}=\bigl(\Delta q_{t'}^o,(\Delta q_{t'}^c)_{c\in\mathcal R}\bigr).
\]
Let $h$ be the release history and $h_t$ its history through date $t$, and fix a strict priority $\pi$ on $I$. Every individual is eligible for open positions. Each earlier recipient retains her position and its category. As in Theorem~\ref{thm:compression}, the outcome in each residual batch is feasible, non-wasteful, and satisfies no justified envy. Assume
\[
|I^c|\geq\sum_{t'=1}^T\bigl(\Delta q_{t'}^o+\Delta q_{t'}^c\bigr)
\qquad\text{for every }c\in\mathcal R.
\]
A member of $I^c$ can receive a position only in category $c$ or the open category. Thus, enough members of $I^c$ remain to fill its reserved positions in every batch, whatever the earlier choices. The same assumption ensures enough demand to fill the open positions. Let
\[
d_{t'}=\min\left\{\Delta q_{t'}^o,\sum_{c\in\mathcal R}\Delta q_{t'}^c\right\}.
\]

\begin{proposition}\label{prop:multigroup}
For any two viable category-locked batch procedures $\Psi,\Psi^{\prime}$ and each date $t=1,\ldots,T$,
\[
\bigl|\widehat\Psi(h_t)\setminus\widehat\Psi^{\prime}(h_t)\bigr|
\leq\sum_{t'=1}^{t}d_{t'}.
\]
In particular, all viable category-locked batch procedures select the same aggregate recipients when $d_{t'}=0$ at each date. For each positive integer $d$, there is an instance with three reserved categories, with at most two positions in any batch, in which the recipient sets obtained by repeated application of the O\&A and minimum guarantee choice rules differ by $d$ recipients and $\sum_{t'=1}^{T}d_{t'}=d$. Moreover, the single-reserve maximum-batch bound in Theorem~\ref{thm:compression} fails even with two reserved categories.
\end{proposition}

\begin{proof}
No justified envy implies that individuals within each eligibility population are selected in priority order. Otherwise, a higher-priority individual who remains unselected would have been available and eligible when a lower-priority individual from the same population received a position. Thus, the number selected from each population determines who is selected. Consider two cumulative selections with population counts $a=(a_v)_{v\in\{\varnothing\}\cup\mathcal R}$ and $a'=(a'_v)_{v\in\{\varnothing\}\cup\mathcal R}$. Within each population, the smaller selection is a subset of the larger, so their symmetric difference has size $|a_v-a'_v|$. Therefore, if the two cumulative selections have the same size, then their replacement distance is
\[
\frac12\sum_{v\in\{\varnothing\}\cup\mathcal R}|a_v-a'_v|.
\]
We prove the following two claims.

\begin{claim}\label{cl:multigroupsamepool}
Two feasible, non-wasteful outcomes with no justified envy for the same residual applicant population in batch $t'$ differ by at most $d_{t'}$ recipients.
\end{claim}
\begin{proof}
Let $p^o=\Delta q_{t'}^o$ and $p^c=\Delta q_{t'}^c$ for each $c\in\mathcal R$, and let $p=p^o+\sum_{c\in\mathcal R}p^c$. Both outcomes fill all positions. Each includes the first $p^c$ remaining members of $I^c$. Otherwise, one of its $p^c$ reserved positions is assigned to a lower-priority member of $I^c$ while a higher-priority member remains unselected, contradicting no justified envy. Since the eligible populations are disjoint, both outcomes select at least $p-p^o$ individuals in common. Moreover, the first $p^o$ applicants in the residual priority order are selected in both outcomes. If one of them were omitted, at least one of the $p^o$ open recipients would have lower priority, again violating no justified envy. These two common sets can overlap, but their union has at least $\max\{p^o,p-p^o\}$ members. The outcomes each have $p$ recipients, so their replacement distance is at most
\[
p-\max\{p^o,p-p^o\}=\min\{p^o,p-p^o\}=d_{t'}.
\]
\end{proof}

\begin{claim}\label{cl:multigroupsequence}
Suppose that two aggregate selections have the same size and each consists of the highest-priority individuals within every eligibility population. Starting from each selection, we use the same order of individual positions, with each position labeled by its category, and assign each position to the highest-priority remaining eligible individual. Assume that when each position is reached, an eligible individual remains in each residual population. Positions of the same category need not be consecutive. Then the replacement distance between the aggregate selections cannot increase.
\end{claim}
\begin{proof}
For a category-$c$ reserved position, $(a_c,a'_c)$ becomes $(a_c+1,a'_c+1)$, and the other counts stay fixed. Thus, all differences in the form $a_v-a'_v$ are unchanged. The same argument applies to an open position when both recipients belong to the same eligibility population.

Suppose instead that, for an open position, the first procedure selects an individual from $I^v$ and the second selects an individual from $I^w$, where $v,w\in\{\varnothing\}\cup\mathcal R$ and $v\neq w$. Only two count differences change: $a_v-a'_v$ increases by one, and $a_w-a'_w$ decreases by one. Each of their absolute values changes by one in magnitude. Hence, their sum can increase only if both absolute values increase, requiring
\[
a_v\geq a'_v,\qquad a_w\leq a'_w.
\]
Let $j_v$ and $j'_v$ be the next remaining individuals in population $I^v$ under the first and second procedures, respectively, and define $j_w,j'_w$ in the same way for population $I^w$. All four individuals exist under these inequalities: the second procedure has selected no more members of $I^v$ than the first, which has just selected another, and the first has selected no more members of $I^w$ than the second, which has just selected another. Because selections within each population follow priority, $j'_v$ has weakly higher priority than $j_v$, and $j_w$ has weakly higher priority than $j'_w$. But the first procedure's open selection requires $j_v\mathrel{\pi}j_w$, while the second requires $j'_w\mathrel{\pi}j'_v$. Together these comparisons contradict strict priority. Hence, the replacement distance cannot increase at an open position either. Applying this argument to each position proves the claim.
\end{proof}

Any feasible, non-wasteful batch outcome satisfying no justified envy can be obtained by processing its recipients in priority order, using their assigned categories. Each is the highest-priority remaining
individual eligible for her position: other remaining recipients rank lower, and a higher-priority unselected eligible individual would violate no justified envy. Conversely, any order of the batch's individual positions, with such a priority choice at each position, gives a viable outcome under the displayed demand assumption.

For this history, choose an order of the individual positions that gives $\Psi$'s date-$t$ batch outcome, and apply this same order starting from each procedure's preceding recipient set. Starting from $\Psi$'s set
reproduces its actual date-$t$ selection.
Starting from $\Psi^{\prime}$'s set gives a possibly different, constructed selection. Claim~\ref{cl:multigroupsequence} bounds the distance between these two cumulative selections by the distance at date $t-1$.

Next, we compare the constructed selection with $\Psi^{\prime}$'s actual selection at date $t$. They have the same earlier recipients, and each of them assigns the date-$t$ batch through a viable outcome for the same residual population. Then, by Claim~\ref{cl:multigroupsamepool}, their sets of new recipients differ by at most $d_t$ individuals, and so do the two cumulative selections, as their earlier recipients are the same.

Now consider an individual in $\widehat\Psi(h_t)\setminus\widehat\Psi^{\prime}(h_t)$. Suppose first that she is not in the constructed selection. By Claim~\ref{cl:multigroupsequence}, the number of such individuals is at most the replacement distance at date $t-1$.
Suppose next that she is in the constructed selection. As she is not in $\widehat\Psi^{\prime}(h_t)$, we showed above that there are at most $d_t$ such individuals.
Therefore,
\[
\bigl|\widehat\Psi(h_t)\setminus\widehat\Psi^{\prime}(h_t)\bigr|
\leq \bigl|\widehat\Psi(h_{t-1})\setminus\widehat\Psi^{\prime}(h_{t-1})\bigr|+d_t.
\]
Both procedures start with no recipients, so we obtain the bound in the proposition by induction on $t$. In particular, if $d_{t'}=0$ at each date, then the two procedures select the same aggregate recipients at each date.

To show that the single-reserve maximum-batch bound need not hold with two reserved categories $x,y \in \mathcal R$, consider priority
\[
\pi:\quad y_1\quad y_2\quad x_1\quad x_2\quad y_3\quad y_4\quad x_3\quad u\quad y_5\quad y_6,
\]
followed by enough lower-priority members of $I^x$ and $I^y$ to satisfy the displayed demand assumption. The two release vectors, with coordinates ordered as $(o,x,y)$, are $(1,0,2)$ and $(1,2,2)$. MG selects $\{x_1,y_1,y_2\}$ in the first batch and $\{x_2,x_3,y_3,y_4,u\}$ in the second. The rule that processes category $x$ first, the open category second, and category $y$ last selects $\{y_1,y_2,y_3\}$ in the first batch and $\{x_1,x_2,y_4,y_5,y_6\}$ in the second. The recipients selected only by the former are $\{x_3,u\}$, and those selected only by the latter are $\{y_5,y_6\}$. Their replacement distance is two, although $d_{t'}=1$ in each batch.

Finally, fix a positive integer $d$. We construct an instance attaining the bound with reserved categories $x,y,z \in \mathcal R$. Let $n_k=k(k+1)/2$ and $n_0=0$. Form the priority order by placing the following blocks in order for $k=1,\ldots,d$:
\[
\pi:\quad y_{2k-1}\quad z_{n_{k-1}+1}\quad\cdots\quad z_{n_k}\quad y_{2k}\quad x_{n_{k-1}+1}\quad\cdots\quad x_{n_k}.
\]
Append enough lower-priority members of each eligible population to meet the displayed demand assumption. Phase $k$ contains one batch with an open position and a category-$y$ reserved position, followed by $k$ category-$z$ positions released separately, then one open position, and, when $k<d$, $k$ category-$x$ positions released separately. Only the first batch of each phase is mixed, so $\sum_{t'=1}^{T}d_{t'}=d$.

At the start of phase $k$, the cumulative eligible-population counts are
\[
\begin{array}{c|ccc}
 &x&y&z\\\hline
\MG&n_{k-1}&2k-2&n_{k-1}+k-1\\
\OA&n_{k-1}+k-1&2k-2&n_{k-1}.
\end{array}
\]
The counts hold at the start of phase 1, when both procedures have selected no one. Suppose they hold at the start of phase $k$. Both have selected the first $2k-2$ category-$y$ individuals. MG has selected $n_{k-1}+k-1=n_k-1$ category-$z$ individuals, while OA has selected $n_{k-1}$. Thus, in the mixed batch, MG selects $y_{2k-1}$ for the reserved position and then $z_{n_k}$ for open. OA selects $y_{2k-1}$ for open and $y_{2k}$ for reserved. Releasing the next $k$ category-$z$ positions increases both category-$z$ counts by $k$.

At the next open position, MG still has $y_{2k}$ available, and she has higher priority than every other remaining applicant. OA has already selected her and every category-$z$ individual in the current priority block. Its category-$x$ count is $n_{k-1}+k-1=n_k-1$, so it selects $x_{n_k}$. If $k<d$, the final $k$ category-$x$ positions increase both category-$x$ counts by $k$. The resulting counts are exactly those displayed for the start of phase $k+1$, proving the induction. In the last phase there is no final release of category-$x$ positions, and the final numbers selected from each eligible population are
\[
\begin{array}{c|ccc}
 &x&y&z\\\hline
\MG&n_d-d&2d&n_d+d\\
\OA&n_d&2d&n_d.
\end{array}
\]
Since each procedure selects the highest-priority individuals within every population, these counts give a replacement distance of exactly $d$. Every batch contains at most two positions, completing the proof.
\end{proof}

In this construction, the two procedures differ by $d$ recipients out of $d^2+3d$ positions. This difference can be arbitrarily large, but its share of total capacity, $1/(d+3)$, tends to zero. If each batch contains only open positions or only reserved positions, all viable category-locked batch procedures select the same recipients. Cumulative implementation of the intended rule may still select different recipients.

\subsection{Failure of the Single-Reserve Bound with Arrivals}\label{app:entryboundary}

\begin{example}\label{ex:entry}
Use the single-reserve notation of Theorem~\ref{thm:compression}. Fix an integer $T\geq2$ and dates $0,1,\ldots,T$. At date 0, no individual has been selected and $2T$ category-$c$-eligible individuals are available. At each date $t'=1,\ldots,T$, one open position and one category-$c$ position become available. Three individuals also arrive: category-$c$-eligible applicants $r_{t'}^1,r_{t'}^2$ and applicant $u_{t'}\notin \mathcal I^c$. Fix a priority order that ranks every later cohort above every earlier cohort and ranks all arriving cohorts above the initial population. Within each cohort, priority is
\[
\pi:\quad r_{t'}^1\quad u_{t'}\quad r_{t'}^2.
\]
Every unselected individual remains available through date $T$, and each earlier recipient keeps her category assignment. At every date $t'$, repeated MG selects $r_{t'}^1,u_{t'}$, whereas repeated OA selects $r_{t'}^1,r_{t'}^2$. Thus, the individuals selected only by MG are $\{u_1,\ldots,u_T\}$, and the replacement distance is $T$, whereas
\[
\max_{t'\leq T}\min\{\Delta q_{t'}^o,\Delta q_{t'}^c\}=1.
\]
The initial population alone contains enough category-$c$-eligible individuals to fill all $2T$ positions. Nevertheless, the two procedures differ by one additional recipient at every date.
\end{example}

\section{History and the Current Allocation Problem}\label{app:accountdetails}

In this section, we give four examples of how the account forms the current problem, as described in Section~\ref{sec:state}. 
In the first, earlier participation affects whether a person can be admitted. In the second, doses owed to earlier recipients reduce the capacity for new ones. In the third, the category of a vacancy depends on the post. 
The fourth is a promotion case. There, a policy requirement restricts the account and the choice rule together.

\subsection{Annual Participation in Medicaid}\label{app:medicaidchoice}
Section~\ref{subsubsec:ledgerinformation} illustrates the distinction between the limit on current enrollment and the annual limit on different participants for California's Assisted Living Waiver (see main-text footnote~\ref{fn:medicaid} for the official sources). Fix a waiver year with limits $\bar q^{\mathrm{occ}}$ and $\bar q^{\mathrm{ann}}$.
At date $t$, the ledger records who has been served so far in the waiver year, $I_t^{\mathrm{yr}}$, which of them are still enrolled, $I_t^{\mathrm{cont}}\subseteq I_t^{\mathrm{yr}}$, and who is waiting and eligible, $I_t^{\mathrm{wait}}$; nobody is both enrolled and waiting. We assume that neither limit has been exceeded, so $|I_t^{\mathrm{cont}}|\le\bar q^{\mathrm{occ}}$ and $|I_t^{\mathrm{yr}}|\le\bar q^{\mathrm{ann}}$.

The account sets $I_t=I_t^{\mathrm{wait}}$ and calculates
\[
q_t^{\mathrm{occ}}=\bar q^{\mathrm{occ}}-|I_t^{\mathrm{cont}}|,
\qquad q_t^{\mathrm{ann}}=\bar q^{\mathrm{ann}}-|I_t^{\mathrm{yr}}|,
\qquad R_t=I_t^{\mathrm{wait}}\medcap I_t^{\mathrm{yr}}.
\]
Fixing the strict priority $\pi$, the current problem is $\sigma_t=(I_t,q_t^{\mathrm{occ}},q_t^{\mathrm{ann}},R_t)$. The new admissions $L\subseteq I_t$ are feasible precisely if
\[
|L|\le q_t^{\mathrm{occ}}\quad\mbox{and}\quad
|L\setminus R_t|\le q_t^{\mathrm{ann}}.
\]

Let $v$ be the single category of new admissions. A priority choice rule subject to these constraints is
\[
C^v_{\mathrm{med}}(\sigma_t)
=\TopSet^{\pi}_{q_t^{\mathrm{occ}}}
\!\left(R_t\medcup\TopSet^{\pi}_{q_t^{\mathrm{ann}}}(I_t^{\mathrm{wait}}\setminus R_t)\right).
\]
The rule considers returning applicants together with the highest-priority new applicants allowed by the annual limit, then selects by priority up to current capacity. Returning does not move an applicant up in $\pi$; it only avoids adding to the annual count.

\subsection{Vaccination with Completion Obligations}\label{app:completiondetails}
Section~\ref{subsec:publicstate} distinguishes honoring an earlier allocation from providing additional resources to complete it. On February 19, 2021, Delaware's weekly update asked providers to complete the courses they had started and noted that supply limitations might restrict new first-dose appointments.\footnote{Delaware Division of Public Health, \href{https://news.delaware.gov/2021/02/19/weekly-covid-19-update-feb-19-2021-new-positive-cases-current-hospitalizations-continue-to-trend-downward/}{\emph{Weekly COVID-19 Update}}, February~19, 2021, ``DPH Update on Second Doses.''} Suppose there is one vaccine product and one completion date, with no intervening deadlines, expiration, spoilage, storage losses, or binding administration constraints at either date. A new course uses a dose today and another one at the completion date. If there are six doses on hand plus two more assured, then the institution can start four new courses (if no one is owed a second dose) or two (if four people are). The earlier commitments reduce the number of new courses from four to two, even though the number of doses available today is unchanged.

Suppose date 1 is today and date 2 is the completion date. Let $q^{\mathrm{now}}$ be the number of usable doses in stock at date 1, $q^{\mathrm{assured}}$ the number of distinct additional doses assured by date 2, and $q^{\mathrm{due}}$ the number of earlier recipients who are due completion by date 2.

Assume $q^{\mathrm{due}}\le q^{\mathrm{now}}+q^{\mathrm{assured}}$, so that the doses on hand and those assured for later delivery suffice to complete the courses already started. Each new course requires two additional doses, and one of them must be on hand today for the first injection. 
So we can start at most $q^{\mathrm{now}}$ new courses today. We also have to set aside $q^{\mathrm{due}}$ doses for the earlier recipients. That leaves $q^{\mathrm{now}}+q^{\mathrm{assured}}-q^{\mathrm{due}}$ doses, and each new course takes two of them. 
Therefore, the \emph{completion-preserving account} has capacity
\[
q^v=q^*(\Lambda)=\min\left\{q^{\mathrm{now}},
\left\lfloor\frac{q^{\mathrm{now}}+q^{\mathrm{assured}}-q^{\mathrm{due}}}{2}\right\rfloor\right\}.
\]
The priority choice rule selects the top-priority waiting applicants up to this capacity. The earlier recipients show up in the problem only through $q^{\mathrm{due}}$; they are not applicants.

\subsection{Post-Based and Vacancy-Based Recruitment}\label{app:postdetails}
A post-based account is one where reservation is applied to the set of posts constituting a cadre, not to the vacancies arising over time. In \emph{R.\ K.\ Sabharwal}, the Supreme Court affirmed that reservation percentages apply to cadre posts and that candidates from reserved groups who have been appointed on merit do not use the reserved quota when filling open posts.\footnote{Supreme Court of India, \href{https://indiankanoon.org/doc/1871744/}{\emph{R.\ K.\ Sabharwal v.\ State of Punjab}}, February~10, 1995, (1995) 2 SCC~745, paragraphs~5--8.} Let post categories be fixed, in which case a post holder who moves out creates a vacancy in the category of her post. Under a vacancy-based sequence, the vacancy created by the next post holder to leave instead takes the next category in the prescribed series.

For example, assume five filled posts. Four of these are open and one is reserved. If the holder of an open post leaves, under a post-based account this leaves an open vacancy. But if the next opening is assigned to the reserved category, as in a vacancy-based series, the result is two reserved appointments out of the five filled posts. This comparison does not depend on the social group of the individual who leaves.

Fixing the category of a post does not necessarily bind the post-holder to it. If posts are interchangeable, an incumbent can move to another eligible post without altering the categorization of any post. The ledger then records which holder is in which post in order to release the right post on her future departure.

\subsection{Earlier Open Appointments and Promotion Eligibility}\label{app:promotionproof}
An employee appointed through open competition may later seek a promotion reserved for her group. In \emph{Jugraj Singh v.\ State of Punjab}, the state refused to consider a Backward Class employee for a reserved promotion because he had initially been appointed in the general category. The Punjab and Haryana High Court ruled against the state. It decided that his earlier appointment in the open category did not bar him from being considered for promotion against the reserved quota.\footnote{Punjab and Haryana High Court, \href{https://www.legitquest.com/case/jugraj-singh-v-state-of-punjab/7A597C}{\emph{Jugraj Singh v.\ State of Punjab}}, 2023, paragraphs~3--8.}

In our framework, the account determines whether the employee is reported as eligible for reserved promotion. If it excludes her because of her earlier open appointment, the O\&A choice rule can promote another member of the same group with lower priority while satisfying non-wastefulness, no justified envy, and maximal reserve protection at the reported inputs.
Taking eligibility as given, these three axioms do not rule out the state's treatment of earlier open appointments.
We are allocating positions for promotion. 
The employee's earlier position is not one of them. It matters only as part of her record.

The no-penalty requirement below takes two records that are the same except for the category of the employee's earlier appointment, and compares how they are treated. The allocation axioms apply to each problem the account generates, one at a time. The no-penalty axiom is different. It also restricts how the account recognizes eligibility.

Let $\mathcal V=\{o\}\cup\mathcal R$, with nonoverlapping reserved eligibility. Fix a finite cohort $J$, and let $J^c=J\cap\mathcal I^c$ for each $c\in\mathcal R$. For each $i\in J^c$, let $\gamma_i\in\{o,c\}$ be her earlier appointment category and let $\eta_i$ denote her other qualification information. The promotion ledger $\Lambda$ consists of these records, the applicant set $I\subseteq J$, capacities $q=(q^v)_{v\in\mathcal V}$, and priority $\pi$. We assume that all applicants meet the promotion requirements other than category eligibility. Let $J^{c,\mathrm{qual}}\subseteq J^c$ be the set of employees whose records $\eta_i$ qualify them for category-$c$ promotion, regardless of $\gamma_i$. No promotion position is subject to an inherited right.

An account rule $A$ reports $\mathcal E^c(\Lambda)\subseteq J^{c,\mathrm{qual}}$ for each $c\in\mathcal R$ and $\mathcal E^o=J$, leaving the remaining inputs unchanged. We impose the following restrictions on $A$:
\begin{enumerate}
\item \emph{Locality:} For each $c\in\mathcal R$ and $i\in J^c$, membership in $\mathcal E^c(\Lambda)$ depends only on $(\gamma_i,\eta_i)$.
\item \emph{Calibration:} If $i\in J^{c,\mathrm{qual}}$ and $\gamma_i=c$, then $i\in\mathcal E^c(\Lambda)$.
\end{enumerate}
These restrictions allow a qualified employee with $\gamma_i=o$ to be excluded from $\mathcal E^c(\Lambda)$. We allow all finite cohorts and employee records, all applicant sets meeting the other promotion requirements, all nonnegative capacities, and all strict priorities.

The \emph{promotion-eligibility-preserving account} recognizes $\mathcal E^c(\Lambda)=J^{c,\mathrm{qual}}$ for every $c\in\mathcal R$, under either earlier appointment category. Maximal reserve protection is defined in Section~\ref{para:allocationmodel}. We formalize the no-penalty requirement as follows.

\begin{definition}\label{def:promotionnopenalty}
An account rule--choice rule pair $(A,\mathrm{C})$ satisfies \emph{no penalty from an earlier open appointment} if, for each ledger, category $c\in\mathcal R$, and qualified applicant $i\in I\cap J^{c,\mathrm{qual}}$, the applicant selected when her earlier appointment is recorded in a reserved category remains selected when only that record is changed to the open category. Let $(\Lambda_{-\gamma_i};v)$ be the ledger with her earlier category replaced by $v$. The requirement is
\[
i\in\widehat{\mathrm{C}}\!\left(A(\Lambda_{-\gamma_i};c)\right)
\quad\Longrightarrow\quad
i\in\widehat{\mathrm{C}}\!\left(A(\Lambda_{-\gamma_i};o)\right).
\]
\end{definition}
In words, a promotion that goes through when her earlier appointment is recorded in a reserved category must also go through when it is recorded in the open category.

\begin{proposition}\label{prop:promotionjoint}
An account rule satisfying the preceding restrictions and a choice rule jointly satisfy non-wastefulness, no justified envy, maximal reserve protection, and no penalty from an earlier open appointment on the promotion domain if and only if the account treats each qualified employee as eligible for her group's reserved category, whether her earlier appointment was in the open category or in that reserved category, and the choice rule agrees with the O\&A choice rule on each problem the account generates.
\end{proposition}

We use the following property of ordinary choice in the proof.

We say a feasible outcome is \emph{admissible} if it satisfies non-wastefulness, no justified envy, and maximal reserve protection as defined in Section~\ref{para:allocationmodel}.

In the following lemma, $I^c$ is the eligible set reported in the current problem; for a promotion ledger, $I^c=I\cap\mathcal E^c(\Lambda)$.
\begin{lemma}\label{lem:ordinaryopen}
Every admissible outcome assigns $O=\TopSet_{q^o}(I)$ to open positions. For each reserved category $c$, let $P_c=\TopSet_{q^c}(I^c\setminus O)$. Every member of $P_c$ is selected, and at most $|P_c|$ individuals are assigned to $c$. If $I^c$ is disjoint from the other reserved eligible sets, exactly $P_c$ is assigned to $c$.
\end{lemma}
\begin{proof}
If an open position is vacant, maximal reserve protection rules out assigning reserved positions, and non-wastefulness requires every applicant to be assigned an open position. Otherwise, if a higher-priority $i$ is not assigned an open position while a lower-priority $j$ is, either $i$ is unselected, violating no justified envy, or holds a reserved position, violating maximal reserve protection. Thus, precisely the individuals in $O$ receive open positions.

Category-$c$ recipients belong to $I^c\setminus O$, so their number is at most $\min\{q^c,|I^c\setminus O|\}=|P_c|$. If some $i\in P_c$ were unselected, non-wastefulness would require $c$ to be full. Some category-$c$ recipient would then have lower priority than $i$, contradicting no justified envy. Hence, every member of $P_c$ is selected. If $I^c$ is disjoint from the other reserved eligible sets, these individuals can only be assigned to $c$, giving equality.
\end{proof}

\begin{proof}[\bfseries Proof of Proposition~\ref{prop:promotionjoint}]
Suppose the account rule and the choice rule jointly satisfy the four axioms. Fix a qualified employee $i$ and her reserved category $c$. By the account restrictions, the account treats her as eligible for category-$c$ promotion if her earlier appointment is in $c$. Toward a contradiction, suppose the account treats her as ineligible for category-$c$ promotion when that appointment is open. Consider a ledger with $I=\{i\}$, one category-$c$ promotion position, and no other positions, holding her qualification information and earlier appointment category fixed. By locality, her recognized eligibility cannot change when we change the other ledger entries. Thus, she is still ineligible for category $c$, and feasibility implies that she cannot be selected. Changing only her earlier appointment category to $c$ makes her eligible and, by non-wastefulness, selected. This contradicts no penalty from an earlier open appointment. Thus, the account treats every qualified employee as eligible for her group's reserved category under either earlier appointment category. Since the account cannot treat an unqualified employee as eligible, it preserves reserve eligibility.

At each generated problem, the three allocation axioms imply admissibility. Since reserved eligibility is nonoverlapping, Lemma~\ref{lem:ordinaryopen} identifies every category's recipients with those of O\&A.

Conversely, suppose the account treats every qualified employee as eligible for her group's reserved category under either earlier appointment category, and the choice rule agrees with O\&A at each generated problem. Suppose we change only an employee's earlier appointment category. By assumption, her eligibility remains unchanged. Every other employee has the same qualification record and earlier appointment category, so locality leaves her eligibility unchanged as well. Since all other inputs also remain unchanged, the account generates the same problem and the choice rule gives the same outcome. Thus, no penalty from an earlier open appointment holds. The other three axioms hold because the choice rule agrees with O\&A at each generated problem.
\end{proof}

The following examples explain why each restriction is needed.

Without calibration, an account that recognizes no one for category-$c$ promotion treats earlier open and reserved appointments alike. The allocation requirements use the eligibility reported by the account. Thus, O\&A satisfies these requirements even when no one is recognized for category-$c$ promotion. Changing an employee's earlier appointment category does not change the eligibility report, so O\&A selects the same outcome. Hence, no penalty also holds, although the account denies the protection entirely. Without the no-penalty requirement, an account can instead recognize only qualified earlier reserved appointees. This account satisfies locality and calibration, and O\&A satisfies the allocation requirements at its reported inputs, but a qualified earlier open appointee can lose her reserved eligibility.

Without locality, first compute O\&A under full qualification recognition. Keep every qualified earlier reserved appointee and every recipient of a reserved position recognized; remove recognition only from qualified earlier open appointees who receive no reserved position. Recall that the open recipients of O\&A are determined by the population, priority, and open capacity. Since these inputs do not change when reported reserved eligibility is reduced, the same individuals are selected for open positions. Next, consider a reserved category. All its recipients under full recognition remain eligible, while every individual whose eligibility is removed was not selected for this category. Hence, O\&A also selects the same recipients in each reserved category when it is reapplied to the reduced eligibility sets.

Therefore, for every ledger, the constructed account gives the same outcome as full recognition. Under full recognition, earlier appointment categories have no effect on eligibility. Thus, changing an employee's earlier appointment category does not change the outcome under the constructed account either. Also, every qualified earlier reserved appointee remains recognized, so calibration is satisfied. However, a qualified earlier open appointee can be excluded: her recognition depends on whether O\&A under full recognition assigns her a reserved position. This assignment can change with the competing applicants, capacities, and priority. Hence, the account does not satisfy locality.

\section{Retention, Continuation, and Other Reserve Policies}\label{app:retentionextensions}
These results extend the continuation analysis of Section~\ref{sec:rules} to arrivals and withdrawals, and establish the properties used for other category orders and coordinated choice in Section~\ref{sec:multiinstitution}.

\subsection{Arrivals and Inherited Appointments}\label{app:arrivalsretention}

We allow arrivals. Earlier recipients retain their appointments while their category assignments can change. Fix a finite potential set of individuals $\mathcal I$, eligibility, a strict priority on $\mathcal I$, and any category precedence. Let $I_0\subseteq\cdots\subseteq I_T\subseteq\mathcal I$ be the populations at successive dates, with priority and eligibility restricted from this potential set. Once an individual arrives, she remains in the population. Capacities satisfy $q_{t+1}\ge q_t$ componentwise. Given a retention-feasible initial inherited profile $\mathrm H_0$ on $I_0$, let $\sigma_t=(I_t,q_t,\mathrm H_0)$, suppressing the fixed inputs.

Fix an RRB choice rule $\CC\in\{\CC_{\parallel},\CC_{\triangleright}\}$, and define
\[
\mathrm G_0=\CC(\sigma_0),\qquad
\mathrm G_t=\CC\!\left(\sigma_t[\widehat{\mathrm G}_{t-1}]\right),
\quad t=1,\ldots,T.
\]
As in Section~\ref{sec:rules}, the bracket replaces the incumbent set and does not lock its category assignments. The corresponding ordinary benchmark at date $t$ is $\CC(\sigma_t[\emptyset])$. For joint RRB, full outcomes use the same fixed attribution convention.

\begin{corollary}\label{cor:arrivalsretention}
The above procedure is well defined and retains every earlier recipient. At each date it satisfies the corresponding choice axioms in Theorem~\ref{thm:flexcharacterization} or~\ref{thm:precedencecharacterization}. For $t\ge1$,
\[
\begin{aligned}
\widehat{\mathrm G}_t\setminus\widehat\CC(\sigma_t[\emptyset])
 &=\widehat{\mathrm G}_{t-1}\setminus\widehat\CC(\sigma_t[\emptyset]),\\
|\widehat\CC(\sigma_t[\emptyset])\setminus\widehat{\mathrm G}_t|
 &\le |\widehat{\mathrm G}_{t-1}\setminus\widehat\CC(\sigma_t[\emptyset])|.
\end{aligned}
\]
If $I_t=I_{t'}$ for every $t\ge t'$, then
\[
\mathrm G_t=\CC\!\left(\sigma_t[\widehat{\mathrm G}_{t'}]\right)
\qquad\text{for every }t\ge t'.
\]
On this interval, $\widehat{\mathrm G}_{t'}\setminus\widehat\CC(\sigma_t[\emptyset])$ is weakly decreasing in $t$.
\end{corollary}
\begin{proof}
The initial problem is retention-feasible, so $\mathrm G_0$ is well defined. Suppose next that $\mathrm G_{t-1}$ has been defined. Each individual selected at date $t-1$ remains in the population at date $t$ and has the same eligibility. Moreover, the positions used by these individuals remain available, since capacities do not decrease. Therefore, we can retain them using their assignments in $\mathrm G_{t-1}$. Hence, $\sigma_t[\widehat{\mathrm G}_{t-1}]$ is retention-feasible, and the corresponding characterization theorem implies that the RRB outcome retains these individuals and satisfies the stated axioms. Starting with date zero and repeating this argument proves the first part of the corollary. For the two displayed comparisons, apply Proposition~\ref{prop:legacybound} to the date-$t$ problem with $\widehat{\mathrm G}_{t-1}$ as the incumbent set. The RRB outcome of this problem is $\mathrm G_t$, and its ordinary benchmark uses the same population and capacities.

Next, suppose there are no arrivals after date $t'$. Since $\mathrm G_{t'}$ includes the incumbents at that date, Lemma~\ref{lem:selectedretention} implies $\CC(\sigma_{t'}[\widehat{\mathrm G}_{t'}])=\mathrm G_{t'}$. Thus, we can take $\widehat{\mathrm G}_{t'}$ as the original incumbent set and consider the subsequent increases in capacities as releases. The population, eligibility, priority, and precedence do not change after date $t'$. By Theorem~\ref{thm:flexrelease}, cumulative implementation at these releases gives $\CC(\sigma_t[\widehat{\mathrm G}_{t'}])$ for every $t\ge t'$. This implementation follows the same recursion as $\mathrm G_t$ and starts from the same outcome $\mathrm G_{t'}$, so the two outcomes coincide at every such date. By part~(i) of the theorem, applied with no incumbents, the ordinary benchmark's recipient set is weakly increasing as capacities increase. Since $\widehat{\mathrm G}_{t'}$ is fixed, the set of its members not selected by the ordinary benchmark is weakly decreasing.
\end{proof}

If a higher-priority individual arrives, ordinary choice may select her in place of an earlier recipient. Thus, while arrivals continue, the set of earlier recipients not selected by ordinary choice need not be weakly decreasing.

Under the condition in the next lemma, ordinary choice itself retains earlier recipients after arrivals.

\begin{lemma}\label{lem:lowerpriorityarrivals}
Let $\sigma$ and $\sigma'$ have populations $I$ and $I'$, respectively, such that $I\subseteq I'$, and capacity vectors $q$ and $q'$, respectively, such that $q'\ge q$, with eligibility, priority, and any category precedence unchanged. If every individual in $I'\setminus I$ has lower priority than every recipient in $\widehat\CC(\sigma[\emptyset])$, then
\[
\widehat\CC(\sigma[\emptyset])\subseteq\widehat\CC(\sigma'[\emptyset]).
\]
\end{lemma}

\begin{proof}
Let $L=\widehat\CC(\sigma[\emptyset])$ and $N=I'\setminus I$. Let $\bar\sigma$ be the problem with population $I'$ and capacities $q$, with the same fixed inputs. We first compare $\CC(\sigma[\emptyset])$ and $\CC(\bar\sigma[\emptyset])$.

If $|I|<q^o$, then $L=I$. Every member of $I$ has higher priority than every member of $N$, so all of $I$ still receives open positions. Suppose instead that $|I|\ge q^o$, and let $O=\TopSet_{q^o}(I)$. Since $O\subseteq L$, the priority assumption gives
\[
\TopSet_{q^o}(I')=\TopSet_{q^o}(I)=O.
\]

For joint RRB, take $i\in L\setminus O$, and let $P_i=\{j\in I\setminus O:j\mathrel{\pi}i\}$. Since $i$ is selected, Lemma~\ref{lem:greedyprefix} implies
\[
\rho_q(P_i\cup\{i\})=\rho_q(P_i)+1.
\]
Every member of $N$ has lower priority than $i$, so
\[
P_i=\{j\in I'\setminus O:j\mathrel{\pi}i\}.
\]
Eligibility and capacities are unchanged, so the ranks of $P_i$ and $P_i\cup\{i\}$ are the same in both problems. By the same lemma, $i$ is selected again. This proves retention of every member of $L\setminus O$.

For precedence RRB, let $R_c$ and $R'_c$ be the category-$c$ recipient sets in the two problems. We show by induction on precedence that $R'_c\cap I=R_c$. Suppose this equality holds for all categories preceding $c$. The old applicants still available at $c$ are then exactly the same in both problems. If $|R_c|=q^c$, every member of $R_c$ precedes every member of $N$, so $R'_c=R_c$. If $|R_c|<q^c$, then $R_c$ contains every remaining eligible old applicant. Each of them precedes every member of $N$, so $R'_c$ contains $R_c$ and can add only members of $N$. This proves the induction, starting from the common open selection $O$.

For either rule, we have established the first inclusion in
\[
L\subseteq\widehat\CC(\bar\sigma[\emptyset])
\subseteq\widehat\CC(\sigma'[\emptyset]).
\]
The second follows from Theorem~\ref{thm:flexrelease}(i), since the population is $I'$ in both problems and $q'\ge q$.
\end{proof}

Consequently, suppose $\mathrm G_0=\CC(\sigma_0[\emptyset])$ and every individual arriving at date $t$ has lower priority than every recipient in $\widehat{\mathrm G}_{t-1}$. The lemma and benchmark recovery in Relation~\eqref{eq:benchmarkrecovery}, applied successively, imply $\mathrm G_t=\CC(\sigma_t[\emptyset])$ at every date. An arriving individual may receive a position that was previously vacant. Hence, the recipient set may increase.

Fix an arrival sequence $(I_t)_{t=0}^T$. We compare two release schedules at the same date, keeping this entire arrival sequence unchanged. At each date, a procedure can use only the observations made by that date. It assigns positions only to individuals who have arrived and cannot exceed the cumulative capacity of any category. Resource-release invariance in this setting requires the two schedules to give the same outcome whenever their cumulative category capacities at the date of comparison are the same.

\begin{proposition}\label{prop:arrivalsimpossibility}
On the domain of all finite arrival and nonnegative capacity-release histories, no such procedure satisfies recipient retention, non-wastefulness at every decision, ordinary no justified envy whenever there are no earlier recipients, and resource-release invariance as defined above. The conflict already arises with one open position.
\end{proposition}
\begin{proof}
At date 1 only $i$ is available; at date 2 a higher-priority individual $j$ arrives. Start with no positions or recipients and release one open position at either date. A date-1 release requires selecting $i$ by non-wastefulness and retaining her at date 2. With a date-2 release, there are no earlier recipients, so non-wastefulness and no justified envy require selecting $j$. Equal arrival sequences and final capacities thus give different recipient sets, contradicting resource-release invariance.
\end{proof}

Once arrivals cease, Corollary~\ref{cor:arrivalsretention} compares future releases from a common population and inherited recipient set.

\subsection{Continuation after Withdrawals}\label{app:rrbwithdrawals}
Fix an RRB rule $\CC\in\{\CC_{\parallel},\CC_{\triangleright}\}$, eligibility, priority, and any category precedence. Individuals may leave permanently; a departing recipient's position remains available. The rights of removed individuals cease. In the following lemma, the inherited profile in the smaller problem is restricted to surviving incumbents; as elsewhere, RRB depends on that profile only through its set of protected individuals.

\begin{lemma}\label{lem:rrbwithdrawals}
Let $\sigma=(I,q,\mathrm H)$ be retention-feasible, let $L=\widehat\CC(\sigma)$, and let $D\subseteq I$. Form $\sigma'=(I\setminus D,q,\mathrm H')$ by restricting eligibility and priority to $I\setminus D$ and setting $\mathrm H'^v=\mathrm H^v\setminus D$ for every category. Then $\sigma'$ is retention-feasible and
\[
L\setminus D\subseteq\widehat\CC(\sigma'),\qquad
\CC\bigl(\sigma'[L\setminus D]\bigr)=\CC(\sigma').
\]
For joint RRB, full-outcome equality uses the same attribution convention.
\end{lemma}

\begin{proof}
Fix the ranking for $\CC$ in Lemma~\ref{lem:fixedrepresentation} on population $I$, including its fixed order of individual--position pairs. Consider only graph matchings that do not assign any individual in $D$. When we compare two such graph matchings, every individual in $D$ is absent from both recipient sets at each comparison. Also, neither graph matching contains a pair involving an individual in $D$. Thus, these individuals and their pairs do not affect the ranking. The ranking restricted to these graph matchings has the form prescribed by the lemma for the remaining population, with priority restricted to that population and the same category precedence.

Let $\beta$ be the highest-ranked feasible graph matching before removal, retaining $K=\widehat{\mathrm H}$, and let $K'=K\setminus D$. Removing from $\beta$ the pairs involving members of $D$ leaves a feasible graph matching that assigns every member of $K'$. Thus, the smaller problem is retention-feasible. Let $\beta'$ be its highest-ranked feasible graph matching retaining $K'$.

Suppose, toward a contradiction, that an individual $i\in L\setminus D$ is unassigned in $\beta'$. Since $\beta'$ assigns every member of $K'$, we have $i\notin K'$. Moreover, $i$ survives the removal, so $i\notin K$ either. Set aside the pairs common to $\beta$ and $\beta'$, and form the disjoint alternating chains and cycles described in the proof of Lemma~\ref{lem:reserveassignment}. The individual $i$ has a pair from $\beta$ and none from $\beta'$, so she is an end node of a chain.

This chain contains no member of $D$. Indeed, a removed individual has no pair from $\beta'$ and thus, if she belongs to the chain, must be another individual end node with a pair from $\beta$. But a chain with two individual end nodes has an even number of alternating edges. Starting at $i$ with a pair from $\beta$, it must end at its other individual end node with a pair from $\beta'$. Consequently, that end node, if present, survives removal, is assigned only by $\beta'$, and lies outside $K$ because $\beta$ assigns every member of $K$.

Switch this chain in $\beta$ to its assignments from $\beta'$. Every internal individual remains assigned, and the only individual who becomes unassigned is $i\notin K$. If the other end node is an individual, she becomes assigned; if it is a position, there is no other individual end node. Thus, the switch retains $K$ and is feasible in the original problem.

Conversely, switch the chain in $\beta'$ to its assignments from $\beta$. This assigns a position to $i$ and keeps all internal individuals assigned. If the other end node is an individual, she is the only person who becomes unassigned, and she lies outside $K$, hence outside $K'$. If the other end node is a position, no individual becomes unassigned. The chain contains no removed individual, all positions remain available, and eligibility is unchanged. Hence, this reverse switch is feasible in the smaller problem and retains $K'$.

As established in the proof of Lemma~\ref{lem:reserveassignment}, the ranking prefers the same assignment subgraph on this chain in both comparisons. At each recipient-set comparison, assignments outside the chain occur on both sides and do not affect the highest-priority individual whose membership differs. If all these comparisons tie, the first differing pair in the fixed pair order also lies on the chain. Since both switches are feasible and run in opposite directions, the representative graph matching using the less-preferred assignment subgraph could be improved. This contradicts its highest rank. Therefore, $L\setminus D\subseteq\widehat\CC(\sigma')$.

Finally, $K'\subseteq L\setminus D$ because every original incumbent belonged to $L$. Lemma~\ref{lem:selectedretention}, applied to the smaller problem, now shows that protecting $L\setminus D$ leaves its outcome unchanged. For joint RRB, the comparison is between two problems on the same smaller population with the same capacities and selected open and reserved recipient sets. The fixed attribution convention does not depend on which of these recipients are protected, so it gives the same full category outcome.
\end{proof}

Successive applications of the lemma allow withdrawals to be interspersed with capacity additions under Theorem~\ref{thm:flexrelease}. With eligibility, priority, and any precedence fixed on the remaining population, protecting surviving recipients gives the same outcome as protecting surviving original incumbents. This does not imply invariance to arrivals or priority changes.

\subsection{Retention under Other Category Precedences}\label{app:otherprecedences}

Fix a finite population, eligibility sets, capacities, and a common strict priority as in Section~\ref{sec:rules}. Let $\triangleright^\ast$ be any strict order on $\mathcal V=\{o\}\cup\mathcal R$, allowing the open category $o$ to occupy any position. Let $\CC_{\triangleright^\ast}$ denote the retention-adjusted sequential choice rule. Its input $\sigma$ includes a retention-feasible incumbent set $\widehat{\mathrm H}$. As before, $\sigma[K]$ replaces this set by $K$ without changing the other inputs.

Process categories in the order $\triangleright^\ast$. At category $v$, let $K$ be the incumbents not yet assigned and let $\bar q$ be the capacities that would remain after assigning one more individual to $v$. Earlier categories have zero remaining capacity; $\bar q$ includes the rest of $v$ and all later categories. Select the highest-priority unassigned individual $i$ eligible for $v$ for whom
\begin{equation}\label{eq:generalcompletion}
|K\setminus\{i\}|-\rho_{\bar q}(K\setminus\{i\})\leq\bar q^o.
\end{equation}
Assign her to $v$ and repeat until $v$ is full or no remaining individual eligible for $v$ satisfies Inequality~\eqref{eq:generalcompletion}. In the latter case, leave its remaining capacity unfilled and proceed to the next category. Here $\rho_{\bar q}$ still counts matchings into \emph{reserved} positions only. Lemma~\ref{lem:retentionfeasibility} therefore says exactly that the remaining incumbents can be assigned to the remaining reserved and open positions. When the open category comes first, Inequality~\eqref{eq:generalcompletion} reduces to Inequality~\eqref{eq:opentest} at the open step. At each subsequent step, $\bar q^o=0$, giving Equality~\eqref{eq:sequentialtest}. Thus, when $\triangleright^\ast$ places the open category first, $\CC_{\triangleright^\ast}$ is the precedence RRB rule for that order.

For sets $A\ne B$, write $A\succ_\pi B$ when the highest-priority member of their symmetric difference belongs to $A$. Compare category-assignment profiles lexicographically by this set order, considering categories in the order $\triangleright^\ast$. We call this the \emph{sequential ranking}. It compares the selections in the first category where two profiles differ.

\begin{proposition}\label{prop:sequentialretention}
At every retention-feasible problem, $\CC_{\triangleright^\ast}$ retains all incumbents and is non-wasteful. It is the highest-ranked feasible retaining outcome under the sequential ranking. Moreover,
\begin{enumerate}[label=(\roman*)]
\item it reproduces ordinary sequential choice with the same category precedence and no protected incumbents if and only if that choice selects every incumbent;
\item if $\widehat{\mathrm H}\subseteq K\subseteq\widehat{\CC}_{\triangleright^\ast}(\sigma)$, then
\[
\CC_{\triangleright^\ast}(\sigma[K])=\CC_{\triangleright^\ast}(\sigma).
\]
\end{enumerate}
\end{proposition}

\begin{proof}
We first show that the procedure assigns every incumbent. Initially, retention feasibility supplies an assignment of all incumbents. Whenever the procedure assigns an individual to a category $v$, she satisfies Inequality~\eqref{eq:generalcompletion}, i.e., the incumbents who are still unassigned can all be assigned to the positions that remain in $v$ and in the later categories. Next, suppose that the procedure leaves category $v$ with some capacity unfilled, and consider such an assignment of the unassigned incumbents. If this assignment placed an incumbent $k$ in category $v$, assigning $k$ to $v$ would leave the other incumbents in their positions under that assignment. The procedure could therefore make another assignment in $v$, contrary to its stopping there with a vacancy. Thus, the unassigned incumbents can all be assigned to positions in the categories after $v$. After the last category is processed, there is no later category, and hence, no incumbent is unassigned.

Let $\mathrm G$ be the outcome of the procedure, which is feasible and retains every incumbent, as we showed. Suppose, toward a contradiction, that a feasible outcome $\mathrm F$ that retains every incumbent is ranked above $\mathrm G$. Let $v$ be their first differing category, and let $i$ be the highest-priority individual in $\mathrm F^v\setminus\mathrm G^v$ or $\mathrm G^v\setminus\mathrm F^v$. Since $\mathrm F$ ranks above $\mathrm G$, we have $i\in\mathrm F^v\setminus\mathrm G^v$. Then every member of $\mathrm G^v$ who precedes $i$ in $\pi$ belongs to $\mathrm F^v$ as well, by the choice of $i$. Moreover, $\mathrm F$ and $\mathrm G$ have the same recipients in each category before $v$, so $i$ is unassigned when the procedure reaches $v$. Consider the first time in category $v$ that the procedure selects an individual with lower priority than $i$ or leaves the category. Every individual selected for $v$ before this time precedes $i$ in $\pi$ and belongs to $\mathrm F^v$. Since $\mathrm F^v$ satisfies the capacity constraint for $v$ and also contains $i$, the category is not filled yet. Moreover, $\mathrm F$ assigns $i$ to $v$ and retains every incumbent while preserving the assignments already made. Thus, we showed that $i$ satisfies Inequality~\eqref{eq:generalcompletion} at this time, so the procedure would select her or a higher-priority individual, leading to a contradiction. As two different category-assignment profiles are never tied under the sequential ranking, $\mathrm G$ is the unique highest-ranked feasible outcome that retains every incumbent.

For non-wastefulness, suppose that an individual $j$ is unselected under $\mathrm G$ while a position is vacant in a category $v$ for which she is eligible. If we add $j$ to $\mathrm G^v$, then we obtain a feasible outcome that retains every incumbent and is ranked above $\mathrm G$, since $\mathrm G^v\cup\{j\}\succ_\pi\mathrm G^v$. This is a contradiction. For part~(i), when there are no incumbents, Inequality~\eqref{eq:generalcompletion} holds for every unassigned individual eligible for the current category. The rule therefore coincides with ordinary sequential choice, whose outcome is the highest-ranked feasible outcome by the argument above. If this ordinary outcome selects every member of $\widehat{\mathrm H}$, then it is also the highest-ranked one among the feasible outcomes that retain every incumbent, so it is equal to $\mathrm G$. Conversely, if $\mathrm G$ is equal to the ordinary outcome, then the latter selects every incumbent, as $\mathrm G$ does.

For part~(ii), let $\widehat{\mathrm H}\subseteq K\subseteq\widehat{\mathrm G}$. Every feasible outcome that retains $K$ also retains $\widehat{\mathrm H}$, and $\mathrm G$ itself retains $K$. Since no feasible outcome that retains $\widehat{\mathrm H}$ is ranked above $\mathrm G$, none of those that retain $K$ can be ranked above it, either. Thus, $\CC_{\triangleright^\ast}(\sigma[K])=\mathrm G$, i.e., the two full outcomes are the same, including their category assignments.
\end{proof}

\paragraph{Choice and Deferred Acceptance.}

Fix a profile of protected sets $H=(H_s)_{s\in\mathcal S}$ and an institution $s$ for which $H_s$ is retention-feasible. Hold its inputs fixed and suppose it uses the rule $\CC_s=\CC_{\triangleright^\ast}$. As in Relation~\eqref{eq:rrbmenuchoice}, for each $J\subseteq I$, write
\[
\widehat{\CC}_s^H(J)=\widehat{\CC}_s\bigl(\sigma_s[J,H_s\cap J]\bigr)
\]
for the set of recipients selected by $s$ from $J$. Restricting a feasible assignment of $H_s$ to $H_s\cap J$ shows that this choice is defined for every $J$.

\begin{lemma}\label{lem:sequentialmenus}
The aggregate choice rule $\widehat{\CC}_s^H$ satisfies substitutes, irrelevance of rejected individuals, the law of aggregate demand, and path independence.
\end{lemma}

\begin{proof}
Label the positions. Rank feasible graph matchings first by the sequential ranking of their category recipient sets and then by a fixed lexicographic order on individual--position pairs, as in Lemma~\ref{lem:fixedrepresentation}. By Proposition~\ref{prop:sequentialretention}, the unique highest-ranked feasible graph matching that retains $H_s\cap J$ represents the rule's outcome on $J$, i.e., it has the same recipient set in every category. We use this same ranking for every set of individuals. When we consider a smaller set of individuals, some pairs become unavailable, but the ranking between two graph matchings that are still feasible does not change.

For two graph matchings, remove common pairs and decompose the remaining edges into alternating chains and cycles, as in the proof of Lemma~\ref{lem:rrbmenus}. Switching one component replaces its edges from one graph matching by its edges from the other. Observe that the sequential ranking prefers the same assignment on this component in both comparisons: The assignments outside the component do not change within each comparison, so the first differing category and the highest-priority differing individual in it are the same in both comparisons. If the recipient sets are the same in every category, then the same first differing pair in the fixed order of pairs determines both comparisons. Consequently, if both switches are feasible and retain the protected individuals in their respective problems, one of the representative graph matchings can be improved, a contradiction. This reasoning uses the order of category comparisons, but does not require open to be first.

Let $\beta$ and $\beta'$ represent choice on $J$ and $J\cup\{i\}$, respectively, where $i\notin J$. If $\beta'$ leaves $i$ unassigned, then $i$ is not protected. Both problems have the same protected set, and each representative is feasible in the other problem. Their highest rank gives $\beta=\beta'$. Otherwise, $\beta'$ assigns $i$. Consider a component that does not contain $i$. Every individual in it belongs to $J$, and every position in it is available in both problems. If a protected individual is in this component, then she is assigned under both graph matchings, so she has one edge from each of them and remains assigned after either switch. Thus, both switches are feasible and retain the protected individuals in their respective problems. By the comparison above, there is no such component.

Individual $i$ has only one edge, which belongs to $\beta'$. Hence, the graph is a single chain with end node $i$. If the other end node is a position, then this position was not used under $\beta$, and $i$ is added without displacing anyone. If the other end node is an individual, then her edge belongs to $\beta$, and she is the only earlier recipient who becomes unassigned. Thus, when we add one individual to the set $J$, no previously rejected individual becomes selected, and the number of selected individuals does not decrease. We obtain substitutes and the law of aggregate demand when we add the individuals one at a time.

For irrelevance of rejected individuals, suppose
$\widehat{\CC}_s^H(J)\subseteq J'\subseteq J$.
Retention implies $H_s\cap J'=H_s\cap J$. Then the representative graph matching on $J$ is feasible on $J'$ as well and retains the same protected set, and every feasible graph matching on $J'$ that retains this set is also feasible on $J$. Hence, it remains the highest-ranked one on $J'$.

Finally, let $X=J\cup J'$ and $X'=\widehat{\CC}_s^H(J)\cup J'$. Substitutes implies that every individual selected from $X$ who belongs to $J$ is selected from $J$. Thus
$\widehat{\CC}_s^H(X)\subseteq X'\subseteq X$.
By irrelevance of rejected individuals, the rule then selects the same recipients from $X$ and from $X'$, i.e., path independence holds.
\end{proof}

Institutions may use these sequential rules, with different precedences, alongside the RRB rules of Section~\ref{sec:rules}. Under the fixed-input and strict-preference assumptions of Section~\ref{subsec:matchingproblem}, Lemmas~\ref{lem:rrbmenus} and~\ref{lem:sequentialmenus} give the properties required for individual-proposing DA. Thus, by the standard DA results that we cite in Section~\ref{subsec:rrbda}, the mechanism is strategy-proof, and its outcome is the individual-optimal stable matching for these institutional choice rules. Moreover, an incumbent who proposes to her inherited institution is never rejected by its choice rule, so she is assigned either to this institution or to one that she prefers. In this way, we extend the construction and its DA properties to an institution with a different policy, and we do not impose on it the over-and-above axioms of Theorem~\ref{thm:rrbdacharacterization}.

\subsection{A Soft-Reserve Extension}\label{app:softsequential}
We now consider a retaining extension to soft reserves, discussed in Section~\ref{app:cpnu}. It uses the construction above with fixed strict category priorities $\pi^v$ and category eligibility that includes any permitted fallback assignments. For a soft reserve, all applicants may be eligible, while its beneficiaries precede other applicants under $\pi^v$; the baseline priority orders individuals within each group. At each category, select the highest-priority unassigned individual under $\pi^v$ whose assignment leaves a matching of every remaining incumbent to an eligible remaining position. This matching may use any remaining positions, including open positions.

For this extension, compare category recipient sets in the prescribed sequence using $\succ_{\pi^v}$ in category $v$. 
With this ranking, the proof of Proposition~\ref{prop:sequentialretention} still applies: each accepted assignment preserves a retaining completion, and any higher-ranked retaining outcome would give a completion for the first person skipped by the procedure. 
The extension is non-wasteful and reproduces the outcome of its ordinary counterpart whenever the counterpart retains all incumbents.
Its outcome also stays the same if some of the recipients it selects are added to the protected set.
Changing the set of incumbents does not alter the category priorities or permitted assignments.

The proof of Lemma~\ref{lem:sequentialmenus} carries over to this extension. Use the fixed order $\pi^v$ to compare recipients in category $v$. The graph now also includes all permitted fallback pairs.
Take an alternating component. 
Look for the first category in which the two outcomes differ on this component, and the highest-priority individual in that category on whom they differ.
The same side of the component is preferred in both comparisons. As in the proof, the two switches preserve the fixed eligibility constraints and the required incumbents.
Substitutes, irrelevance of rejected individuals, the law of aggregate demand, and path independence therefore still hold.

The proof of Proposition~\ref{prop:dabenchmark} gives benchmark recovery for these choice rules too. Suppose that ordinary DA honors all inherited appointments.
Consider a proposal set reached by ordinary DA. None of the incumbents who propose to their inherited institution in that run is rejected there, since any later proposal would be to an institution they prefer less. So at each proposal set reached by ordinary DA, ordinary choice retains all incumbents who have proposed.
The single-institution benchmark property established above makes retaining choice identical on each such set. Induction over proposals gives the same DA outcome. The converse follows because retaining DA honors every inherited appointment. Finally, individual optimality gives the set-inclusion comparison in Corollary~\ref{cor:daharm} among stable matchings for these soft-reserve choice rules: anyone who prefers ordinary DA to her individual-optimal stable assignment also prefers it to her assignment under every other stable matching.

\section{Institutional Details for the Applications}\label{app:additionalaccounts}
We provide institutional details and further analysis for the examples in Section~\ref{sec:rules} and the applications in Sections~\ref{sec:casestudies} and~\ref{sec:multiapplications}.

\subsection{Judicial Corrections: Resources and Category Assignments}\label{app:correctiondetails}

\subsubsection{Resources for the BSNL Correction}\label{app:bsnldetails}

The Indian Supreme Court considered both over-and-above protection and irrevocable appointments in \emph{Bharat Sanchar Nigam Ltd. v. Sandeep Choudhary}. The case concerned appointments to identical technical-assistant positions in Ajmer. There were five open positions and four positions reserved for Other Backward Classes (OBC). The five appointees in the open category were not eligible for the OBC reserved category. After the mark requirements were relaxed,
two of the four appointees in the OBC reserved category had scored high enough to be placed in the open category. However, those appointments had been counted in the reserved category. The court ruled that those individuals should be counted as open-category appointees and the claimant should be appointed in the reserved category. It further ordered that the two previous appointees in the open category who would otherwise have lost their appointments should keep their positions.\footnote{Supreme Court of India, \href{https://indiankanoon.org/doc/41636927/}{\emph{Bharat Sanchar Nigam Ltd. v. Sandeep Choudhary}}, 2022, paragraphs~2 and~9--11. The dispute concerns open and OBC positions; the separately advertised SC and ST positions are outside this comparison. A related direction to reattribute appointments to open while protecting earlier open appointees appears in the Lahore High Court's \href{https://pakistancaselaw.com/case/153852/muhammad-iqbal-shahid-vs-province-of-punjab-through-incharge-2015-plc-c-s-1267}{\emph{Muhammad Iqbal Shahid v. Province of Punjab}}, 2015, paragraphs~3 and~5--7. That judgment mentions vacancies without specifying the capacity available for its remedy.} The case shows that honoring the earlier appointments and carrying out the court's correction make different resource demands.

At the advertised capacities, include the claimant among the applicants and retain all nine prior appointments. Retention is feasible: the existing assignment accommodates five incumbents in open positions and four in OBC positions.

Under ordinary O\&A, the five open positions go to the two higher-scoring OBC incumbents and three of the five incumbents who were not eligible for the reserved category. RRB must instead retain all five reserve-ineligible incumbents in the open category, with the four OBC incumbents in the reserved category. Proposition~\ref{prop:openadjustment} thus implies that the minimal adjustment necessary for retention involves replacing two people in ordinary open selection. This is a shift in who occupies an open position, not a requirement for two more positions. Nor does it require removal of the two OBC incumbents;
under RRB they retain their prior reserved appointments. The two reserve-ineligible incumbents rejected by ordinary choice are the additional recipients protected by RRB. Proposition~\ref{prop:legacybound} implies that at most two recipients selected by ordinary choice are displaced, with equality when the two rules select the same number of individuals.

At these capacities, there are no positions available for a new candidate. The court's ruling goes further. 
Maintaining all nine appointments and admitting the claimant requires at least ten appointments; retaining all five reserve-ineligible incumbents while also counting the two named OBC incumbents in the open category requires seven open positions instead of five. 
Retention feasibility therefore does not imply feasibility of the court's requested reassignment and new appointment. The judgment does not specify how to provide the additional resources.\footnote{For any real number $x$, write $[x]_+=\max\{x,0\}$. If the target is a set of appointments without further category restrictions, the shortfall $[|K|-\rho_q(K)-q^o]_+$ is the minimal addition to open capacity that allows feasibility. If we hold the four reserved positions fixed, maintaining these nine incumbents and admitting the claimant requires one extra open position if no incumbent's category is prescribed. Requiring that both named OBC incumbents count as open instead requires two extra open positions.}

\subsubsection{Category Locking in Rajiv Mittal}\label{app:rajivdetails}

The treatment of an earlier category assignment as settled is illustrated by \emph{Rajiv Mittal v. Maharshi Dayanand University}. Rohtak Medical College distributed open and backward-class seats according to rank in a common entrance examination. 
Sunil Yadav (ranked~62) secured a seat in the reserved category through the first round of counseling. After open-category seats became vacant, his rank was reached in the second round, but he was barred from reconsideration since he already had an admission. The open seat went to Rajiv Mittal (ranked~66). 
The next backward-class candidate challenged the outcome, arguing that if Sunil's admission had been counted in the open category, a reserved seat would have been available for the claimant.

The High Court ordered that Sunil's admission be counted in the open category and that the claimant be admitted, transferring Rajiv to another college. The Supreme Court overturned this remedy and upheld the treatment of Sunil's earlier category assignment as settled.\footnote{Supreme Court of India, \href{https://indiankanoon.org/doc/513378/}{\emph{Rajiv Mittal v. Maharshi Dayanand University}}, November~28, 1997, (1998) 2 SCC~402. The judgment endorses open consideration before reserved consideration within a counseling round; the dispute concerns reconsideration across rounds.}

The timing of the decision matters for the appointments that must be honored. Just before the challenged offer in the second round, Sunil was an incumbent at Rohtak while Rajiv had no appointment right there.
If Sunil's admission alone was protected, RRB could count it in the open category without altering his college or offering him a second admission, and offer the reserved position to the next eligible applicant. In comparison to locking Sunil's category, this correction admits the reserve-eligible applicant instead of Rajiv, with the same total number of admissions. But once Rajiv has been admitted, protecting his admission adds another 
incumbent whose Rohtak appointment cannot be honored by transferring him elsewhere. The High Court's retrospective remedy therefore cannot be equated with the prospective RRB correction.

\subsubsection{Reattribution and New Appointments in Uttar Pradesh}\label{subsec:otherretention}
The 1,252 reattributed incumbents in the \emph{Pramod Kumar Singh} recruitment in Section~\ref{app:chilecontinuation} included 1,118 OBC, 132 SC, and two ST appointees. These changes were distinct from the 3,295 new appointments made in the same exercise. The new OBC, SC, and ST appointments exceeded the additional opportunities originally assigned to those categories by exactly the numbers reattributed to open. Counting the reattributions as new appointments would count the incumbents twice.\footnote{\emph{Pramod Kumar Singh}, paragraphs~2(G)--(H), cited in main-text footnote~\ref{fn:pramod}. The 3,295 new appointments comprised 654 open, 1,849 OBC, 733 SC, and 59 ST appointees.}

\subsection{University of Otago}\label{app:otagodetails}
The Division of Health Sciences at the University of Otago in New Zealand allows an applicant to accept one program's offer while remaining on 
other programs' waiting lists; accepting a later offer requires canceling the earlier place.\footnote{University of Otago, \href{https://www.otago.ac.nz/healthsciences/students/professional/faqs-guidelines-for-admission-to-health-sciences-professional-programmes}{\emph{FAQs: Guidelines for Admission to Health Sciences Professional Programmes}}, under ``How do the waitlists work?''} Committees determine pathway numbers and may determine their sequence under the divisional policy. The General pathway is restricted to applicants not applying through a Priority pathway, so there is no universally available category and the formal open capacity is zero.\footnote{University of Otago, \href{https://www.otago.ac.nz/healthsciences/students/admissions-policy-health-sciences-professional-programmes}{\emph{Admissions Policy for Health Sciences Professional Programmes}}, revised November~26, 2025, definitions and selection procedures.} Radiation Therapy ranks suitable applicants across admission categories by academic assessment after its interview threshold. Dental Technology uses a common ranking 
combining academic assessment and assessed personal suitability. Both publish an equity-first sequence ending with General applicants.\footnote{University of Otago, admission guidelines for \href{https://www.otago.ac.nz/healthsciences/students/professional/radiation-therapy}{\emph{Radiation Therapy}} and \href{https://www.otago.ac.nz/healthsciences/students/professional/dental-technology}{\emph{Dental Technology}}, under ``Selection criteria.''}

\end{document}